\documentclass[]{fairmeta}
\usepackage{amsthm,amsmath,amssymb}
\usepackage{graphicx}
\usepackage{natbib}
\usepackage[hidelinks,breaklinks]{hyperref}
\usepackage{subfig}
\newtheorem{theorem}{Theorem}[]

\newtheorem{remark1}[theorem]{Remark}

\DeclareMathOperator{\E}{\mathop{}\mathbb{E}}

\title{Cumulative differences between paired samples}
\author[1]{Isabel Kloumann}
\author[1]{Hannah Korevaar}
\author[1]{Chris McConnell}
\author[1]{Mark Tygert}
\author[1]{Jessica Zhao}
\affiliation[1]{Fundamental Artificial Intelligence Research at Meta}
\date{\today}
\correspondence{Mark Tygert at \email{mark@tygert.com}}

\begin{document}

\abstract{The simplest, most common paired samples consist of observations
from two populations, with each observed response from one population
corresponding to an observed response from the other population
at the same value of an ordinal covariate.
The pair of observed responses (one from each population)
at the same value of the covariate is known as a ``matched pair''
(with the matching based on the value of the covariate).
A graph of cumulative differences between the two populations
reveals differences in responses as a function of the covariate.
Indeed, the slope of the secant line connecting two points on the graph
becomes the average difference over the wide interval of values
of the covariate between the two points; that is, slope of the graph
is the average difference in responses.
(``Average'' refers to the weighted average if the samples are weighted.)
Moreover, a simple statistic known as the Kuiper metric summarizes
into a single scalar the overall differences over all values of the covariate.
The Kuiper metric is the absolute value of the total difference in responses
between the two populations, totaled over the interval of values
of the covariate for which the absolute value of the total is greatest.
The total should be normalized such that it becomes the (weighted) average
over all values of the covariate when the interval over which the total
is taken is the entire range of the covariate (that is,
the sum for the total gets divided by the total number of observations,
if the samples are unweighted, or divided by the total weight,
if the samples are weighted).
This cumulative approach is fully nonparametric and uniquely defined (with only
one right way to construct the graphs and scalar summary statistics),
unlike traditional methods such as reliability diagrams or parametric
or semi-parametric regressions, which typically obscure significant differences
on account of their parameter settings.}

\maketitle

\vspace{-.1em}

\section{Introduction}
\label{intro}

Analyzing the differences between two populations often involves data
of paired observations: each response and corresponding value
of an ordinal covariate, both observed for one of the populations,
are accompanied by a response from the other population for the same value
of the covariate. Such a pair of responses at the same value for the covariate
is known as a ``matched pair.'' This paper presents graphical methods
and scalar summary statistics for detecting differences.
The graphs reveal differences as a function of the covariate.
The summary statistics condense the graphs into scalar metrics
measuring the overall differences across all values of the covariate.
Appendix~\ref{multidim} reviews methodology from~\cite{tygert_multidim}
for extending the analysis from a single scalar covariate
to multiple covariates (the earlier work concerned unpaired samples,
but turns out to apply just as well to the case of paired samples
treated in the present paper).
Matched pairs are common both in observational studies,
as reviewed by~\cite{rubin}, \cite{caliendo-kopeinig}, \cite{austin},
and others, and in designed experiments such as randomized controlled trials,
as reviewed by Section 3.3 of~\cite{cox-reid}, \cite{bai-romano-shaikh},
and others.

A classic method for analyzing paired samples is to calculate
the weighted average difference in responses between the populations;
see, for example, Chapter 4, ``Analysis of matched pairs using sample means,''
of~\cite{brown-hollander}. (The weighted average is just the usual arithmetic
average if the samples are unweighted or uniformly weighted.) The methods
of the present paper are an extension. Specifically, the scalar metric proposed
below that generalizes the classical statistic of~\cite{kuiper} is always
greater than or equal to the absolute value of the weighted average difference
in responses, so can detect differences to which the weighted average alone
is blind. The proposed Kuiper metric is simply the absolute value
of the total weighted difference in responses, totaled over the interval
of values of the covariate for which the total has the greatest magnitude.
Of course, if the interval of values of the covariate for which the total
has the greatest magnitude is the entire range of values of the covariate,
then the Kuiper metric is the same as the absolute value
of the weighted average difference. (The totals taken here should be normalized
by the total number of observations, in the case that the observations
are unweighted, or by the total weight, in the case that the samples
are weighted.)
Using the interval for which the total has the greatest magnitude
avoids cancellation between positive and negative differences
that can occur in the weighted average over the entire range of values
of the covariate. Such cancellation can arise when differences
at high values of the covariate trade off against differences at low values
of the covariate, for instance.

The present paper assumes that the covariate is ordinal, that is,
that the values of the covariate have a natural ordering.
The ordering need not be perfect, but the ordering should be better
than purely random.
The weighted average difference in responses ignores the ordering,
as do several other classical methods, such as the sign test
or the signed-rank test of~\cite{wilcoxon};
details are available, for example, from~\cite{hollander-wolfe-chicken}
or Chapter 12, ``Analysis of matched pairs using ranks,''
of~\cite{brown-hollander}.

Classic methods that can take advantage of the ordering of the values
for the ordinal covariate include regression with parametric models,
as reviewed, for example, by~\cite{draper-smith}
or Chapter 7, ``Logistic regression for matched case-control studies,''
of~\cite{hosmer-lemeshow-sturdivant}.
In contrast, the methods of the present paper are fully nonparametric.
Parametric models tend to be more statistically powerful than nonparametric
methods when the models are known to be valid, true, and correct.
However, the nonparametric methods of the present paper are always valid
and have no parameters to tune, making the methods especially easy to use
and trust, with no possibility for fudging results (whether intentionally
or inadvertently). The methods of the present paper are thus complementary
to parametric regression.

Section~\ref{results} below extensively compares the nonparametric methods
of Section~\ref{methods} to what are the most popular
among all classical methods that take advantage of the ordering:
the traditional semi-parametric methods known as ``reliability diagrams''
or ``calibration plots'' reviewed in Subsection~\ref{reliadia}.
Reliability diagrams histogram the responses of both populations separately,
partitioning the values of the covariate into ``bins''
(also known as ``buckets''). Being semi-parametric, reliability diagrams
are subject to less manipulation and outright fudging
than parametric regression. Yet the comparisons of Section~\ref{results}
illustrate many problems with reliability diagrams relative
to the proposed cumulative graphs.
The traditional approach depends on the choice of bins,
and that choice often turns out to influence the inferred results dramatically.
The cumulative approach of Section~\ref{methods} below avoids the problems.
Section~\ref{results} demonstrates the advantages
over the simple weighted average discussed above, too.

The advantages of the cumulative approach have been pointed out earlier,
including by~\cite{arrieta-ibarra-gujral-tannen-tygert-xu}
and~\cite{tygert_multidim,tygert_full,tygert_two,tygert_pvals}.
The present paper merely realizes the advantages of the cumulative statistics
for paired samples, whereas these earlier works focused on unpaired data.
Much earlier, the work of~\cite{delgado} introduced an important, archetypal
special case of part of what the present paper proposes; namely,
\cite{delgado} investigated the Kolmogorov-Smirnov scalar summary statistic
for paired samples.

The rest of the paper has the following structure:
Section~\ref{methods} develops the cumulative methods
and reviews the traditional reliability diagrams in Subsection~\ref{reliadia}.
Section~\ref{results} illustrates the performance of the methods
on both synthetic and measured data.\footnote{Permissively licensed
open-source software implementing the methods in Python --- software that
also reproduces all figures and statistics reported below --- is available
at \url{https://github.com/facebookresearch/metapaired}}
Section~\ref{conclusion} draws some conclusions.
Appendix~\ref{multidim} reviews the reduction via space-filling curves
of the analysis with multiple covariates to the simpler case
of a single scalar covariate.

\section{Methods}
\label{methods}

This section presents the methodology of the present paper.
Subsection~\ref{notation} sets notational conventions used throughout.
Subsection~\ref{graphs} constructs cumulative graphical methods for analyzing
the differences in responses between two populations,
controlled for a scalar score.
Subsection~\ref{metrics} summarizes the graphs of Subsection~\ref{graphs}
into scalar statistics.
Subsection~\ref{significance} discusses the statistical significance
of the graphs of Subsection~\ref{graphs} and of the metrics
of Subsection~\ref{metrics}.
Subsection~\ref{reliadia} reviews the classical graphical methods
known as ``reliability diagrams.''
The following section, Section~\ref{results}, provides several examples
of the methods of the present section applied to various data sets.

Appendix~\ref{multidim} reviews previously developed methods that enable
the direct application of the methods of the present section
to the analysis of differences in responses
while controlling for multiple covariates
(rather than for only a single scalar covariate).
The reduction of the problem of conditioning on multiple covariates
to the problem of conditioning on a single scalar score
relies on space-filling curves.
The reduction decouples from the methodology for analyzing the case
of a single scalar covariate; Appendix~\ref{multidim} focuses on the reduction,
while the present section focuses on the scalar case,
specifically for data sets with paired samples
(whereas earlier works studied only unpaired data).

\subsection{Notation}
\label{notation}

We consider real numbers $S_1$, $S_2$, \dots, $S_m$,
called ``scores,'' that are the values of a covariate.
We assume without loss of generality that $S_1 < S_2 < \dots < S_m$.
For each score, say score $S_j$, we consider real numbers
$R_j^{(1)}$, $R_j^{(1)}$, \dots, $R_j^{(n_j)}$ and
$Q_j^{(1)}$, $Q_j^{(1)}$, \dots, $Q_j^{(n_j)}$,
called ``responses,'' that are the values of outcomes for two populations.
Thus, the total sample size is
\begin{equation}
n = \sum_{j=1}^m n_j,
\end{equation}
where again $n_j$ is the number of pairs of responses
$(Q_j^{(1)}, R_j^{(1)})$, $(Q_j^{(2)}, R_j^{(2)})$, \dots,
$(Q_j^{(n_j)}, R_j^{(n_j)})$ associated with score $S_j$.
For each score, say score $S_j$, we also consider positive real numbers
$W_j^{(1)}$, $W_j^{(2)}$, \dots, $W_j^{(n_j)}$, called ``weights,''
that weight the contribution of the corresponding pairs
$(Q_j^{(1)}, R_j^{(1)})$, $(Q_j^{(2)}, R_j^{(2)})$, \dots,
$(Q_j^{(n_j)}, R_j^{(n_j)})$ associated with score $S_j$.
If the samples are unweighted, then we set $W_j^{(k)}$ to be a constant
independent of the values of the indices $j$ and $k$.

{\it We assume that the scores and weights are not random
and that the responses are random and all stochastically independent.}

For $j = 1$, $2$, \dots, $m$, the weighted average responses at score $S_j$ are
\begin{equation}
\tilde{R}_j = \frac{\sum_{k=1}^{n_j} R_j^{(k)} \, W_j^{(k)}}
                   {\sum_{k=1}^{n_j} W_j^{(k)}}
\end{equation}
and
\begin{equation}
\tilde{Q}_j = \frac{\sum_{k=1}^{n_j} Q_j^{(k)} \, W_j^{(k)}}
                   {\sum_{k=1}^{n_j} W_j^{(k)}},
\end{equation}
and the corresponding total weight is
\begin{equation}
\tilde{W}_j = \sum_{k=1}^{n_j} W_j^{(k)}.
\end{equation}

We form the cumulative differences
\begin{equation}
\label{cumdiff}
C_k = \frac{\sum_{j=1}^k (\tilde{Q}_j - \tilde{R}_j) \, \tilde{W}_j}
           {\sum_{j=1}^m \tilde{W}_j}
\end{equation}
for $k = 1$, $2$, \dots, $m$.

\subsection{Graphs}
\label{graphs}

We define abscissae to be the accumulated weights
\begin{equation}
A_k = \frac{\sum_{j=1}^k \tilde{W}_j}{\sum_{j=1}^m \tilde{W}_j}
\end{equation}
for $k = 1$, $2$, \dots, $m$.

In a plot of $C_k$ versus $A_k$ for $k = 1$, $2$, \dots, $m$,
the expected value of the slope from $k = j - 1$ to $k = j$ is
\begin{equation}
\E\left[ \frac{C_j - C_{j-1}}{A_j - A_{j-1}} \right]
= \E\left[ \tilde{Q}_j - \tilde{R}_j \right],
\end{equation}
which is simply the expected difference in the weighted average responses
corresponding to score $S_j$.
Thus, {\it the slope of a secant line connecting two points
on the graph of $C_k$ versus $A_k$ becomes the weighted average difference
in responses between the two populations over the long range of scores
between the two points}.

\subsection{Scalar metrics}
\label{metrics}

We define the Kuiper metric to be the range of the graph,
\begin{equation}
D = \max_{0 \le j \le m} C_j - \min_{0 \le j \le m} C_j,
\end{equation}
where $C_0 = 0$ and $C_j$ is defined in~(\ref{cumdiff})
for $j = 1$, $2$, \dots, $m$.
A mathematically equivalent definition of the Kuiper metric is
\begin{equation}
\label{intuitive}
D = \max_{1 \le i \le k \le m}
    \left| \frac{\sum_{j = i}^k (\tilde{Q}_j - \tilde{R}_j) \, \tilde{W}_j}
                {\sum_{j=1}^m \tilde{W}_j} \right|,
\end{equation}
that is, {\it the Kuiper distance $D$ is the absolute value
of the total weighted difference between the two populations' responses,
with the total taken over the interval of indices (or scores)
for which the magnitude of the total is greatest}.

Another popular statistic is the Kolmogorov-Smirnov metric,
\begin{equation}
E = \max_{1 \le j \le m} |C_j|,
\end{equation}
which unfortunately lacks the intuitive interpretation of the Kuiper metric
that~(\ref{intuitive}) expresses.

The average weighted difference over all the scores (also known as
the ``average treatment effect'') is the value of $C_m$. Thus,
the average weighted difference is never greater than the Kuiper metric
and never less than the negative of the Kuiper metric. And
the average weighted difference is never greater
than the Kolmogorov-Smirnov metric and never less than the negative
of the Kolmogorov-Smirnov metric.
Cancellation of positive and negative differences
in the average weighted difference can mask significant variations
that the Kuiper and Kolmogorov-Smirnov statistics can capture.

\subsection{Statistical significance}
\label{significance}

Under the null hypothesis that the expected value of $\tilde{Q}_j$
is equal to the expected value of $\tilde{R}_j$
for all $j = 1$, $2$, \dots, $m$,
the sequence $C_1$, $C_2$, \dots, $C_m$ is a driftless random walk
with variance for the increment $C_j - C_{j-1}$ being the expected value of
$(\tilde{Q}_j - \tilde{R}_j)^2 \, (\tilde{W}_j)^2
/(\sum_{k=1}^m \tilde{W}_k)^2$.
The expected variance of $C_m$ is then
\begin{equation}
\label{expectedvar}
\sum_{j=1}^m \E\left[
\frac{(\tilde{Q}_j - \tilde{R}_j)^2 \, (\tilde{W}_j)^2}
     {(\sum_{k=1}^m \tilde{W}_k)^2} \right]
= \E\left[ \sum_{j=1}^m
           \frac{(\tilde{Q}_j - \tilde{R}_j)^2 \, (\tilde{W}_j)^2}
                {(\sum_{k=1}^m \tilde{W}_k)^2} \right].
\end{equation}
The right-hand side of~(\ref{expectedvar}) shows that an unbiased estimator
for the variance of $C_m$ based on the observations is
\begin{equation}
\sigma^2 = \sum_{j=1}^m \frac{(\tilde{Q}_j - \tilde{R}_j)^2 \, (\tilde{W}_j)^2}
                             {(\sum_{k=1}^m \tilde{W}_k)^2};
\end{equation}
displaying at the origin of the plots of Subsection~\ref{graphs}
a triangle whose tip-to-tip height is $4\sigma$ gives
an indication of the variations that can be expected
at roughly the 95\% confidence level when assuming the null hypothesis
that there is no difference in the expected responses
between the populations being compared.

\subsection{Review of reliability diagrams}
\label{reliadia}

The classical graphical method is known as the ``reliability diagram,''
as discussed, for example, by~\cite{brocker-smith}.
A reliability diagram considers $\ell$ bins with boundaries
$B_0 < B_1 < B_2 < \dots < B_{\ell}$,
where $B_0 = -\infty$ and $B_{\ell} = \infty$
(whereas $B_1$, $B_2$, \dots, $B_{\ell - 1}$ are finite real numbers).
We form the weighted averages for each bin:
\begin{equation}
\bar{S}_i = \frac{\sum_{j\,:\,B_{i-1} < S_j \le B_i}
                  \sum_{k=1}^{n_j} S_j \, W^{(k)}_j}
                 {\sum_{j\,:\,B_{i-1} < S_j \le B_i}
                  \sum_{k=1}^{n_j} W^{(k)}_j},
\end{equation}
\begin{equation}
\label{averager}
\bar{R}_i = \frac{\sum_{j\,:\,B_{i-1} < S_j \le B_i}
                  \sum_{k=1}^{n_j} R^{(k)}_j \, W^{(k)}_j}
                 {\sum_{j\,:\,B_{i-1} < S_j \le B_i}
                  \sum_{k=1}^{n_j} W^{(k)}_j},
\end{equation}
and
\begin{equation}
\label{averageq}
\bar{Q}_i = \frac{\sum_{j\,:\,B_{i-1} < S_j \le B_i}
                  \sum_{k=1}^{n_j} Q^{(k)}_j \, W^{(k)}_j}
                 {\sum_{j\,:\,B_{i-1} < S_j \le B_i}
                  \sum_{k=1}^{n_j} W^{(k)}_j}
\end{equation}
for $i = 1$, $2$, \dots, $\ell$.
The reliability diagram is then the graph of $\bar{Q}_i$ versus $\bar{S}_i$
plotted together with the graph of $\bar{R}_i$ versus $\bar{S}_i$,
both for $i = 1$, $2$, \dots, $\ell$.

The traditional choice for the boundaries of the bins is to require
$B_2 - B_1 \approx B_3 - B_2 \approx \dots \approx B_{\ell-1} - B_{\ell-2}$.
Another sensible choice is to ensure that the sum of the squares of the weights
associated with the observations in any of the bins,
divided by the square of the sum of the weights in that bin,
be similar for all bins. The latter choice ensures that noise
due to statistical variations in the responses gets averaged away
about the same amount in every bin when forming the averages
in~(\ref{averager}) and~(\ref{averageq}).
Calculating the boundaries of the bins for this latter choice is possible
via the algorithm described in Remark~5 of~\cite{tygert_full}.

\section{Results and discussion}
\label{results}

This section illustrates the methods of the previous section
by analyzing several data sets using the methods.\footnote{Permissively
licensed open-source Python scripts that reproduce all figures and statistics
reported here are available at
\url{https://github.com/facebookresearch/metapaired}}
Subsection~\ref{synthetic} applies the methods to synthetic data
for which the ground truth is known by construction.
Subsection~\ref{kddcup} applies the methodology to a classic data set
from a direct-mail fundraising campaign, combining the methods
of the previous section with those reviewed in Appendix~\ref{multidim}.
Subsection~\ref{acs} applies the methodology to data
from the United States Census Bureau, again leveraging Appendix~\ref{multidim}
in combination with the methods of Section~\ref{methods}.

\subsection{Synthetic examples}
\label{synthetic}

This subsection presents Figures~\ref{ex0}--\ref{ex2}.
These figures display the results of processing data for which the ground truth
is known by construction:

In Figures~\ref{ex0}--\ref{ex2}, the top rows of plots
display the graphs of cumulative differences from Subsection~\ref{graphs}.
In these top rows, the graphs on the left use the random observations,
while the graphs on the right use the exact expected values
of the corresponding observations (thus displaying the ground truth
with no noise from random observations).
The second rows of plots display reliability diagrams each with 10 bins.
The third rows of plots display reliability diagrams each with 100 bins.
The reliability diagrams on the left use the classical binning,
with bins roughly equispaced along the scores.
The reliability diagrams on the right choose bins such that the ratio
of the sum of the squares of the weights in the bin
to the square of the sum of the weights in the bin
is roughly the same for every bin (this ensures that the variance
in each bin is similar).
The fourth, bottom rows of plots are reliability diagrams
constructed from the exact expected values of the corresponding observations
(thus displaying the ground truth for the reliability diagrams).

Figures~\ref{ex0}--\ref{ex2} vary the number $m$ of distinct values
among all the scores ($S_1$, $S_2$, \dots, $S_m$), with $m =$ 1,000
in Figure~\ref{ex0}, $m = n =$ 4,000 in Figure~\ref{ex1},
and $m =$ 400 in Figure~\ref{ex2}. Figures~\ref{ex0}--\ref{ex2}
all set the sample size $n$ to be 4,000, while varying the number $m$
of unique values among the scores for the $n$ pairs of observations.

The present subsection uses the method of Section~2.4 of \cite{tygert_pvals},
which is an alternative to ensuring uniqueness of the scores via perturbing
the scores at random. Subsection~\ref{notation} above summarizes the method.
This method used in the present subsection constructs a new weighted data set
such that the associated graphs of cumulative weighted differences coincide
with the graphs for the randomly perturbed version at the corresponding values
of the scores.

In every figure, the reliability diagrams obscure variations
that are readily apparent in the cumulative graphs;
the captions to the figures discuss in greater detail.
The reliability diagrams with fewer bins are much less noisy
yet smooth over important variations.
The cumulative graphs manage to capture everything all at once, in contrast.
The average weighted difference is the value of $C_m$,
which is the vertical coordinate at the greatest (rightmost) score
in the cumulative plots. In Figures~\ref{ex0} and~\ref{ex2},
the magnitude of the average weighted difference is small even though there are
significant variations that the Kuiper statistics have no trouble detecting.
In Figures~\ref{ex0} and~\ref{ex1}, the scalar statistics
of Kuiper and of Kolmogorov and Smirnov successfully reflect
the steep slopes in the cumulative graphs;
in Figure~\ref{ex2}, oscillations dampen the effectiveness
of the scalar metrics somewhat.
The captions to the figures discuss the results further.

\subsection{KDD Cup 1998}
\label{kddcup}

This subsection presents Figures~\ref{ex02}--\ref{ex012}.
Appendix~\ref{multidim} reviews the Hilbert curve used
in the present subsection and the following subsection.
Figures~\ref{ex02}--\ref{ex012} consider data from an experiment
that solicited donations during a fundraising campaign:

The 1998 Knowledge Discovery and Data-mining (KDD) Cup was a competition
based on a data set regarding donations to a national veterans organization
in the U.S.\footnote{The data from the 1998 KDD Cup is available
at \url{https://kdd.org/kdd-cup/view/kdd-cup-1998/Data}}
The data set reports the donations made in different years,
including 1995 and 1996; we total all the donations that each individual made
in each year and retain the 63,826 individuals who donated at least once.
We exclude all those with missing ages (``AGE'' in the data),
those with missing mean household incomes for the Census block
in which the donor lives (``IC3''), and those missing the fractions
of householders in the corresponding Census block who are married (``MARR1''),
so that we can use these three as covariates.
In the data set, ages are integers, average household incomes
are integer multiples of a hundred dollars, and fractions married
are integer percentages. All covariates get normalized to the unit interval
$[0, 1]$ prior to ordering their values via the Hilbert curve
reviewed in Appendix~\ref{multidim}.
We perturbed the incomes by about one part in a hundred million
in order to ensure that they would be distinct;
the previous subsection omits such random perturbation
in order to illustrate use of the method of Section~2.4 of \cite{tygert_pvals},
which is an alternative to random perturbations.
The random perturbations guarantee that the scores are all unique,
so the total number of scores is equal to the number of individuals, that is,
$m = n =$ 63,826.

The responses are the total amounts in dollars that every individual donated
in each year (1995 and 1996), with those who never donated in either year
removed. To account for inflation and focus on variation with respect
to the covariates, we normalized every individual's total donation in each year
by dividing by the mean total donation, with the mean averaging
over all individuals in that year. In each year, the mean total donation,
with the mean averaging over all individuals in that year,
is of course equal to 1 following this normalization.
The mean difference between 1995 and 1996
in the individuals' normalized donations,
with the mean taken over all individuals, is therefore 0 (as $1 - 1 = 0$).

Figures~\ref{ex02}, \ref{ex20}, and \ref{ex012} condition on the covariates
(1) income and age, (2) age and income,
and (3) age, fraction married, and income, respectively.
(Controlling for covariates via the Hilbert curve displays some dependence
on the order in which we condition on the covariates,
resulting in Figures~\ref{ex02} and~\ref{ex20} being different,
despite conditioning on the same pair of covariates, age and income.)
The top rows of the figures plot the cumulative differences in responses
between 1995 and 1996 (with each year normalized such that the mean response,
averaged over all individuals, is 1).
The top rows of Figures~\ref{ex02} and~\ref{ex20} also display
points in the two-dimensional plane of the covariates,
where the points correspond to the individuals' values for the covariates.
The grayscale shading of each point indicates the score assigned
via parameterization with the Hilbert curve. Interactive display
of the covariates' values corresponding to given scores is available
in the software accompanying the present paper.\footnote{Open-source software
with interactive plots is available at
\url{https://github.com/facebookresearch/metapaired}}
The middle rows of Figures~\ref{ex02}--\ref{ex012} give reliability diagrams
each with 10 bins.
The bottom rows of the figures give reliability diagrams each with 100 bins.
The leftmost reliability diagrams employ the standard binning,
partitioning the scores roughly equally, with bins equispaced along the scores.
The rightmost reliability diagrams bin such that the ratio
of the sum of the squares of the weights in the bin
to the square of the sum of the weights in the bin
is roughly the same for every bin, so that the variance in each bin is similar.

All three figures reveal roughly the same differences between responses
in 1995 and in 1996. In principle Figure~\ref{ex012}
could have been very different from Figures~\ref{ex02} and~\ref{ex20},
since Figure~\ref{ex012} conditions on an extra covariate,
but apparently the extra conditioning made only so much difference.
The reliability diagrams on the right look more helpful
than those on the left; those on the left use the traditional binning,
with the widths of the intervals of scores
that the bins span being roughly the same for all bins.
The cumulative diagrams appear to be the most helpful,
or at least the clearest.
Moreover, the mean difference is the value of $C_m$,
which is the vertical coordinate at the greatest (rightmost) score
in the cumulative diagrams. In Figures~\ref{ex02}--\ref{ex012},
the mean difference is zero (due to the normalization that adjusts
for inflation and other effects), despite the existence
of significant variations that the Kuiper metrics easily identify.
More detailed discussion is available in the captions to the figures.

\subsection{American Community Survey}
\label{acs}

This subsection presents Figures~\ref{san_mateo} and~\ref{counties}.
Appendix~\ref{multidim} reviews the Hilbert curve used
in the present subsection and the previous subsection.
Figures~\ref{san_mateo} and~\ref{counties} consider data
from the 2019 American Community Survey (ACS) of the U.S.\ Census Bureau:

The ACS reports responses to a survey of households from across the United
States together with weights that make the sample be more
representative.\footnote{The data from the 2019 American Community Survey
of the United States Census Bureau is available for download at
\url{https://www.census.gov/programs-surveys/acs/microdata.html}}
As the sampling in the ACS is weighted, our data analysis restricts attention
to only households whose weights (``WGTP'' in the microdata) are strictly
positive; this also filters out group quarters. For covariates, we consider
the number of the householder's own children and
the logarithm of the adjusted household personal income
(the adjusted income is the product of ``HINCP'' and ``ADJINC,''
divided by one million, since ``ADJINC'' omits the decimal point
in the integer-valued microdata).
To use the adjusted household personal income,
we retain only those households for which the personal income (``HINCP'')
is strictly positive and for which the adjustment factor to the income
(``ADJINC'') has not flagged the income data as missing;
the latter also filters away vacant addresses.
We normalized each covariate to range from 0 to 1 before parameterization
via the Hilbert curve in Appendix~\ref{multidim}.

Figure~\ref{san_mateo} processes the households from the ACS that are in
California's San Mateo County. Figure~\ref{counties} considers a variety
of counties from California, ranging from those with small populations
(such as Napa) to those with large populations (such as Los Angeles).
Both figures compare the presence of a laptop or desktop computer
in the household to the presence of a smartphone. The response from a household
is 1 if the respective computational device is present or 0 if the device is
absent. Thus, a household that has a smartphone but no laptop or desktop
computer contributes $1$ to the sum of cumulative differences,
a household that has a laptop or desktop computer but no smartphone
contributes a $-1$ to the sum of cumulative differences,
and households that either have both a smartphone and a laptop or desktop
or have no computational device at all contribute $0$.
(Technically, the contribution of a household is weighted by the weight
associated with the household in the ACS data and the sums are cumulative
weighted differences.)

Figures~\ref{san_mateo} and~\ref{counties} both condition on the logarithm
of the adjusted household personal income together with the number
of the householder's own children.
The top row of Figure~\ref{san_mateo} plots the cumulative weighted differences
as well as the points in the two-dimensional plane of the covariates,
where the points correspond to the households' values for the covariates.
In the latter, two-dimensional plot, the grayscale shade of each plotted point
indicates the score associated with the Hilbert curve's parameterization.
The other rows in Figure~\ref{san_mateo} present reliability diagrams.
The leftmost diagrams partition the scores roughly equally,
with bins equispaced along the scores.
The rightmost reliability diagrams bin such that the ratio
of the sum of the squares of the weights in the bin
to the square of the sum of the weights in the bin
is about the same for every bin; this ensures that the variance for each bin
be similar to the variance for the other bins.

As in Subsection~\ref{synthetic}, Figures~\ref{san_mateo} and~\ref{counties}
illustrate use of the method of Section~2.4 of \cite{tygert_pvals}, which is
an alternative to ensuring uniqueness of the scores from the Hilbert curve
via random perturbations. The previous subsection, Subsection~\ref{kddcup},
considers the approach via random perturbations. The accompanying software
package\footnote{Open-source software that can automatically reproduce
all results reported throughout the present paper is available at
\url{https://github.com/facebookresearch/metapaired}} specifies at the top
of the associated Python scripts, via a constant Boolean variable,
``RANDOMIZE,'' whether to perturb the Hilbert-curve scores at random.
For the script of the present subsection, RANDOMIZE = False,
while for the script of the previous subsection, RANDOMIZE = True.

The Python script for the present subsection also supports substituting
the number of people in a household for the number of the householder's
own children, as well as including both the number of people and the number
of the householder's children as covariates (all in addition to the other
covariate, namely, the logarithm of the adjusted household personal income).
Since the results look similar for the other combinations of covariates
(albeit not quite as sharp for the other combinations),
the presentation here omits display of the associated extra plots.
Still, the similarity may indicate that the number of people
in the household could be a respectable though imperfect proxy
for the number of the householder's own children;
the latter is what turns out to matter most for discerning the association
with the prevalence of smartphones over laptop or desktop computers.

Indeed, interactive exploration of the plots of cumulative weighted differences
indicates that the prevalence of smartphones over laptop and desktop computers
occurs mainly for lower-income households
with multiple children.\footnote{Open-source software with interactive plots
is available at \url{https://github.com/facebookresearch/metapaired}}
High-income households and households without children have essentially no
higher prevalence of smartphones than of laptops or desktops.
And every imbalance seems to skew toward more smartphones than laptops
or desktops (as opposed to more laptops or desktops than smartphones).

Overall, Figure~\ref{san_mateo} illustrates the advantages of the plot
of cumulative weighted differences over the classical reliability diagrams.
The cumulative plot exhibits a clear, steep jump at scores
corresponding to low-income households with children.
Figure~\ref{counties} shows that similar steep inclines appear
in counties of California other than just San Mateo County, too.
The steep jumps show that a low-income household with children
is more likely to have a smartphone than a laptop or desktop computer,
at least when excluding households which either have both a smartphone
and a laptop or desktop or have none of these computational devices at all.
The captions to the figures discuss the results further.

\section{Conclusion}
\label{conclusion}

Earlier publications demonstrated many advantages of the cumulative statistics
over conventional methods for analyzing differences between unpaired samples.
As shown above, these advantages hold for matched pairs, too.
For instance, the Kuiper metric avoids the cancellation
of positive and negative differences that often afflicts
the simplistic weighted average.
(The Kuiper metric is the absolute value of the total weighted difference
in responses, totaled over the interval of values of the covariate
for which the absolute value is greatest. The simplistic weighted average is
the same as the total taken over the entire range of values of the covariate;
the totals here refer to sums normalized such that they become
the weighted average when totaled over the entire range.
If the samples are unweighted, then the weighted average is just the usual
arithmetic average.) The Kuiper statistic can detect differences
that the weighted average averages away.

The cumulative methods have no parameters to set that the analyst could bungle
or fudge; the cumulative graphs always reveal all significant features
in the differences between the pair of populations being analyzed,
unlike the canonical reliability diagrams. In contrast, the choices of bins
for reliability diagrams often obscure important features,
as illustrated above with both measured data sets and synthetic examples
for which the ground truth is known by construction.
Furthermore, the methods detailed above for a single scalar covariate extend
to the case of multiple covariates (as illustrated already
in Section~\ref{results} above); the following appendix presents
this multivariate extension.

\section*{Acknowledgements}

We would like to thank Kamalika Chaudhuri and Khalid El-Arini
for their support.

\newlength{\vertsep}
\setlength{\vertsep}{.085in}
\newlength{\imsize}
\setlength{\imsize}{.365\textwidth}

\begin{figure}
\begin{centering}

\parbox{\imsize}{\includegraphics[width=\imsize]
                {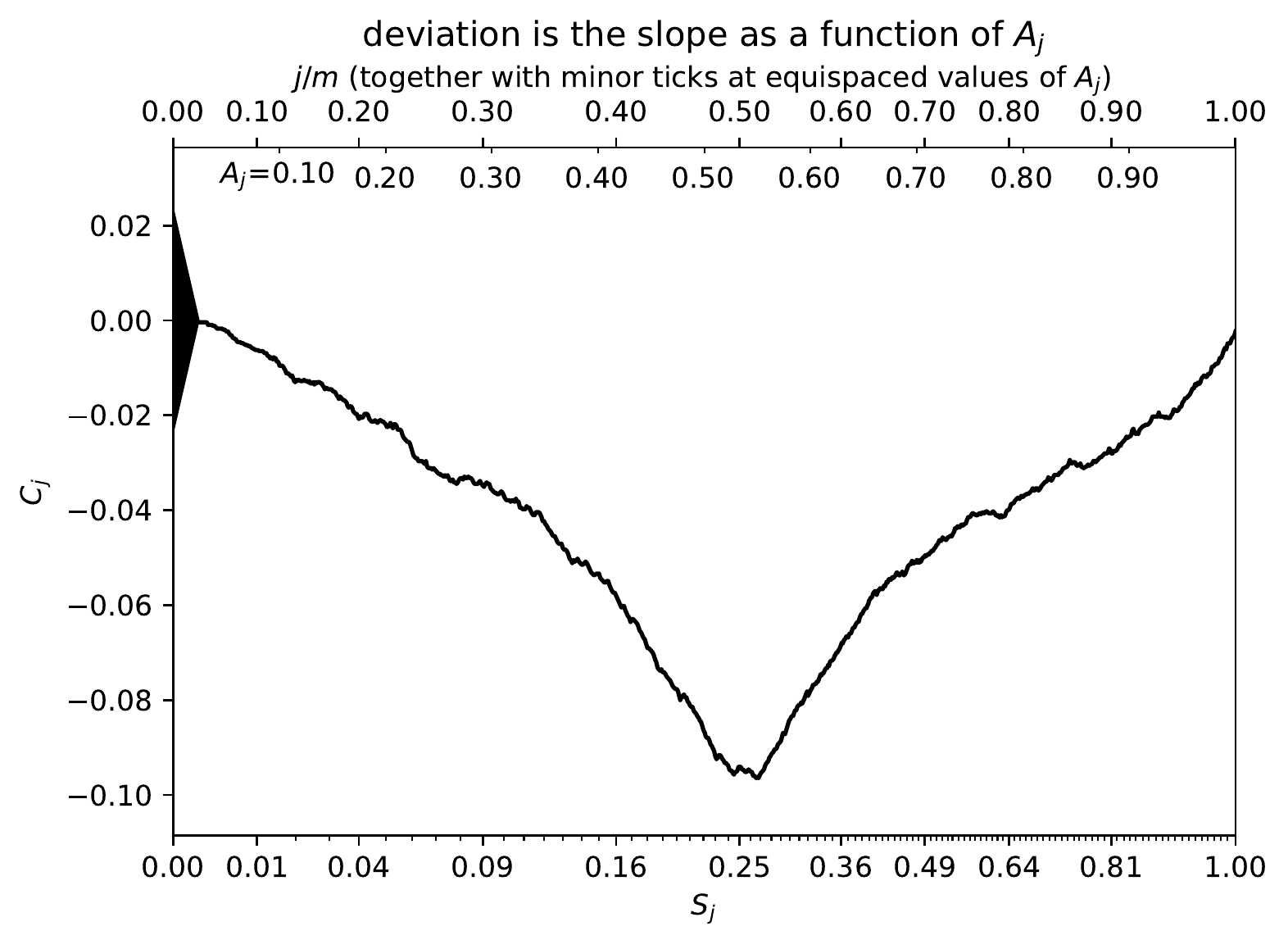}}
\quad\quad
\parbox{\imsize}{\includegraphics[width=\imsize]
                {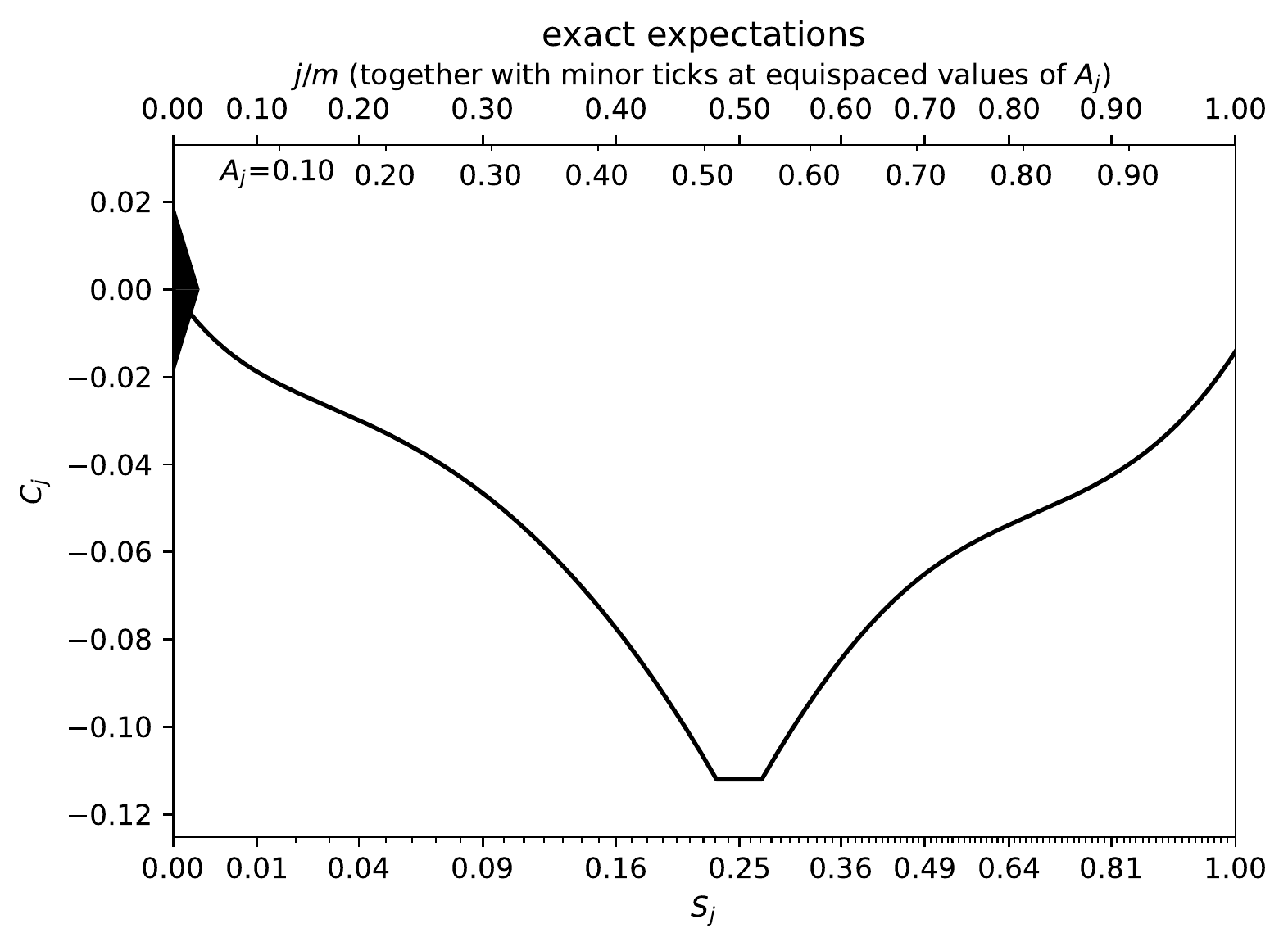}}

\vspace{\vertsep}

\parbox{\imsize}{\includegraphics[width=\imsize]
                {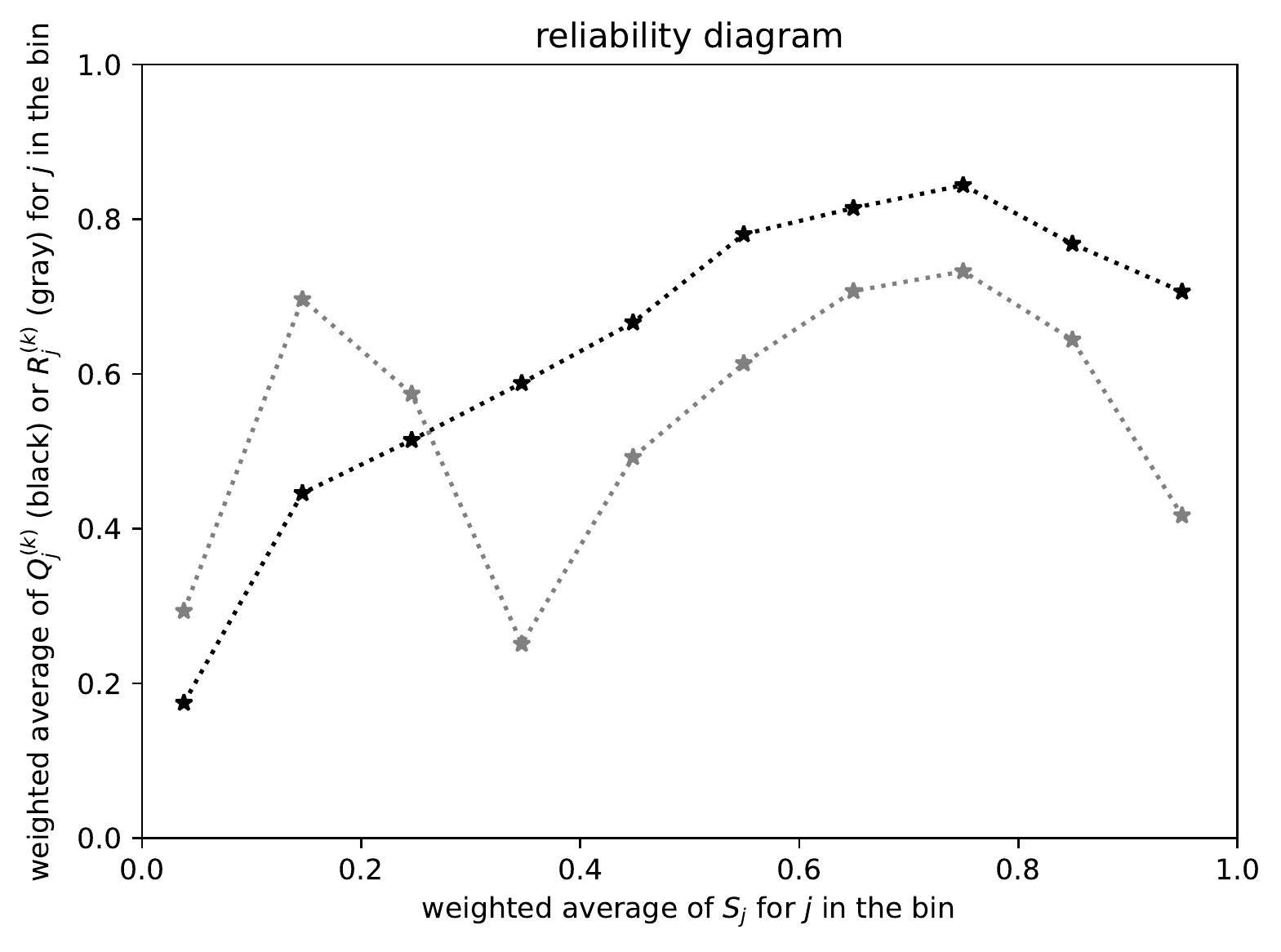}}
\quad\quad
\parbox{\imsize}{\includegraphics[width=\imsize]
                {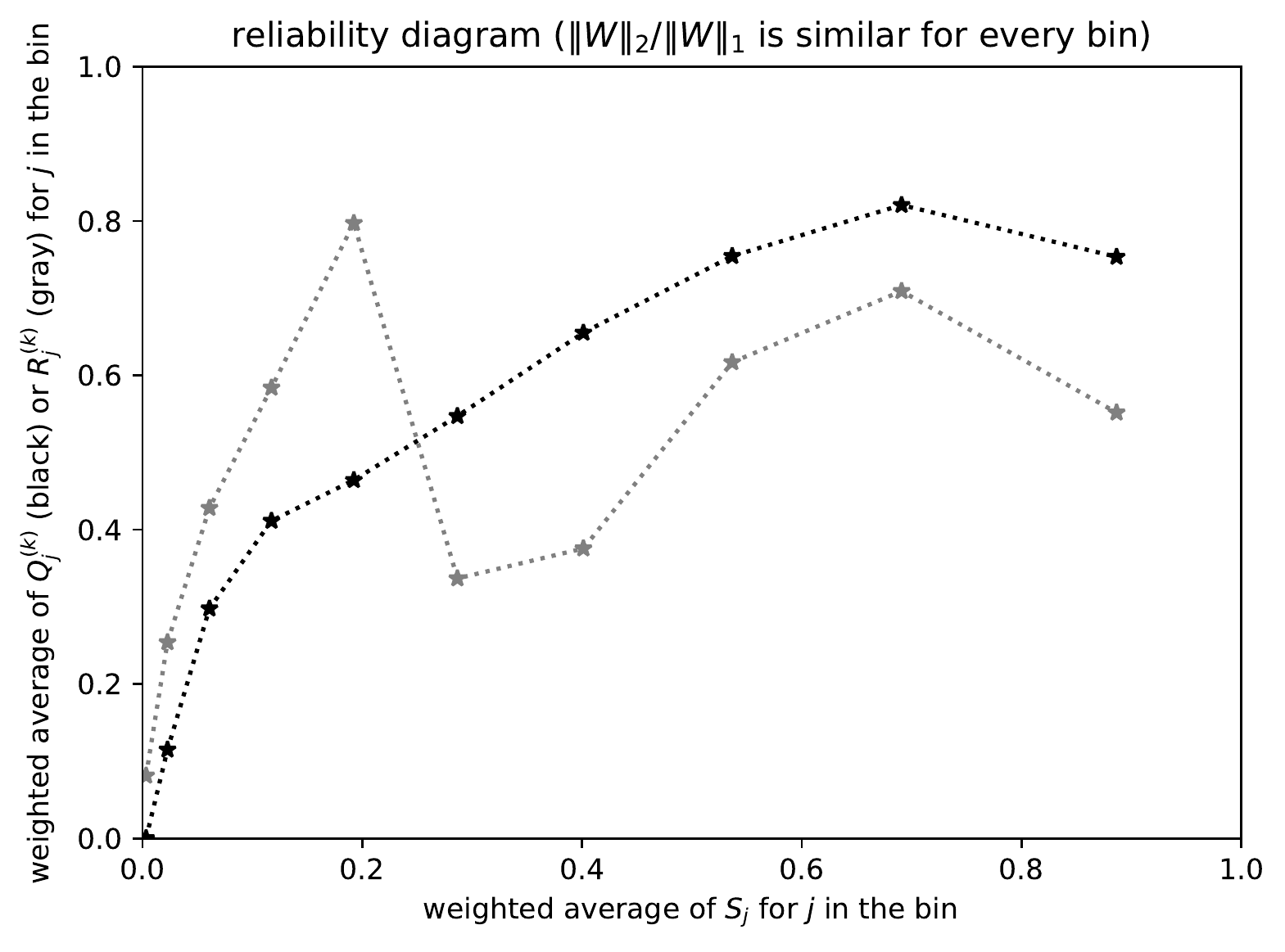}}

\vspace{\vertsep}

\parbox{\imsize}{\includegraphics[width=\imsize]
                {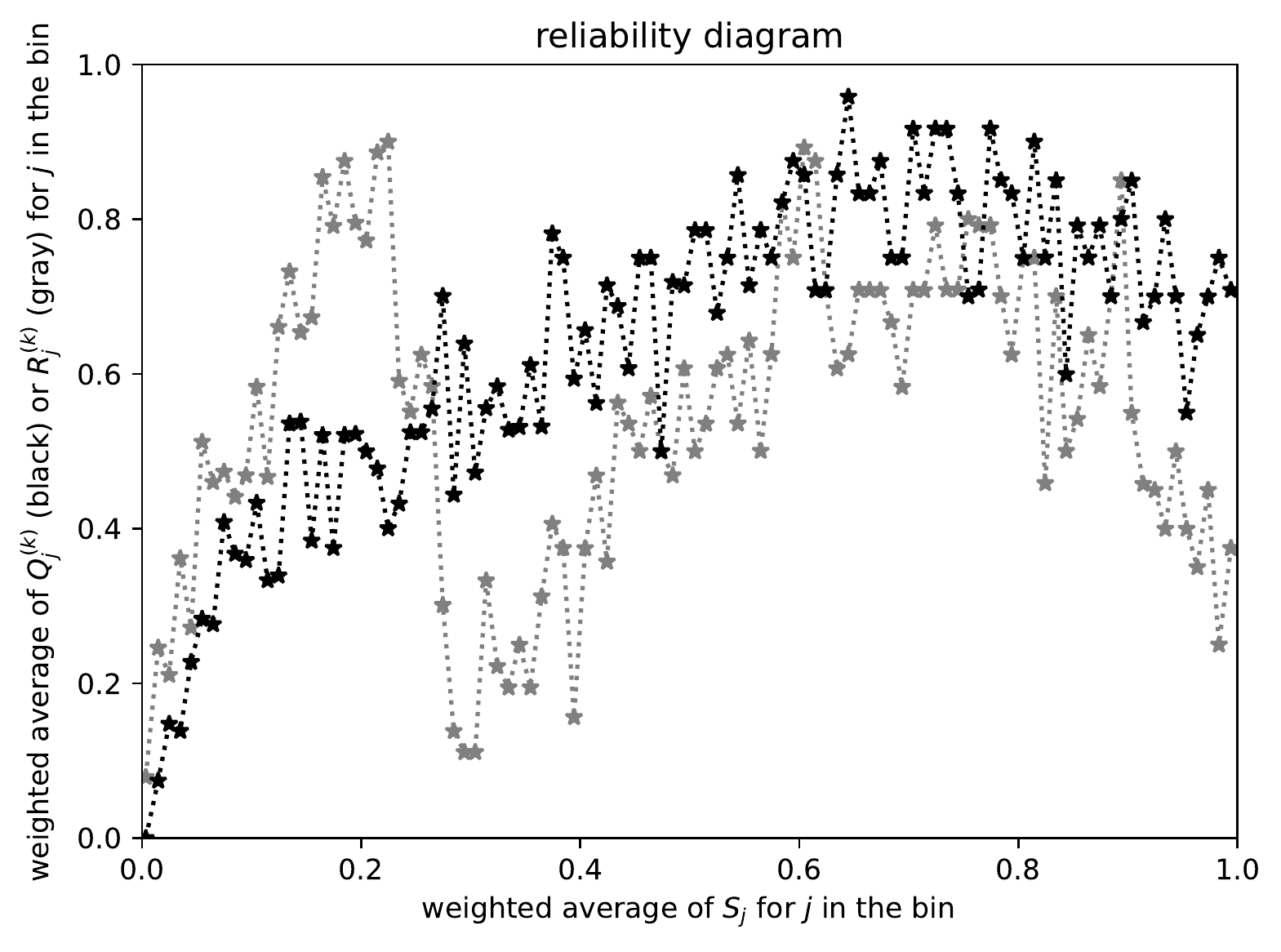}}
\quad\quad
\parbox{\imsize}{\includegraphics[width=\imsize]
                {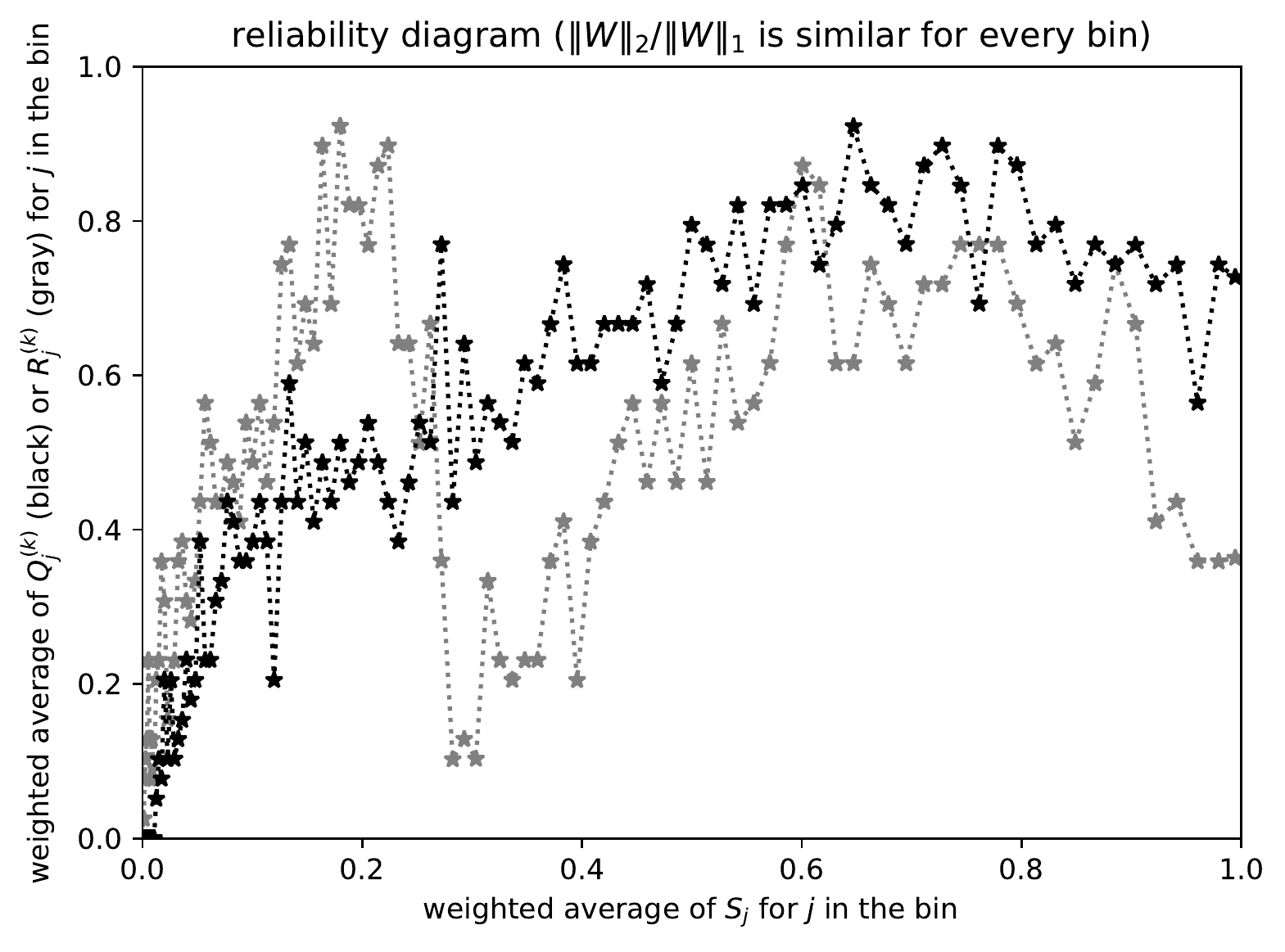}}

\vspace{\vertsep}

\parbox{\imsize}{\includegraphics[width=\imsize]
                {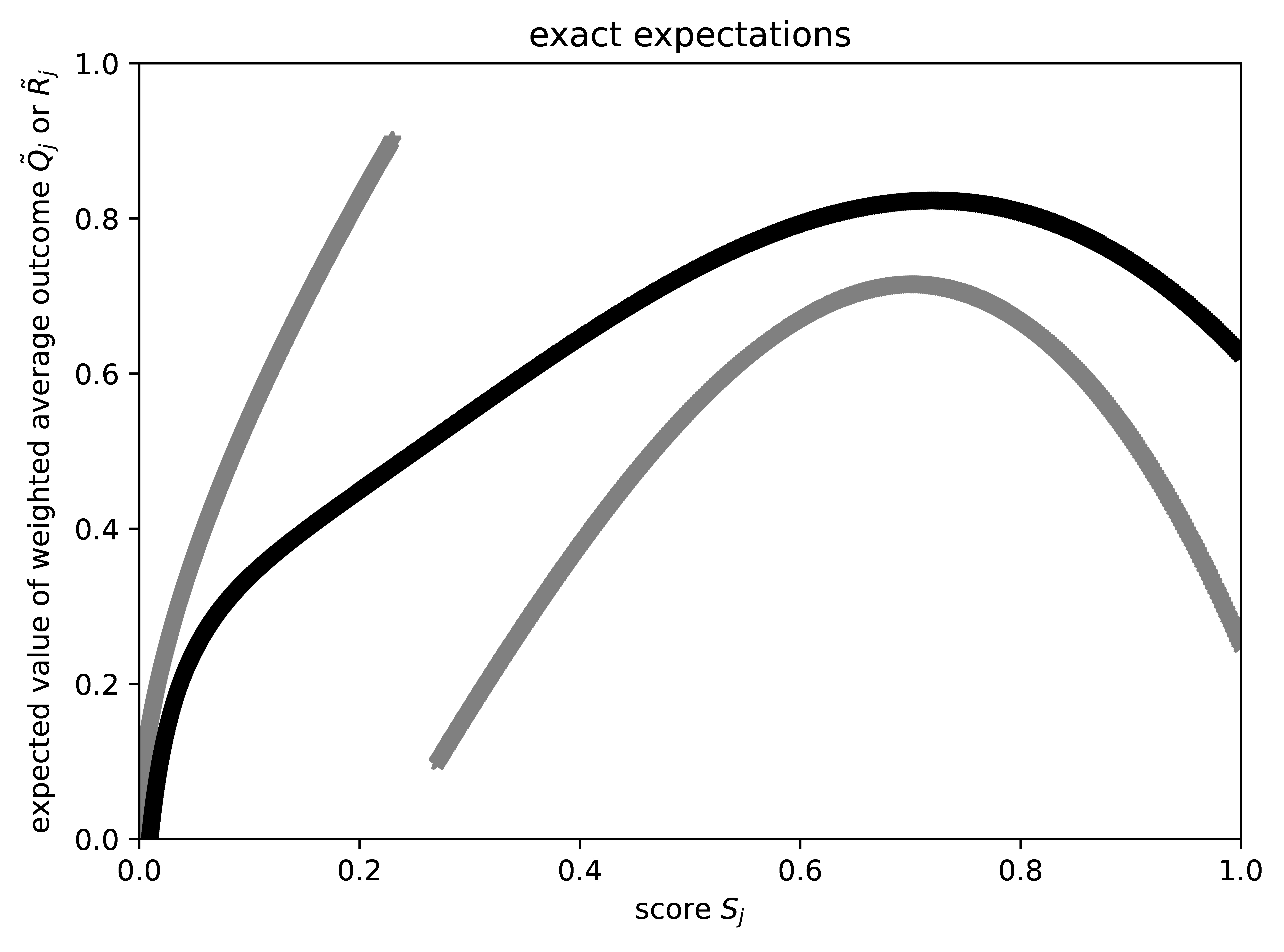}}

\end{centering}
\caption{$m =$ 1,000, $n =$ 4,000;
         Kuiper's statistic is $0.09816 / \sigma = 7.784$,
         Kolmogorov's and Smirnov's is $0.09403 / \sigma = 7.456$.
         The cumulative graphs clearly reveal a narrow range of scores
         (right around the median score) that is flat, that is, where there is
         little difference in the responses between the populations.
         In contrast, discerning the lack of difference in responses
         between the populations in this narrow range is very challenging 
         using only the reliability diagrams --- the cumulative graphs
         are far clearer. The weighted average difference in responses
         is the vertical coordinate $C_m$ at the greatest (rightmost) score
         $S_m$ in the cumulative plots, which is clearly much smaller
         (closer to 0) than the full vertical range of the graph;
         the vertical range
         ($\max_{0 \le j \le m} C_j - \min_{0 \le j \le m} C_j$) is the value
         of the Kuiper metric. Overall, the empirical cumulative graph
         closely matches the exact expected ground-truth cumulative graph.
}
\label{ex0}
\end{figure}

\begin{figure}
\begin{centering}

\parbox{\imsize}{\includegraphics[width=\imsize]
                 {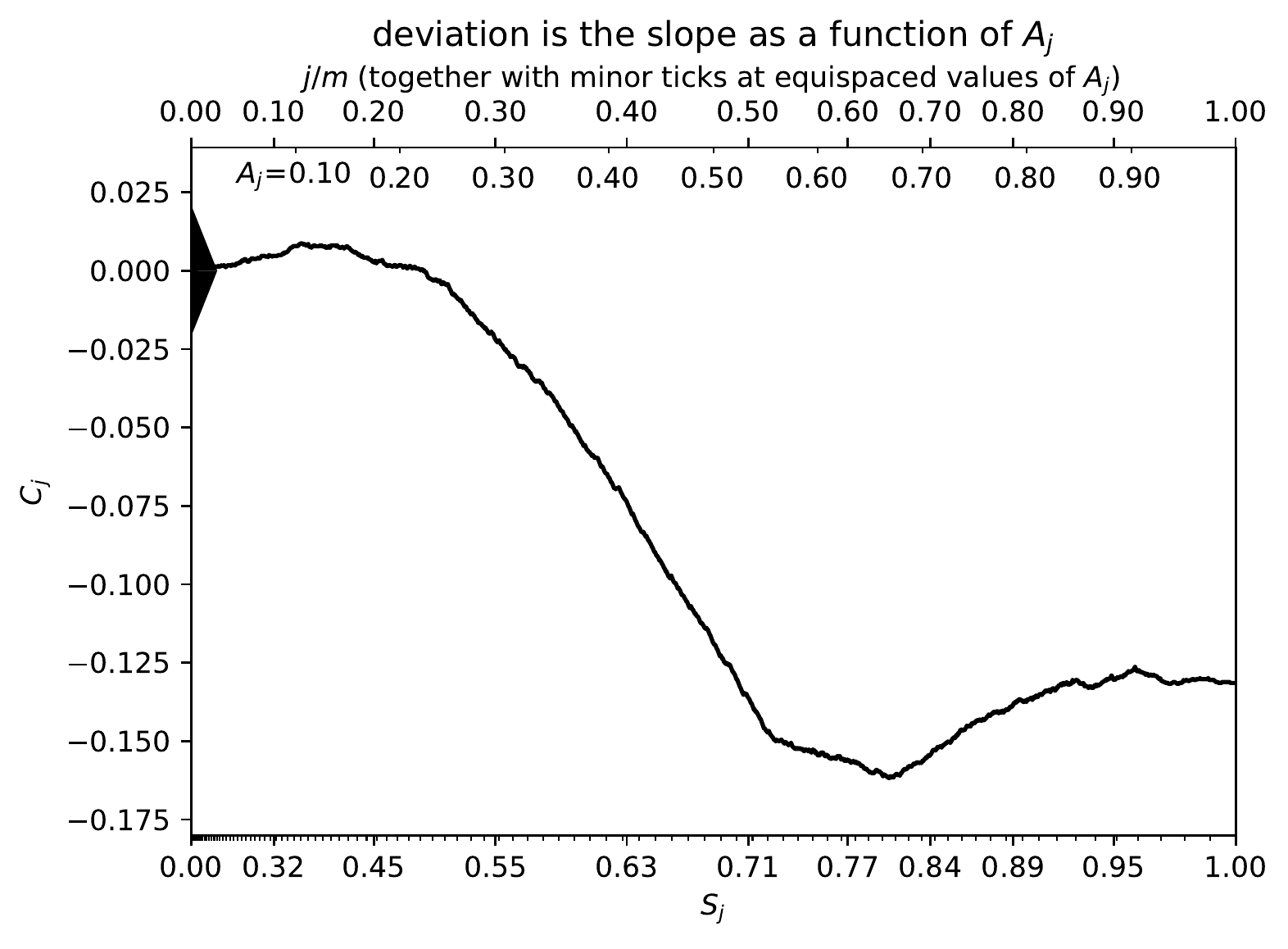}}
\quad\quad
\parbox{\imsize}{\includegraphics[width=\imsize]
                 {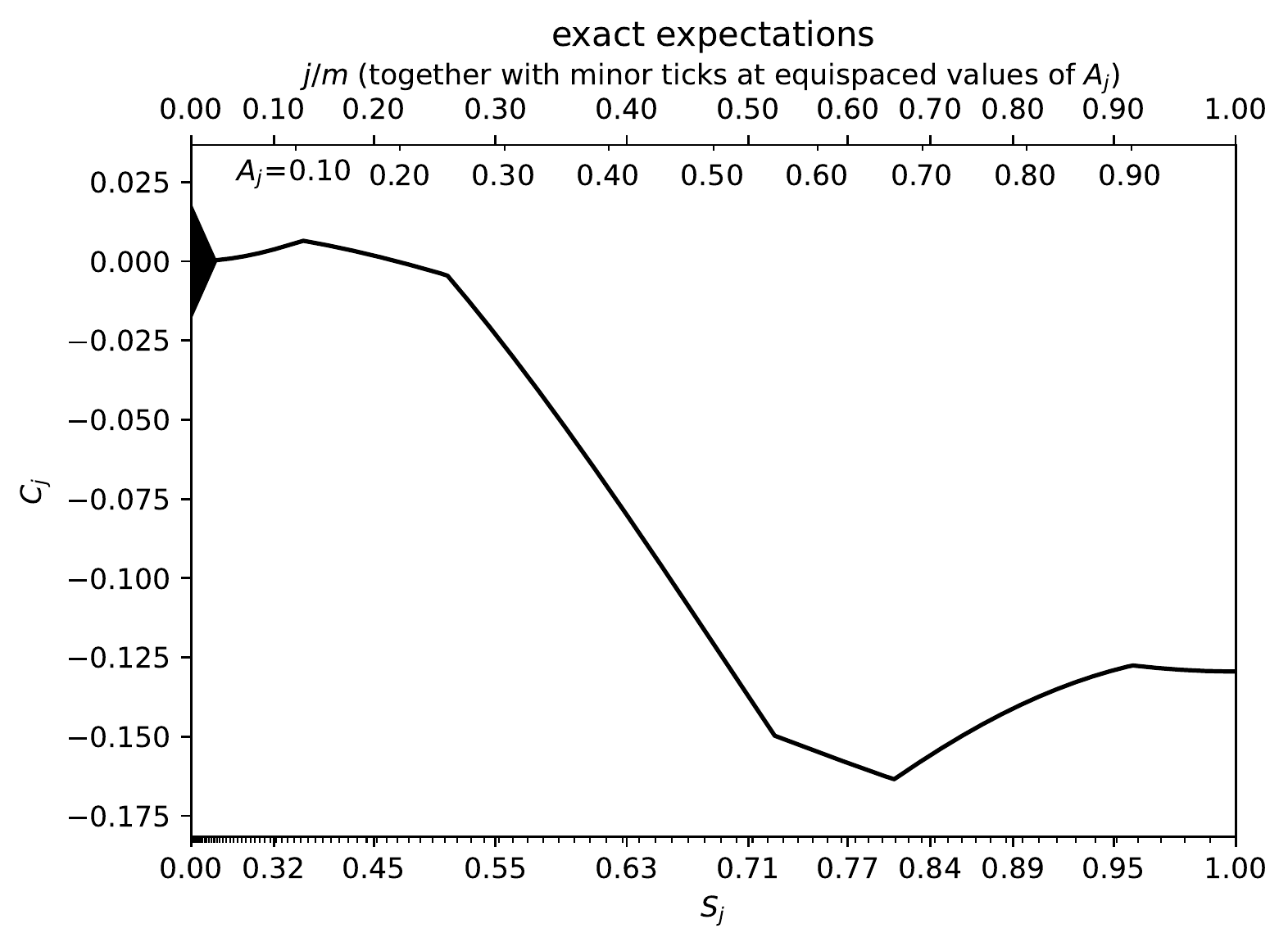}}

\vspace{\vertsep}

\parbox{\imsize}{\includegraphics[width=\imsize]
                 {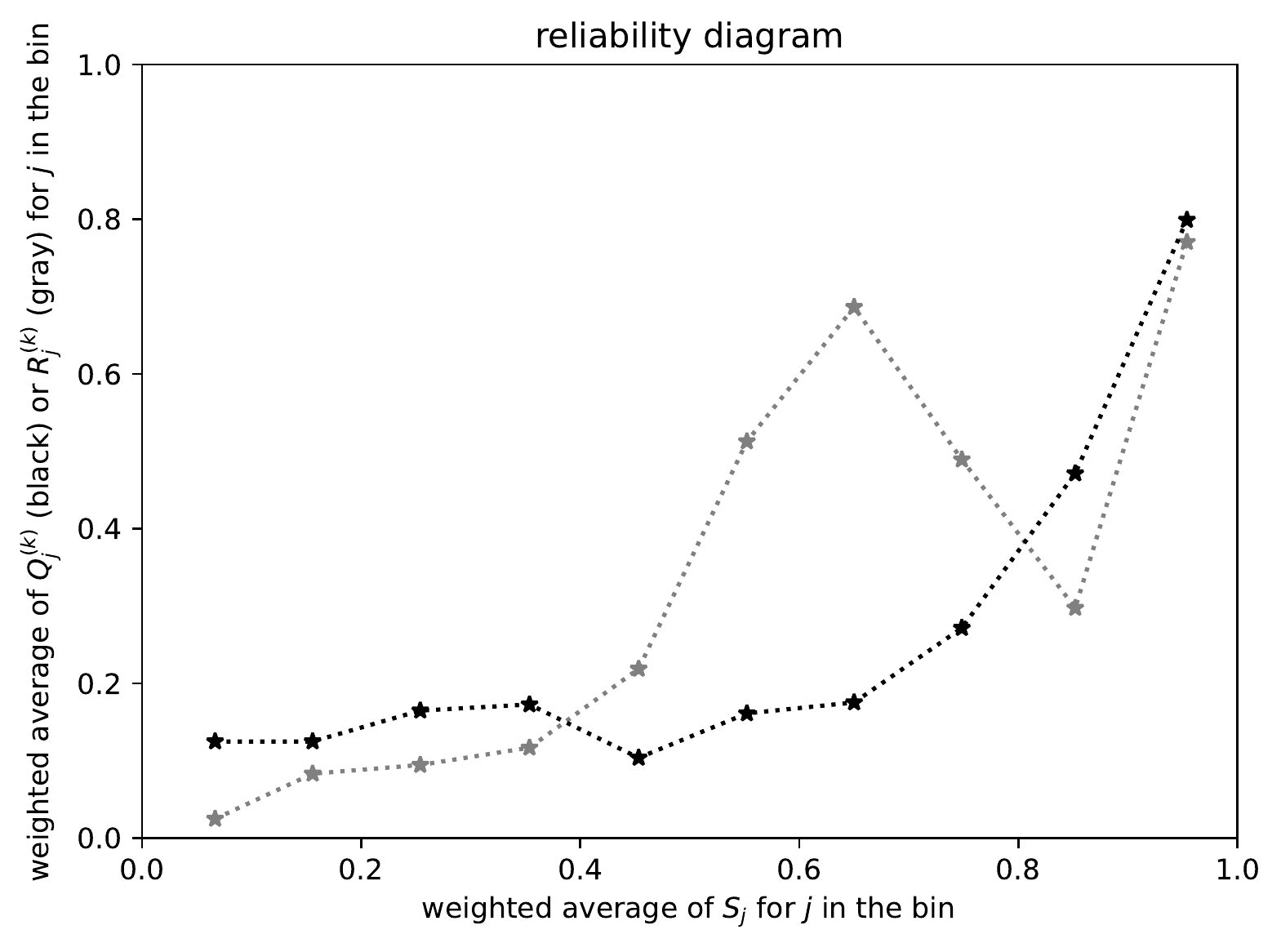}}
\quad\quad
\parbox{\imsize}{\includegraphics[width=\imsize]
                 {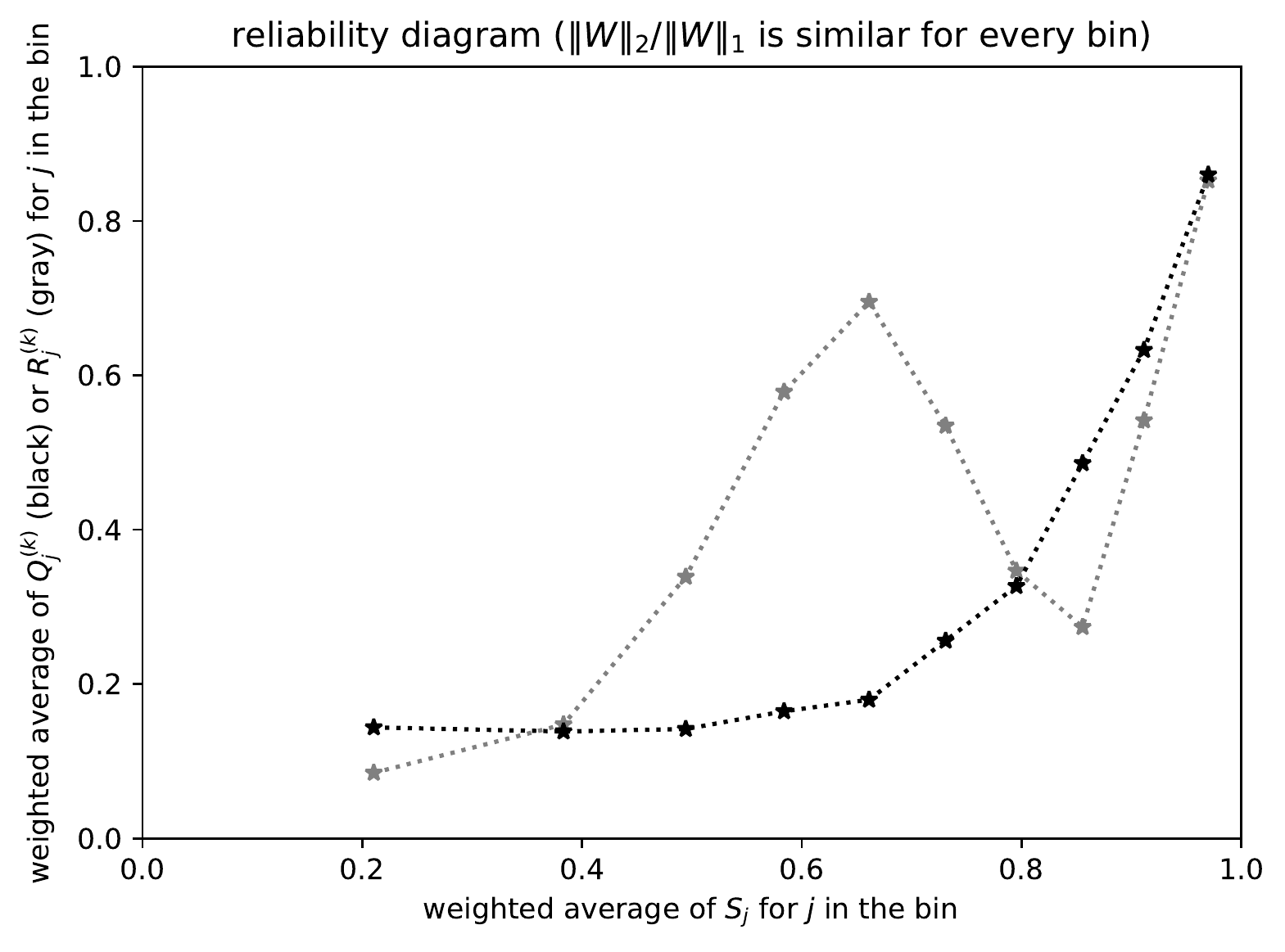}}

\vspace{\vertsep}

\parbox{\imsize}{\includegraphics[width=\imsize]
                 {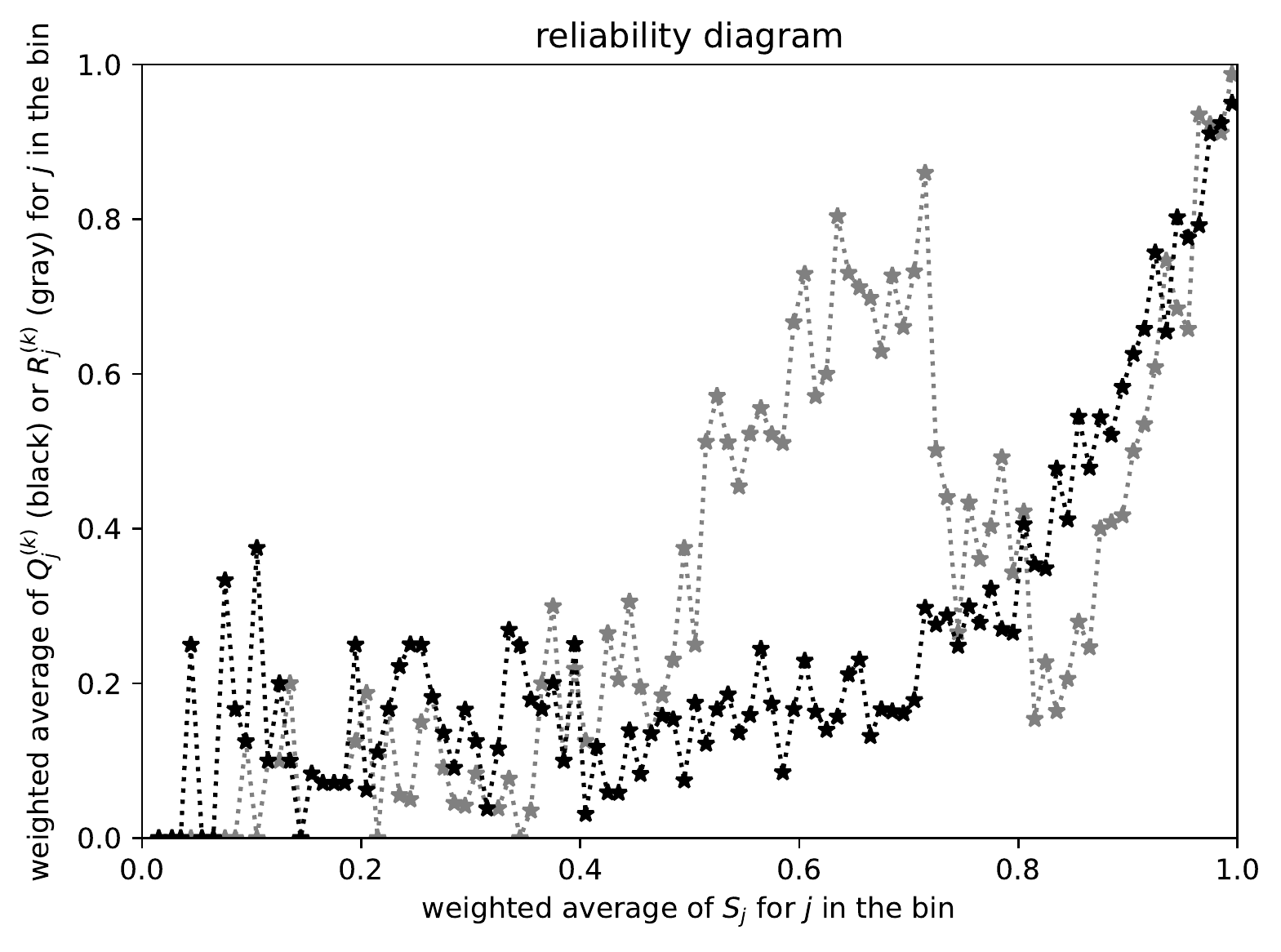}}
\quad\quad
\parbox{\imsize}{\includegraphics[width=\imsize]
                 {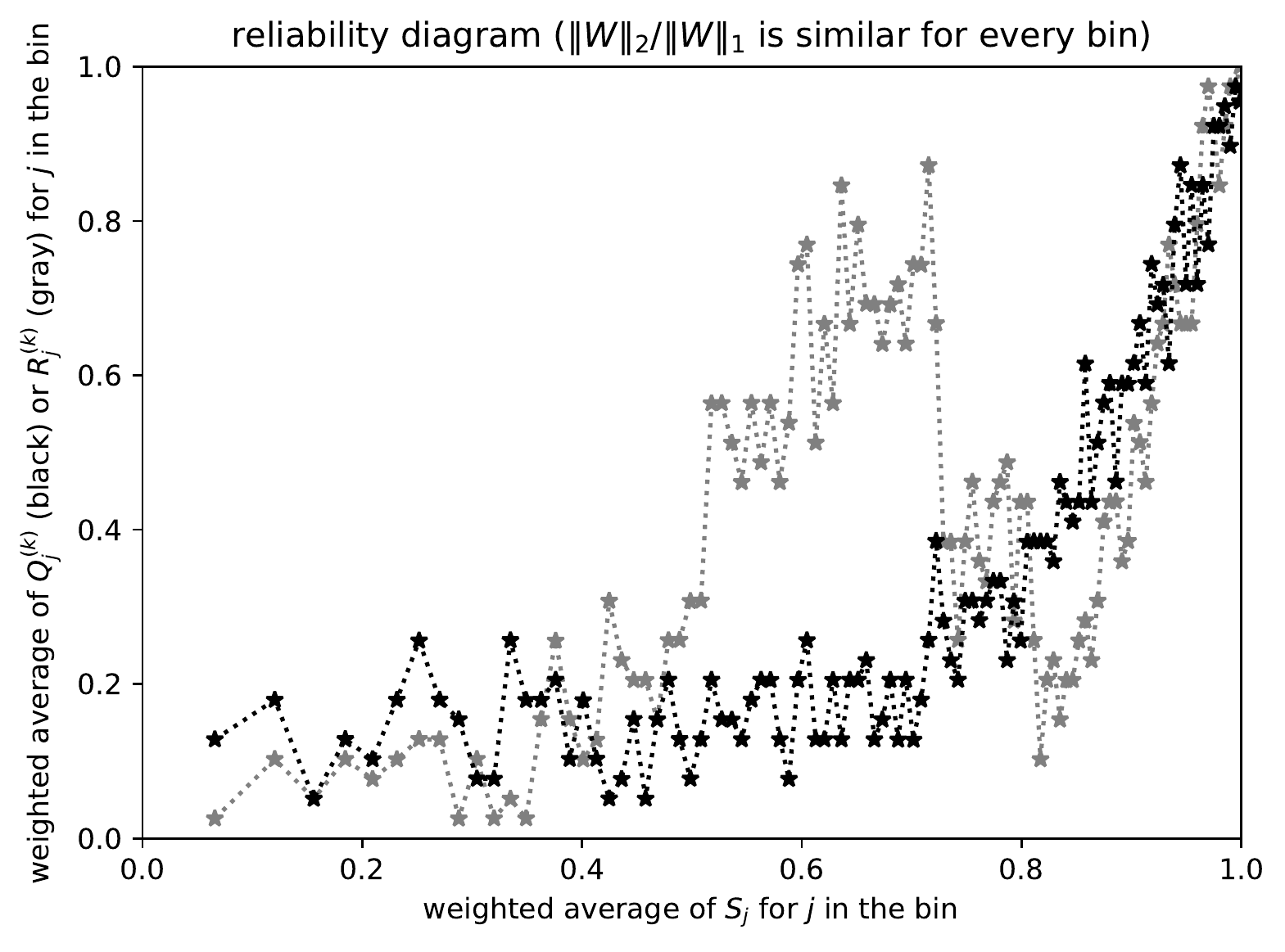}}

\vspace{\vertsep}

\parbox{\imsize}{\includegraphics[width=\imsize]
                 {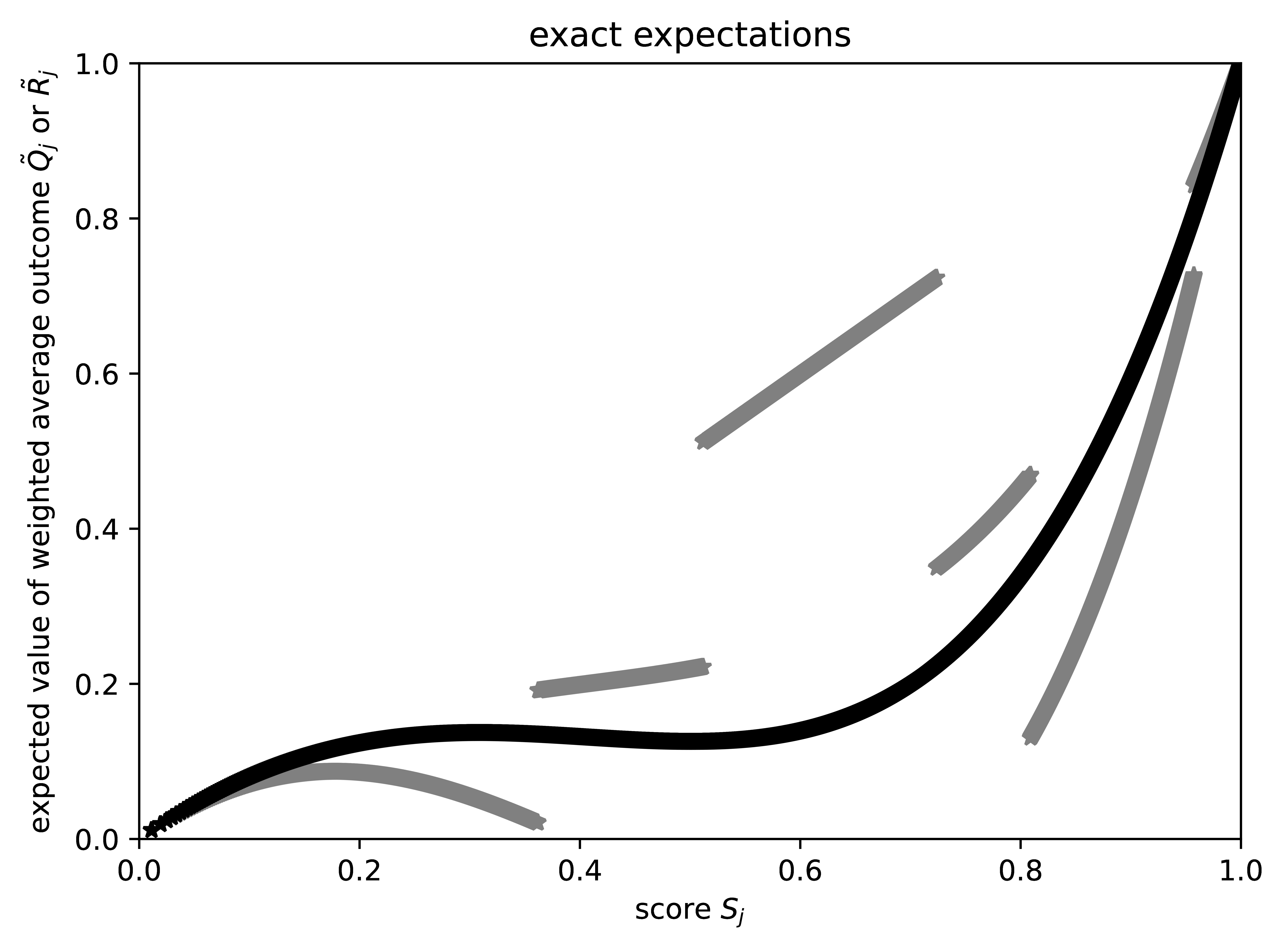}}

\end{centering}
\caption{$m =$ 4,000, $n =$ 4,000;
         Kuiper's statistic is $0.1686 / \sigma = 16.20$,
         Kolmogorov's and Smirnov's is $0.1571 / \sigma = 15.09$.
         The cumulative graphs have fairly clear kinks
         at the scores where the reliability diagrams ideally would jump
         discontinuously in order to match the lowermost plot.
         However, detecting discontinuous jumps
         in the reliability diagrams based on the actual random observations
         is very hard. When the bins are narrow enough to resolve the jumps,
         the noise on the weighted average response in each bin is large,
         with the averages jumping all over due to the noise
         (in addition to the jumps in the underlying expected distribution
         depicted in the lowermost plot).
         For the most part, the cumulative graph based on random observations
         closely matches the exact expected ground-truth cumulative graph.
}
\label{ex1}
\end{figure}

\begin{figure}
\begin{centering}

\parbox{\imsize}{\includegraphics[width=\imsize]
                {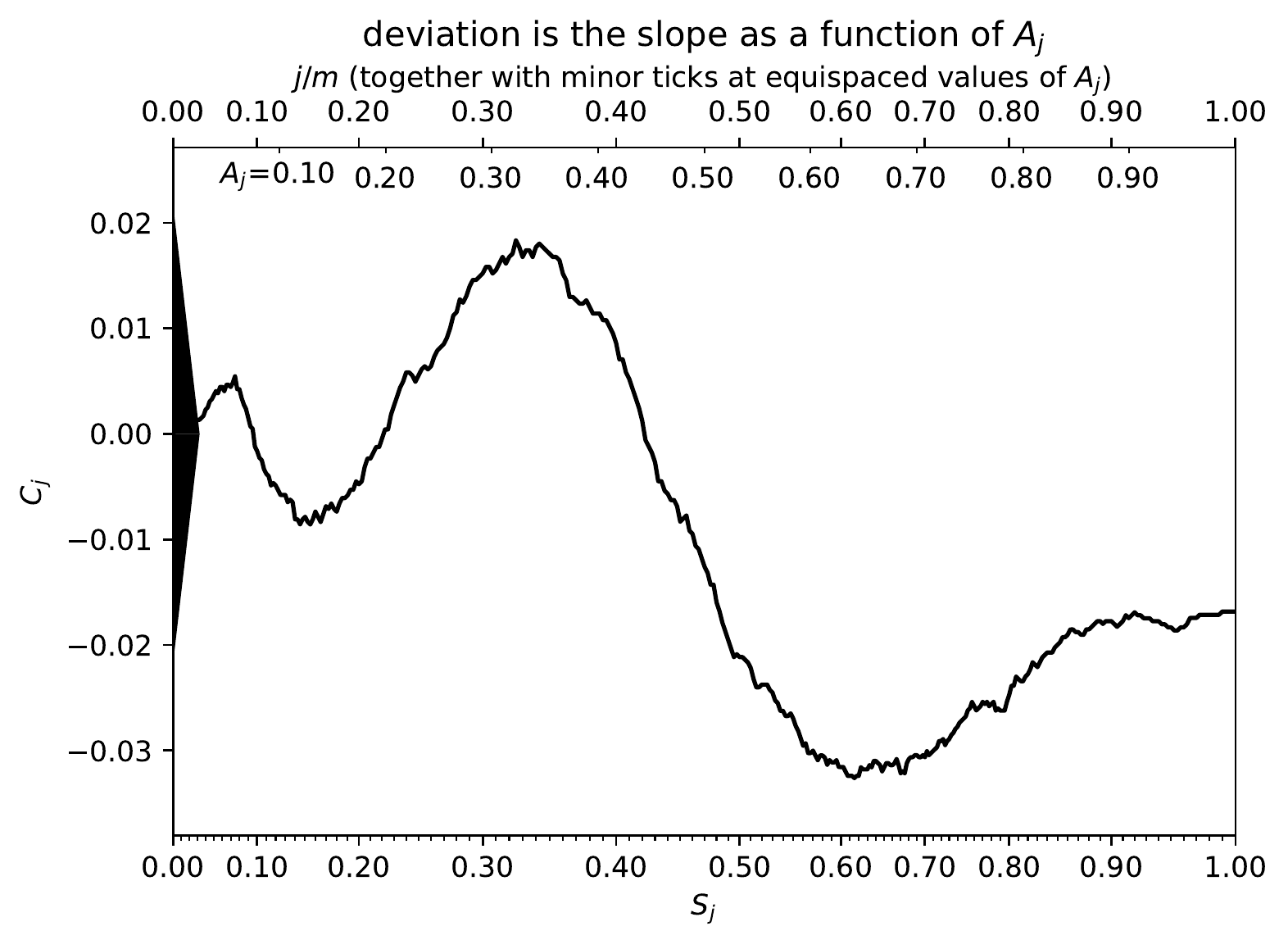}}
\quad\quad
\parbox{\imsize}{\includegraphics[width=\imsize]
                {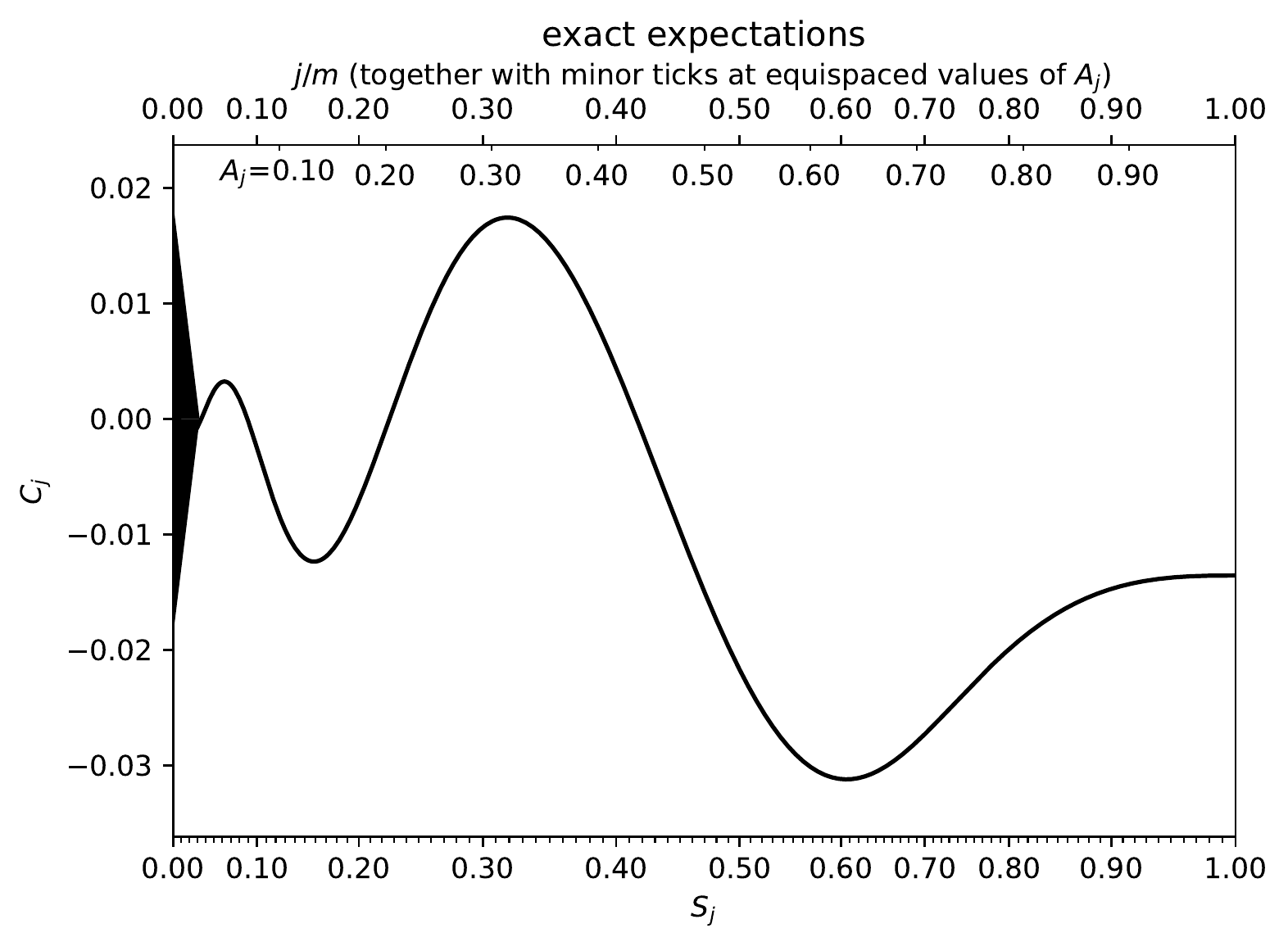}}

\vspace{\vertsep}

\parbox{\imsize}{\includegraphics[width=\imsize]
                {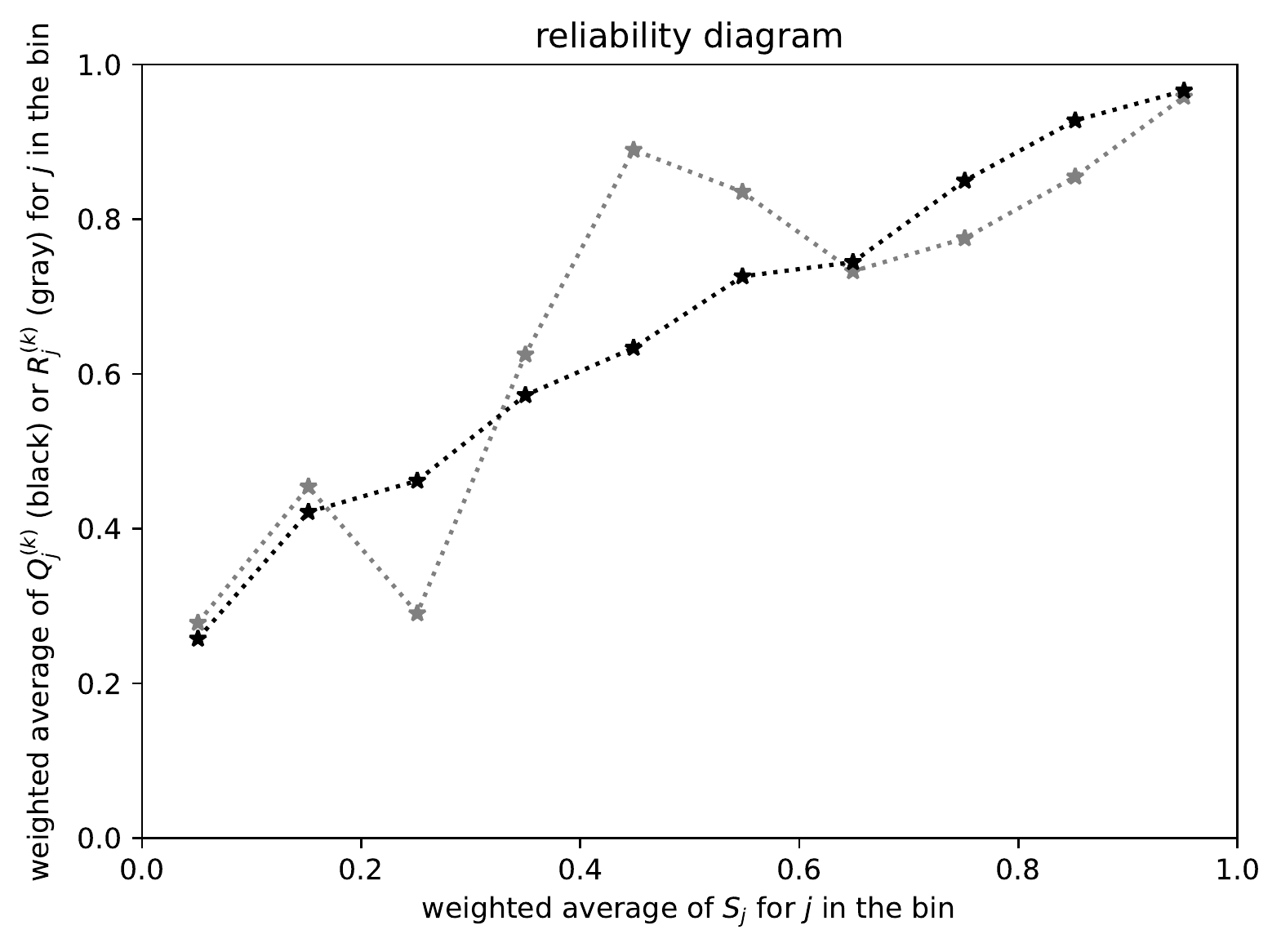}}
\quad\quad
\parbox{\imsize}{\includegraphics[width=\imsize]
                {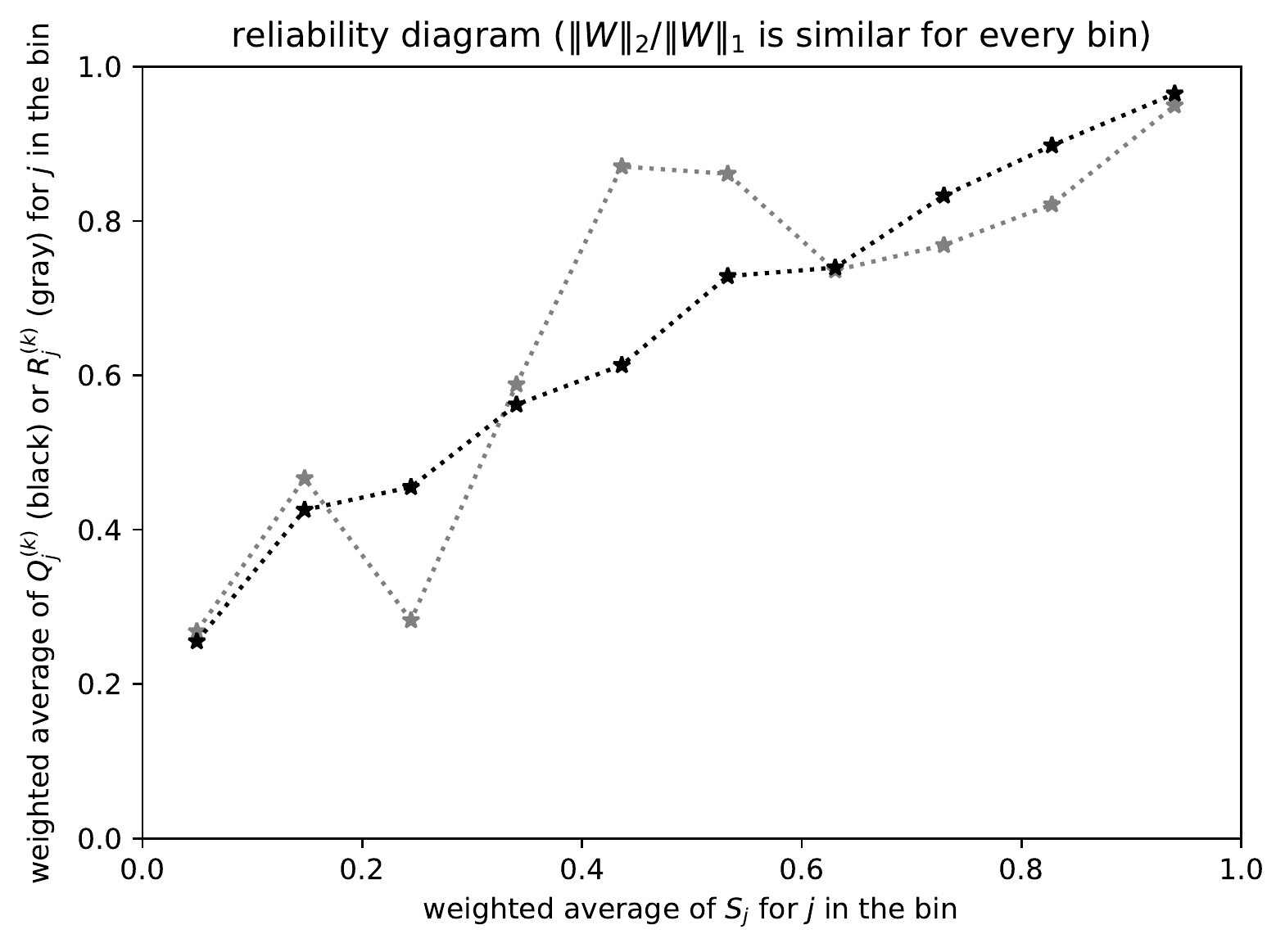}}

\vspace{\vertsep}

\parbox{\imsize}{\includegraphics[width=\imsize]
                {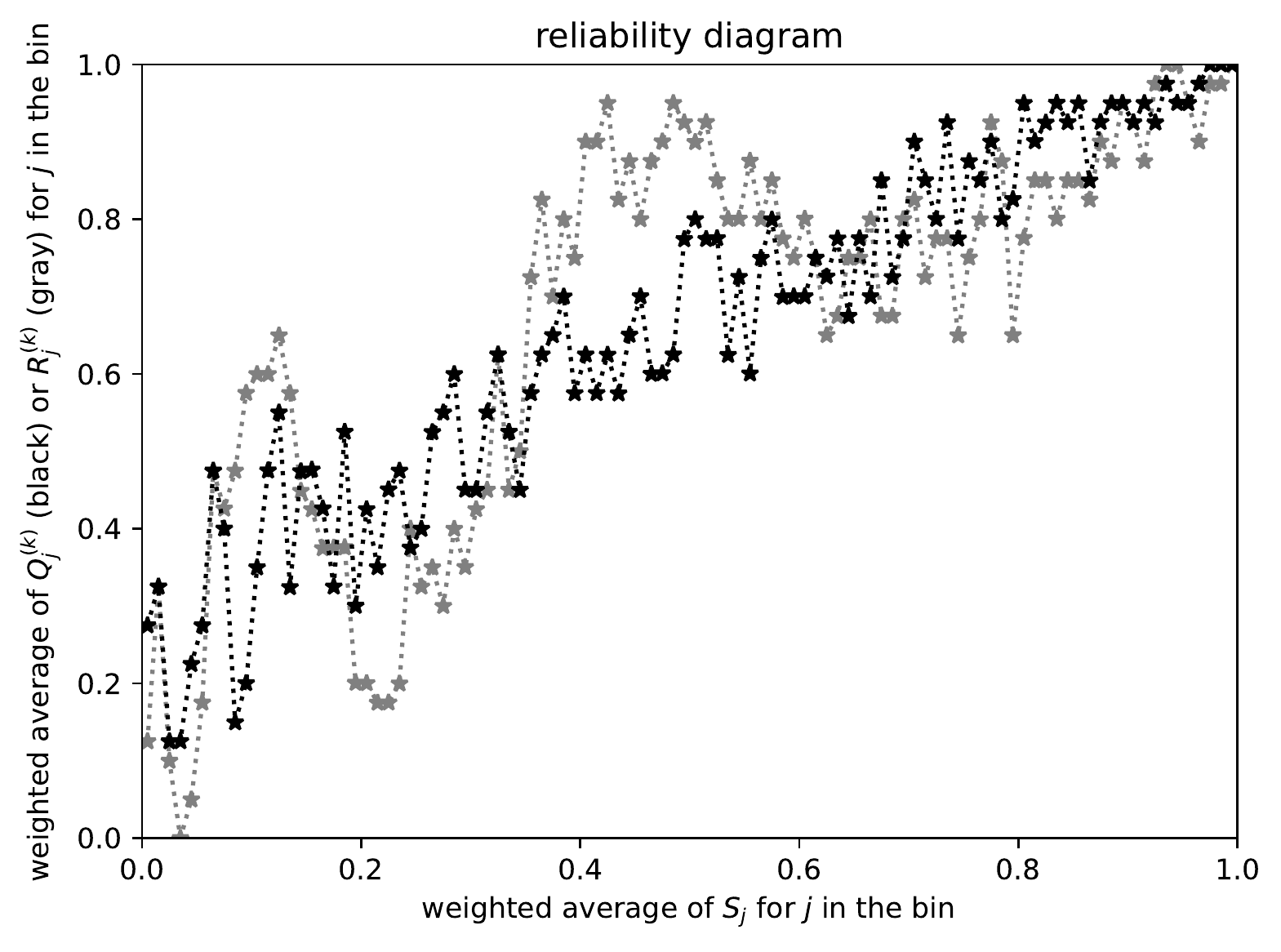}}
\quad\quad
\parbox{\imsize}{\includegraphics[width=\imsize]
                {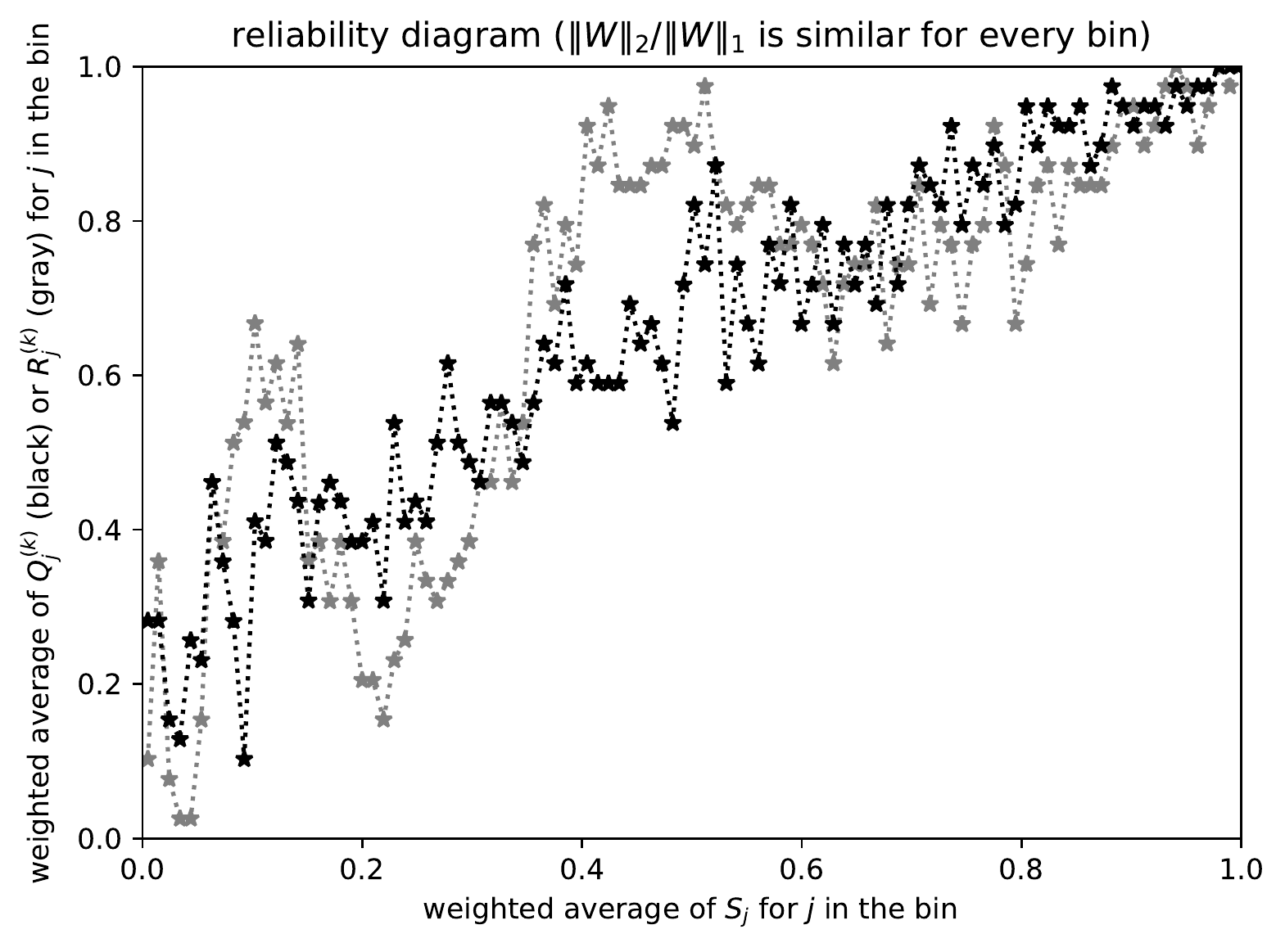}}

\vspace{\vertsep}

\parbox{\imsize}{\includegraphics[width=\imsize]
                {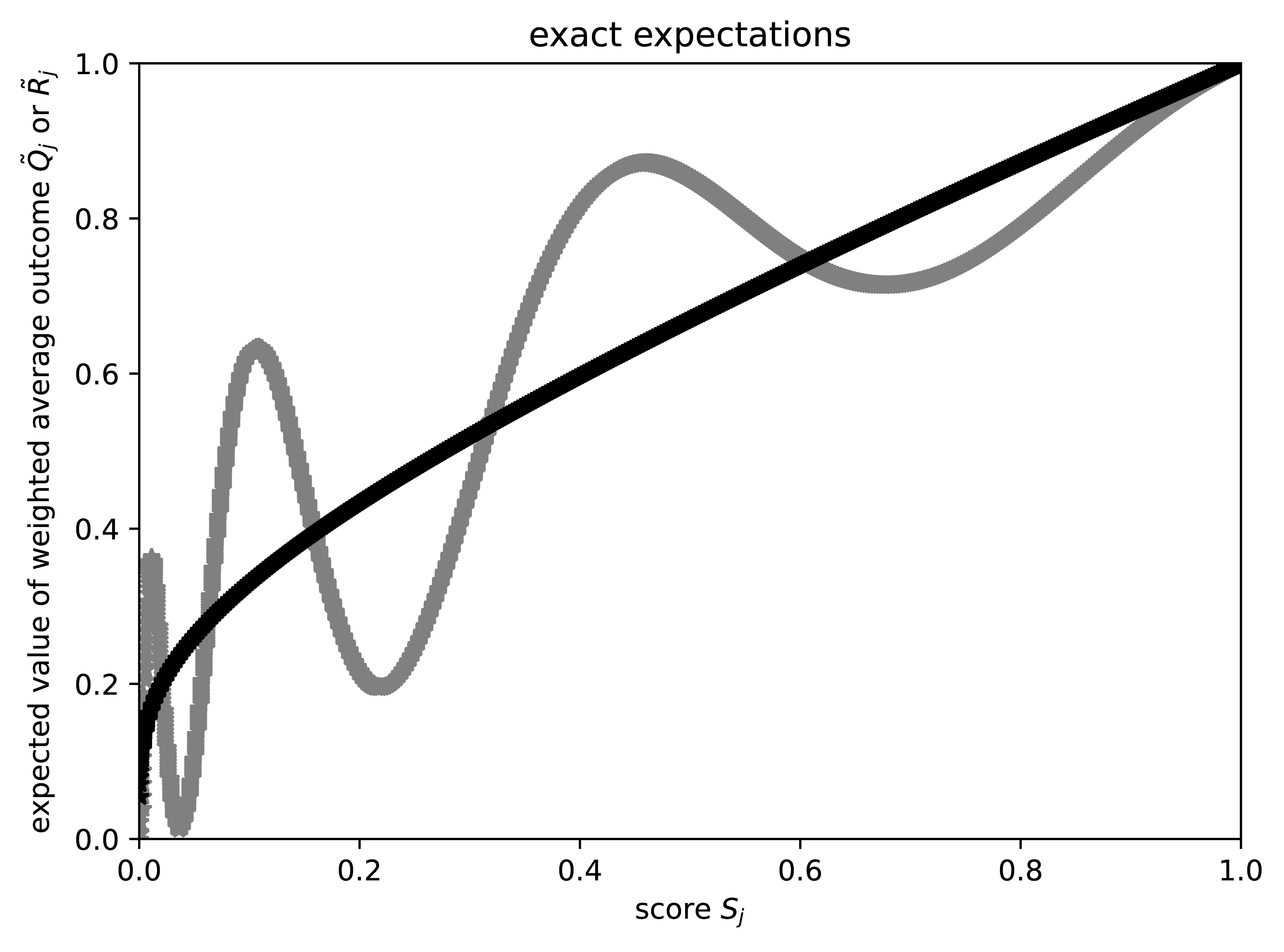}}

\end{centering}
\caption{$m =$ 400, $n =$ 4,000;
         Kuiper's statistic is $0.04235 / \sigma = 3.676$,
         Kolmogorov's and Smirnov's is $0.02328 / \sigma = 2.021$.
         Recall that the weighted average difference in responses
         is the vertical coordinate $C_m$ at the greatest (rightmost) score
         $S_m$ in the cumulative plots. This value is significantly smaller,
         that is, nearer to 0, than the vertical range of the graph;
         the vertical range
         ($\max_{0 \le j \le m} C_j - \min_{0 \le j \le m} C_j$) is the value
         of the Kuiper statistic.
         Whereas any one of the empirical reliability diagrams obscures
         at least some aspect of the exact ground-truth diagram
         displayed in the lowermost plot, the empirical cumulative graph
         closely resembles the exact expected ground-truth cumulative graph.
}
\label{ex2}
\end{figure}

\begin{figure}
\begin{centering}

\parbox{\imsize}{\includegraphics[width=\imsize]
                {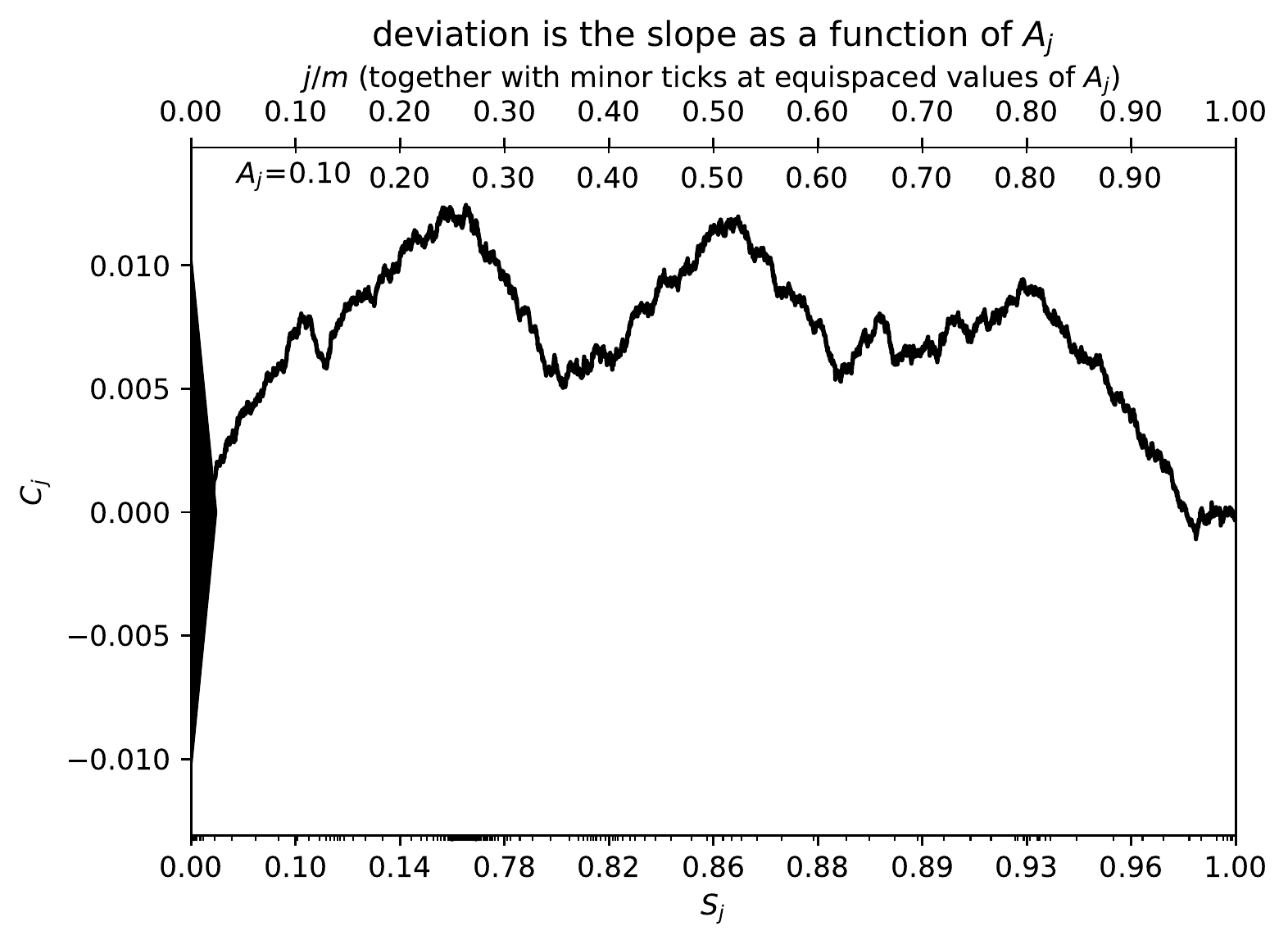}}
\quad\quad
\parbox{\imsize}{\includegraphics[width=\imsize]
                {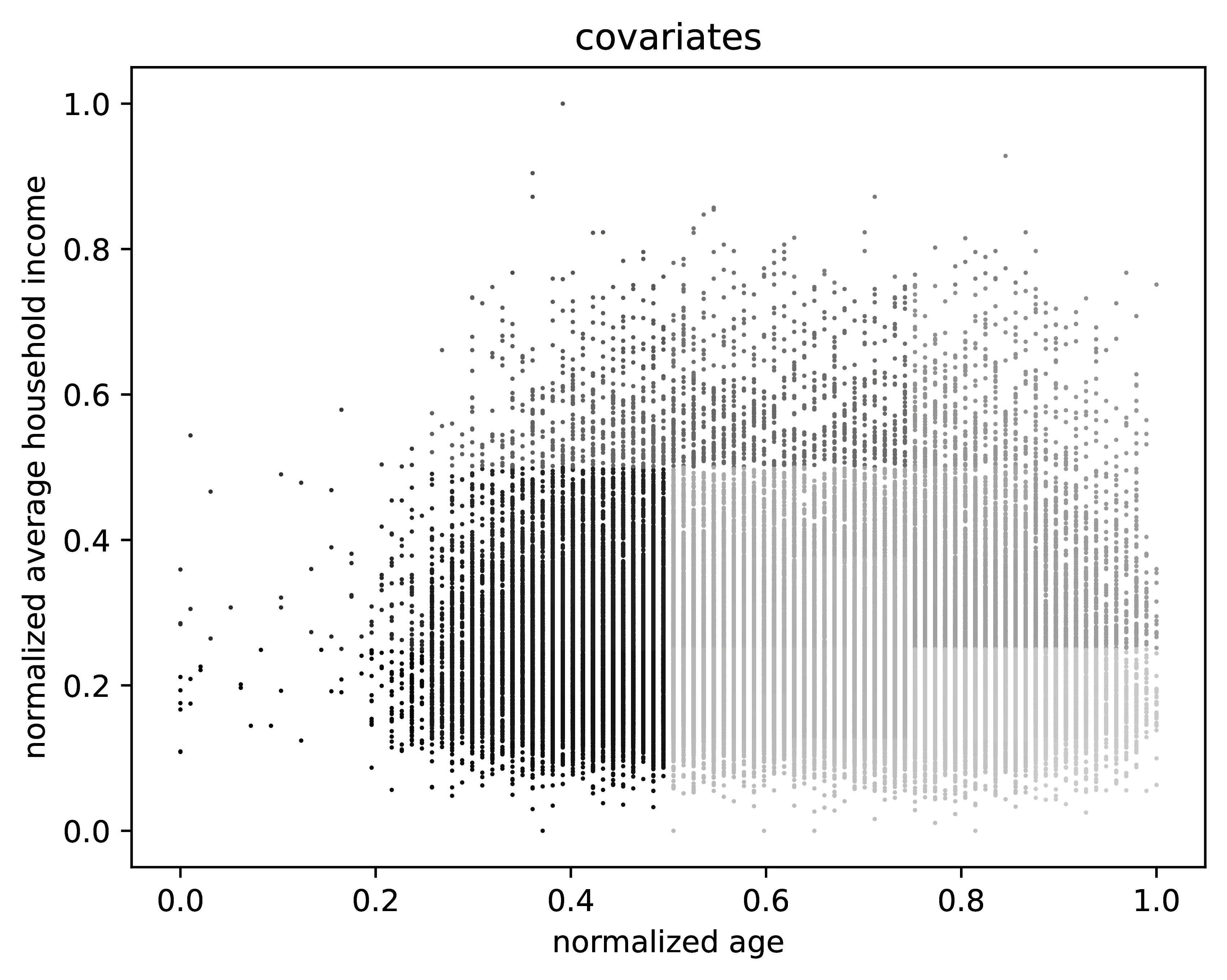}}

\vspace{\vertsep}

\parbox{\imsize}{\includegraphics[width=\imsize]
                {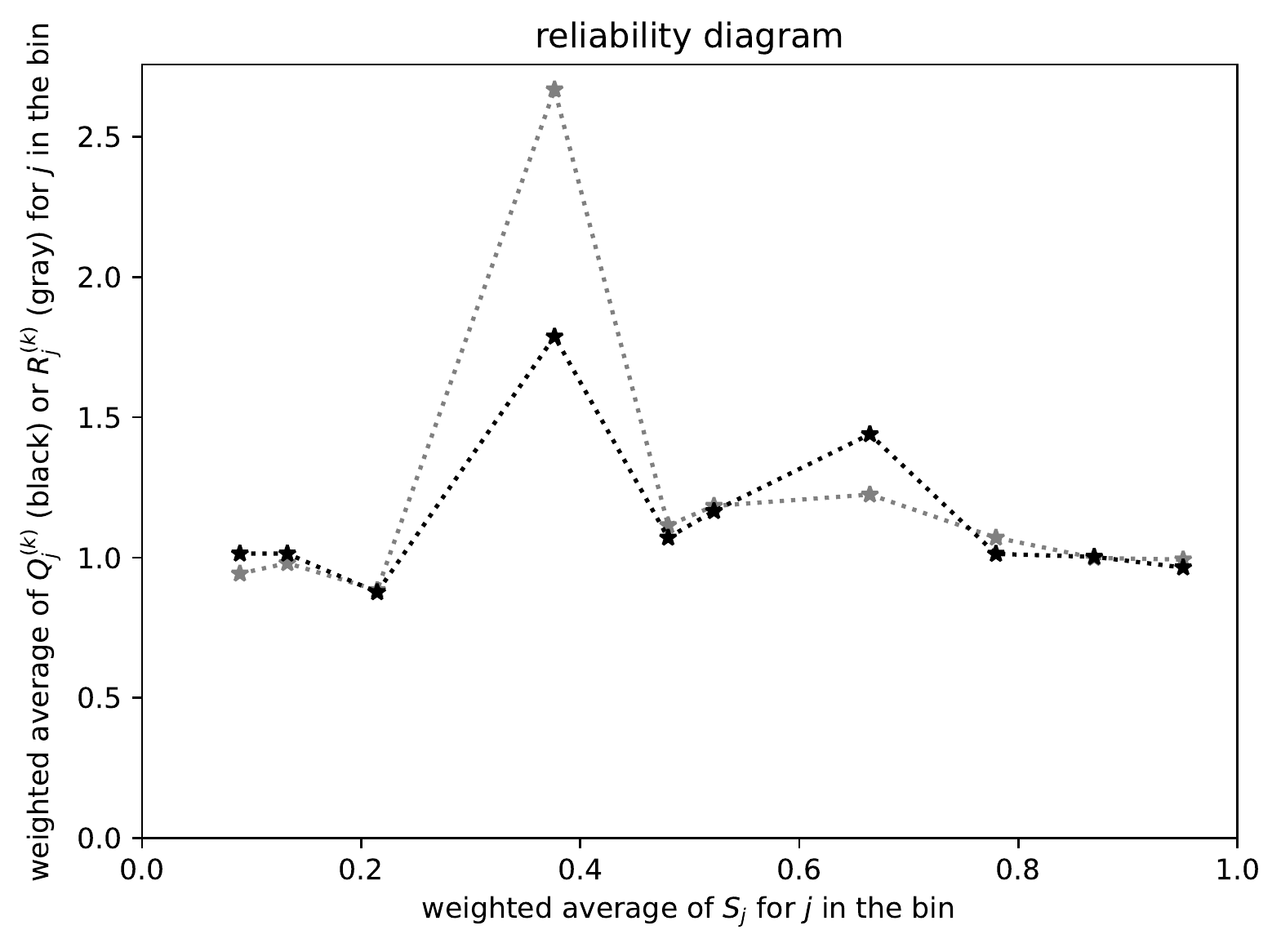}}
\quad\quad
\parbox{\imsize}{\includegraphics[width=\imsize]
                {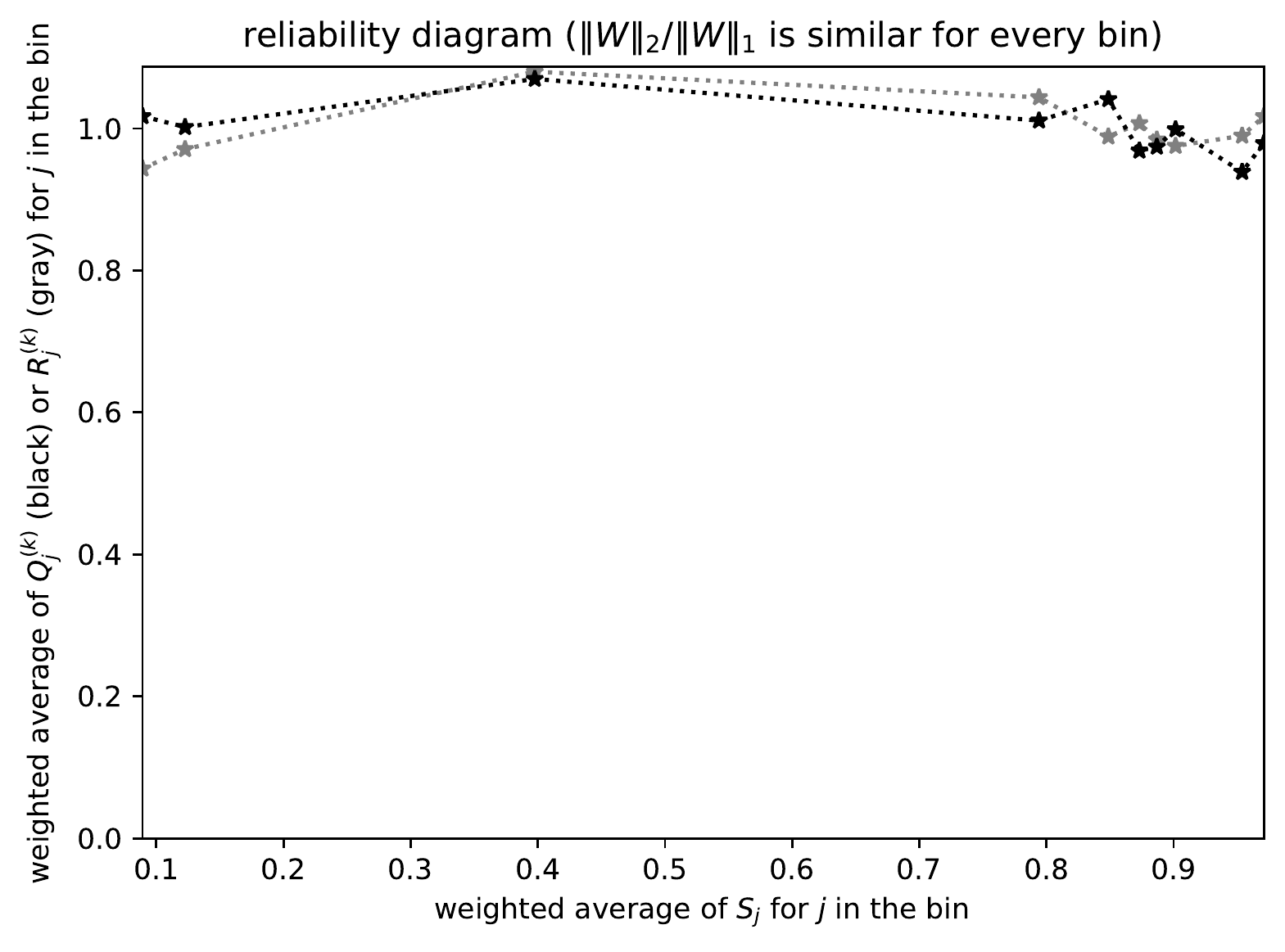}}

\vspace{\vertsep}

\parbox{\imsize}{\includegraphics[width=\imsize]
                {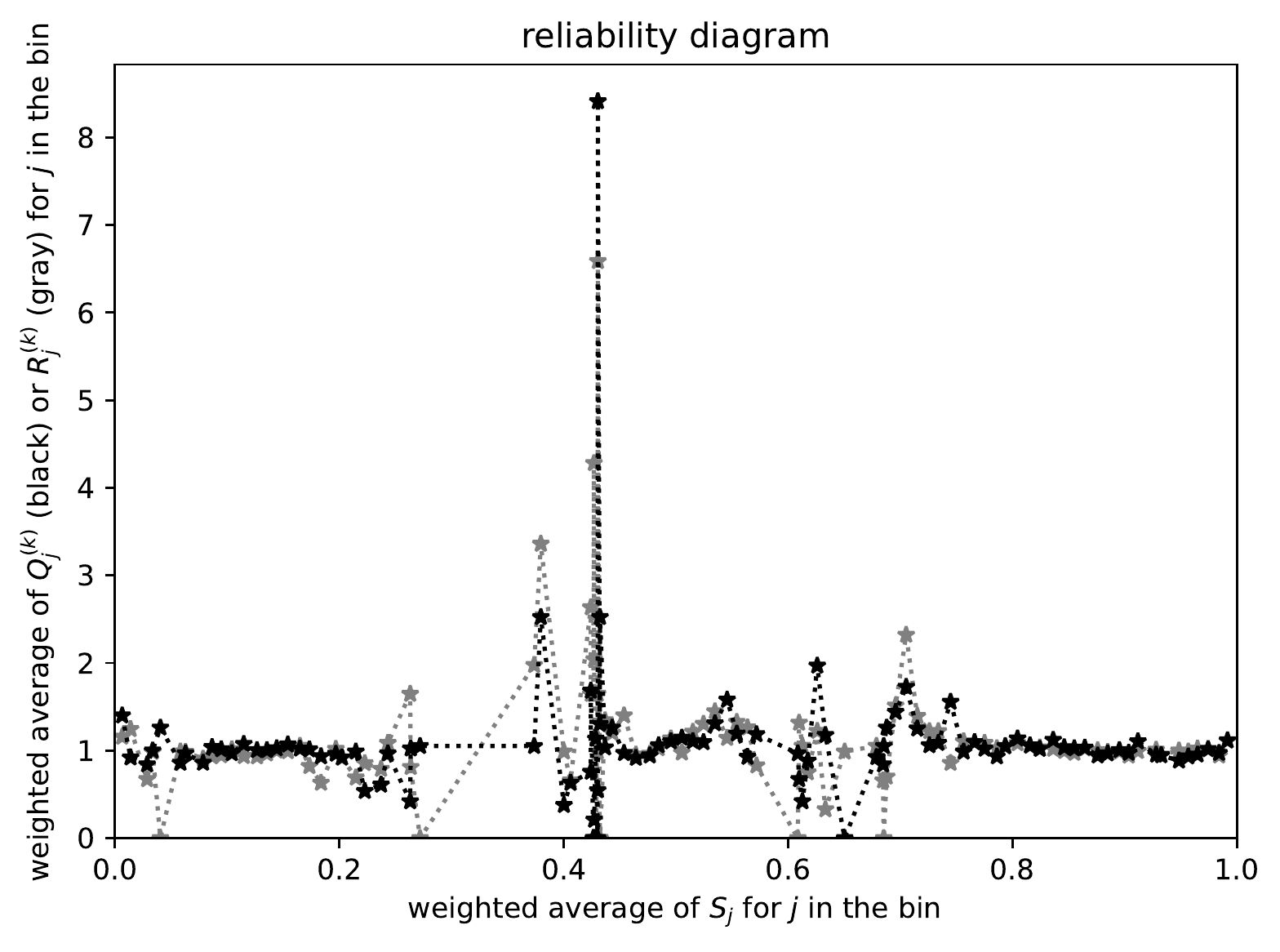}}
\quad\quad
\parbox{\imsize}{\includegraphics[width=\imsize]
                {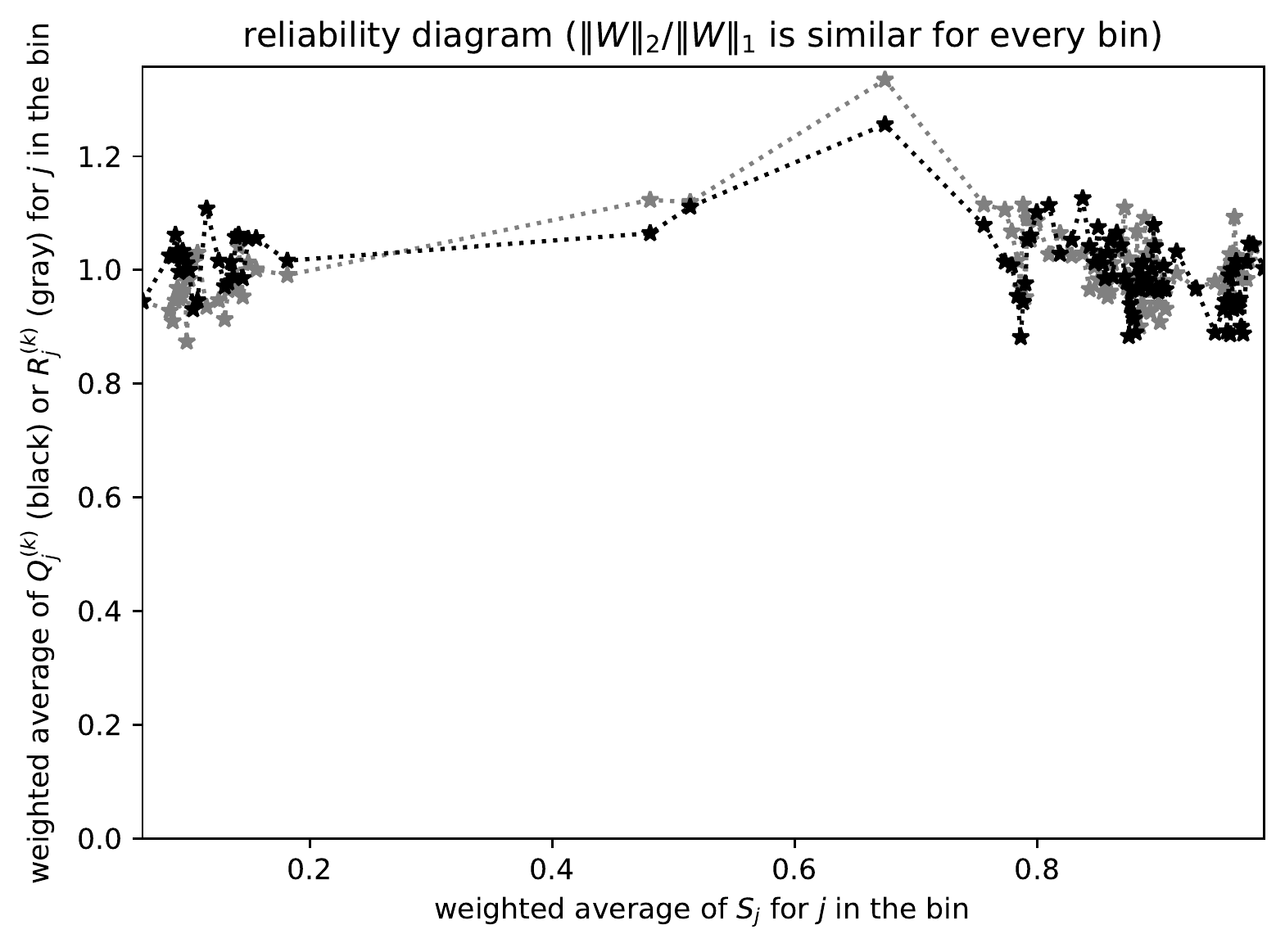}}

\end{centering}
\caption{$m = n =$ 63,826;
         Kuiper's statistic is $0.01354 / \sigma = 2.514$,
         Kolmogorov's and Smirnov's is $0.01245 / \sigma = 2.312$.
         Notice how these scalar statistics miss the very steep slopes
         in the cumulative graph, as the graph oscillates rapidly.
         The following figure, Figure~\ref{ex20}, fixes this failure
         by reversing the order of the covariates in the parameterization
         via the Hilbert curve. The present figure and the following figure
         display similarly high slopes, but re-ordering the scores
         as in the following figure prevents the oscillation that
         the scalar metrics cannot capture.
}
\label{ex02}
\end{figure}

\begin{figure}
\begin{centering}

\parbox{\imsize}{\includegraphics[width=\imsize]
                {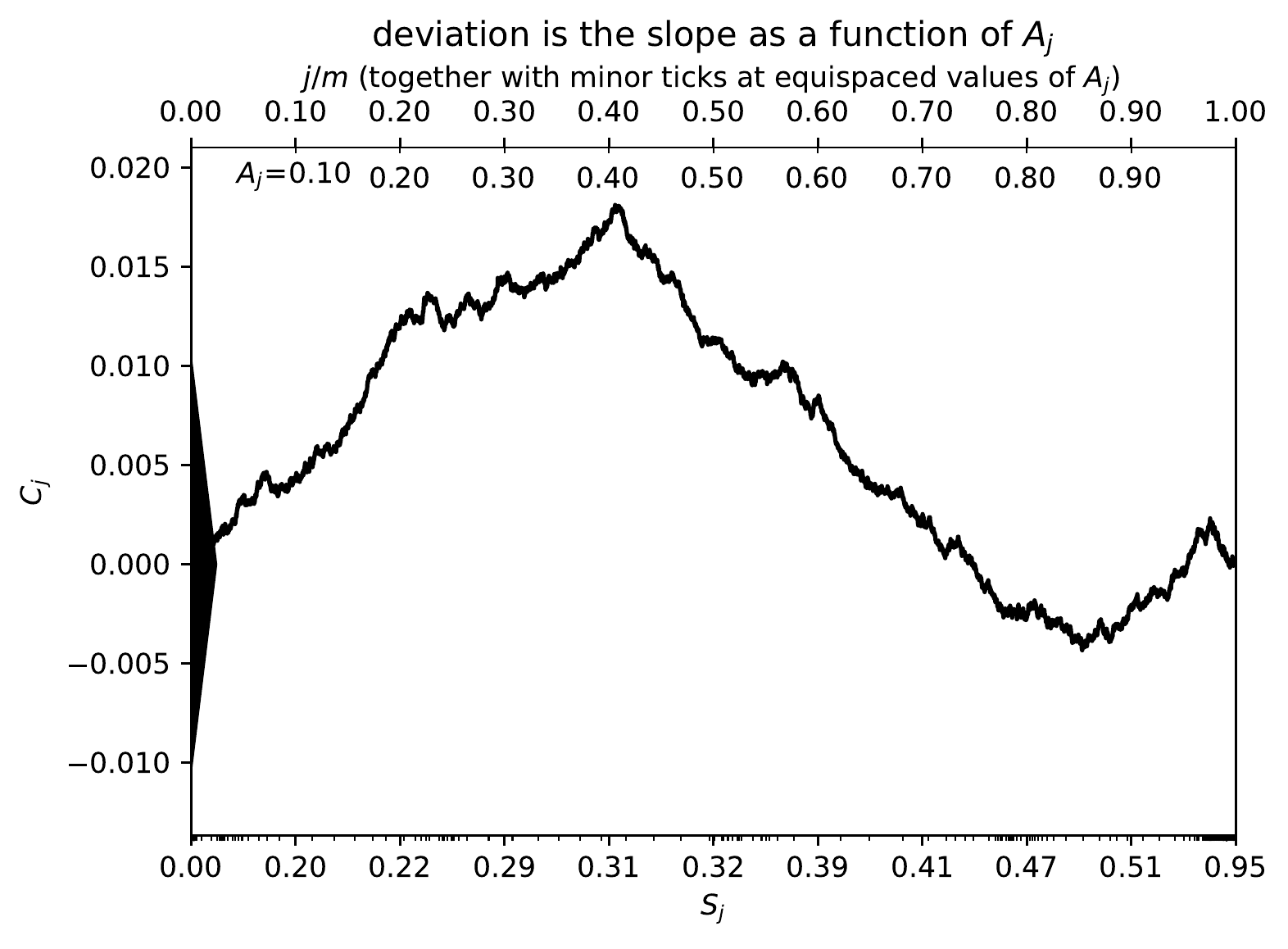}}
\quad\quad
\parbox{\imsize}{\includegraphics[width=\imsize]
                {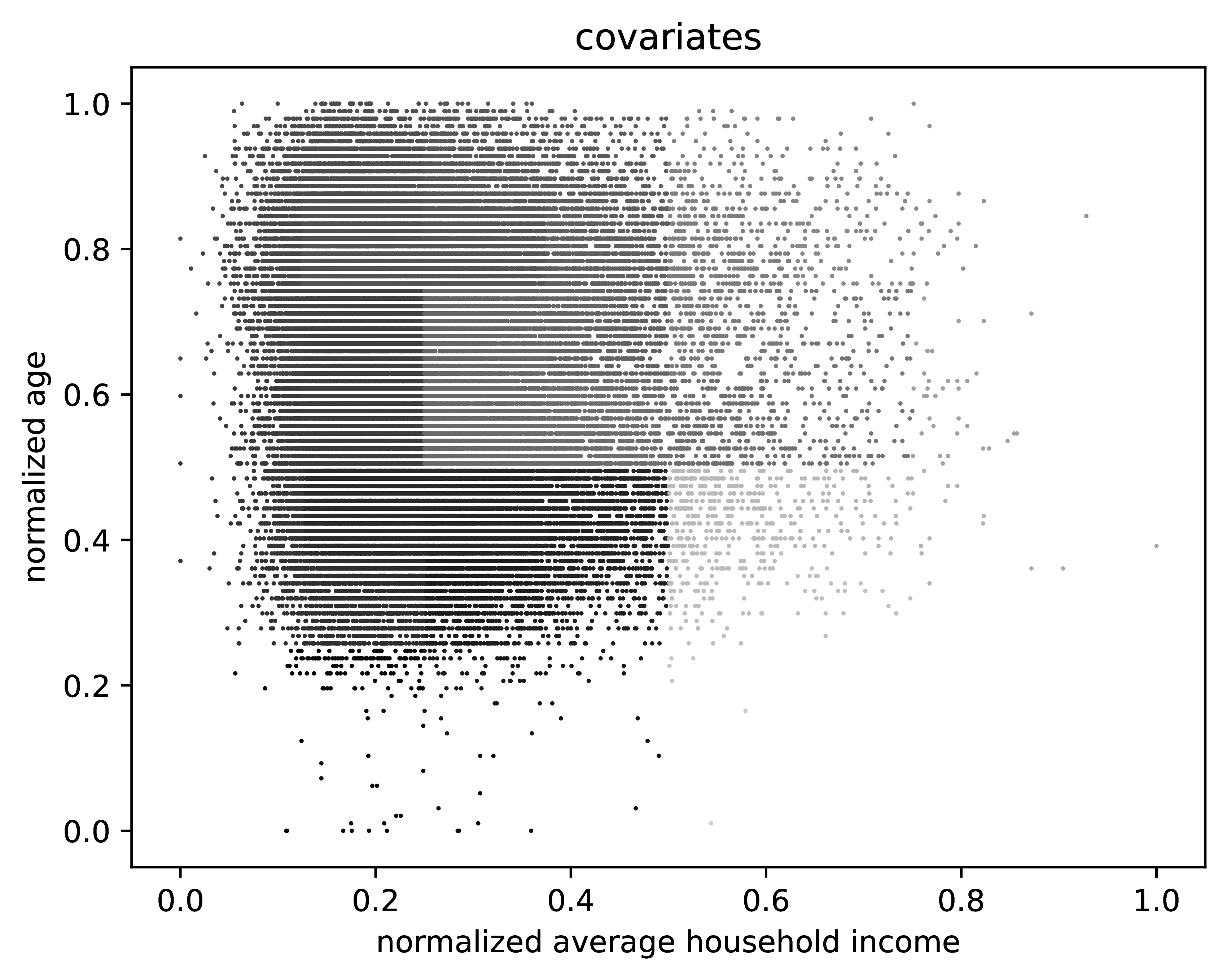}}

\vspace{\vertsep}

\parbox{\imsize}{\includegraphics[width=\imsize]
                {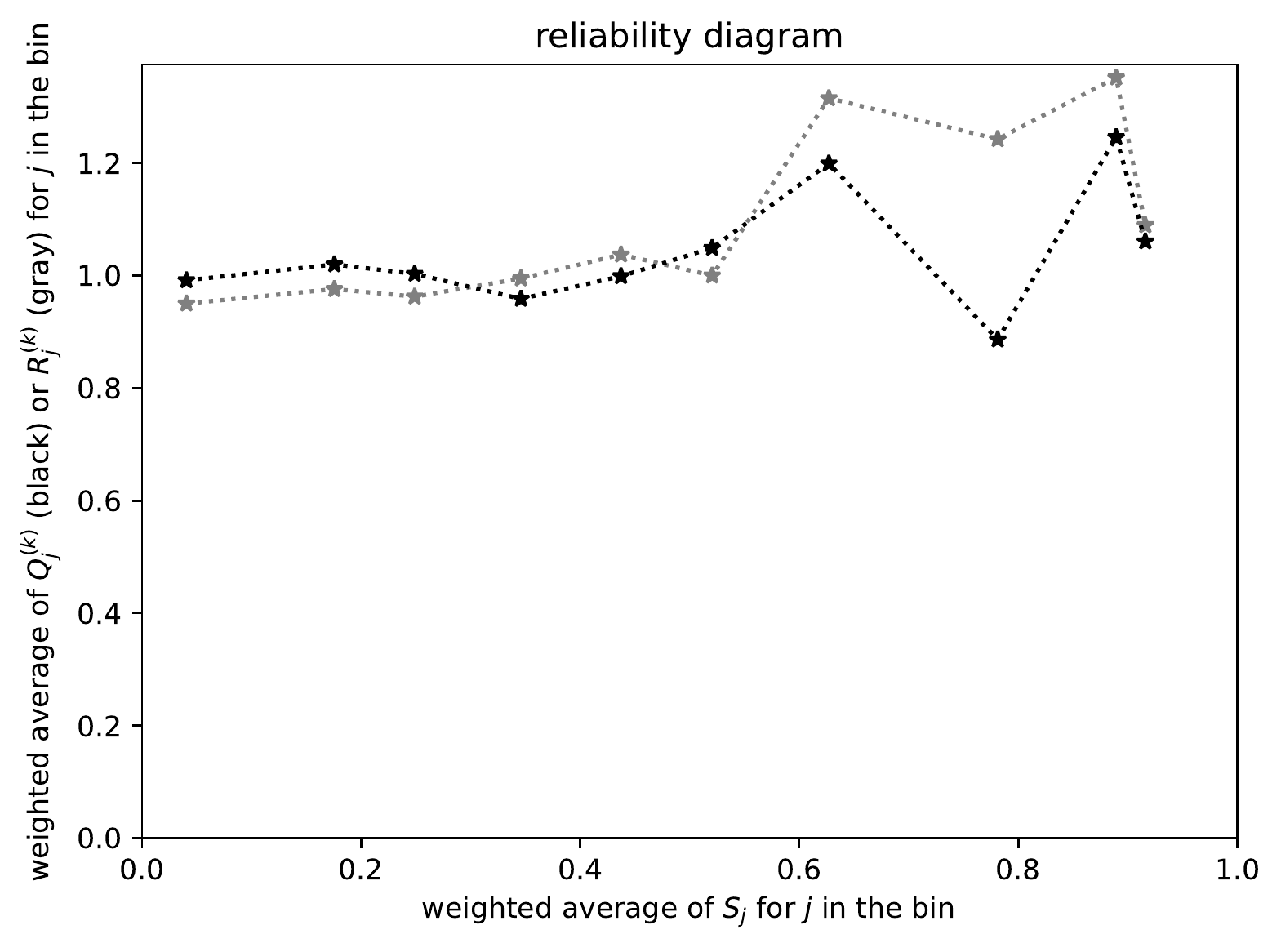}}
\quad\quad
\parbox{\imsize}{\includegraphics[width=\imsize]
                {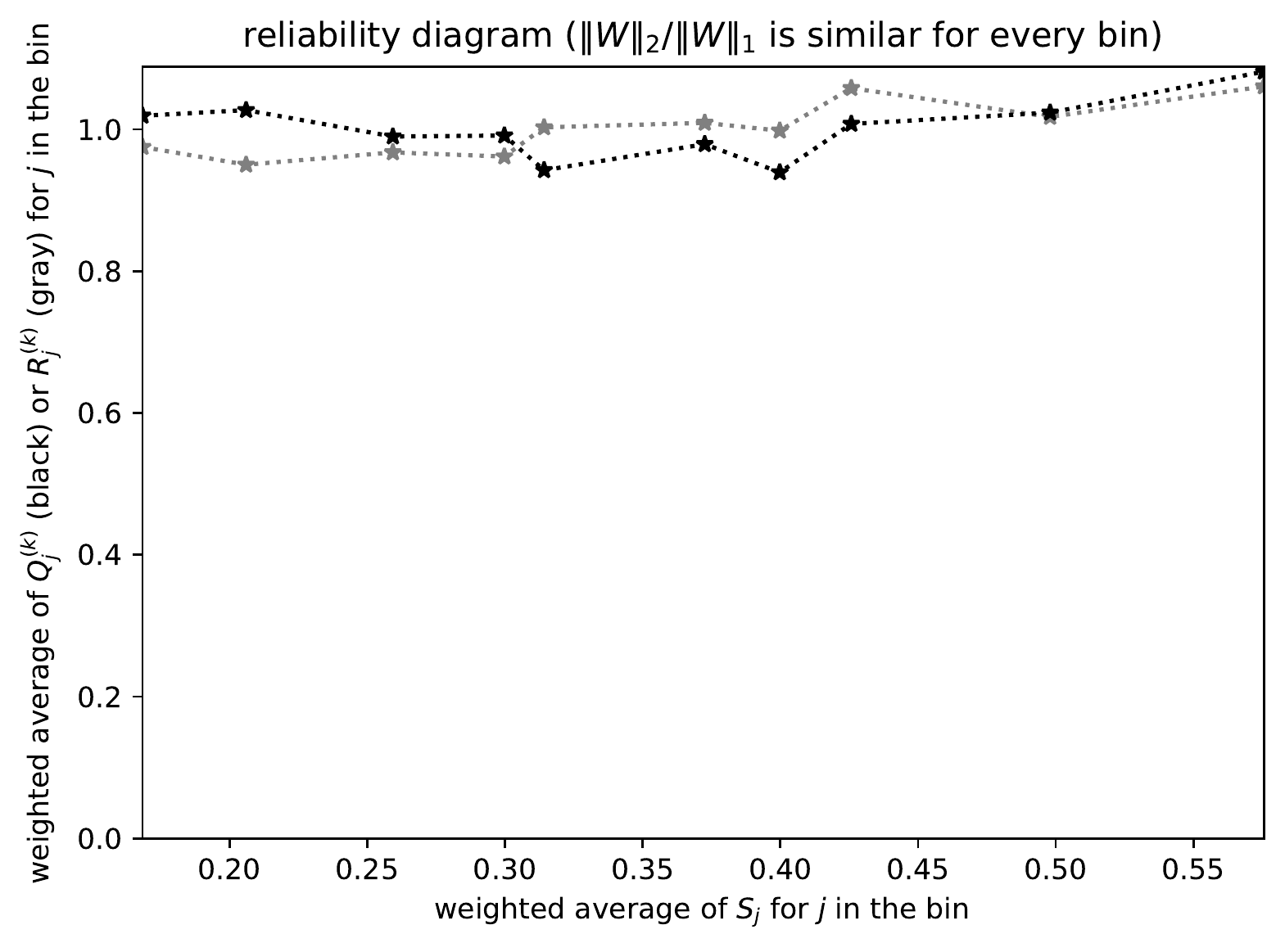}}

\vspace{\vertsep}

\parbox{\imsize}{\includegraphics[width=\imsize]
                {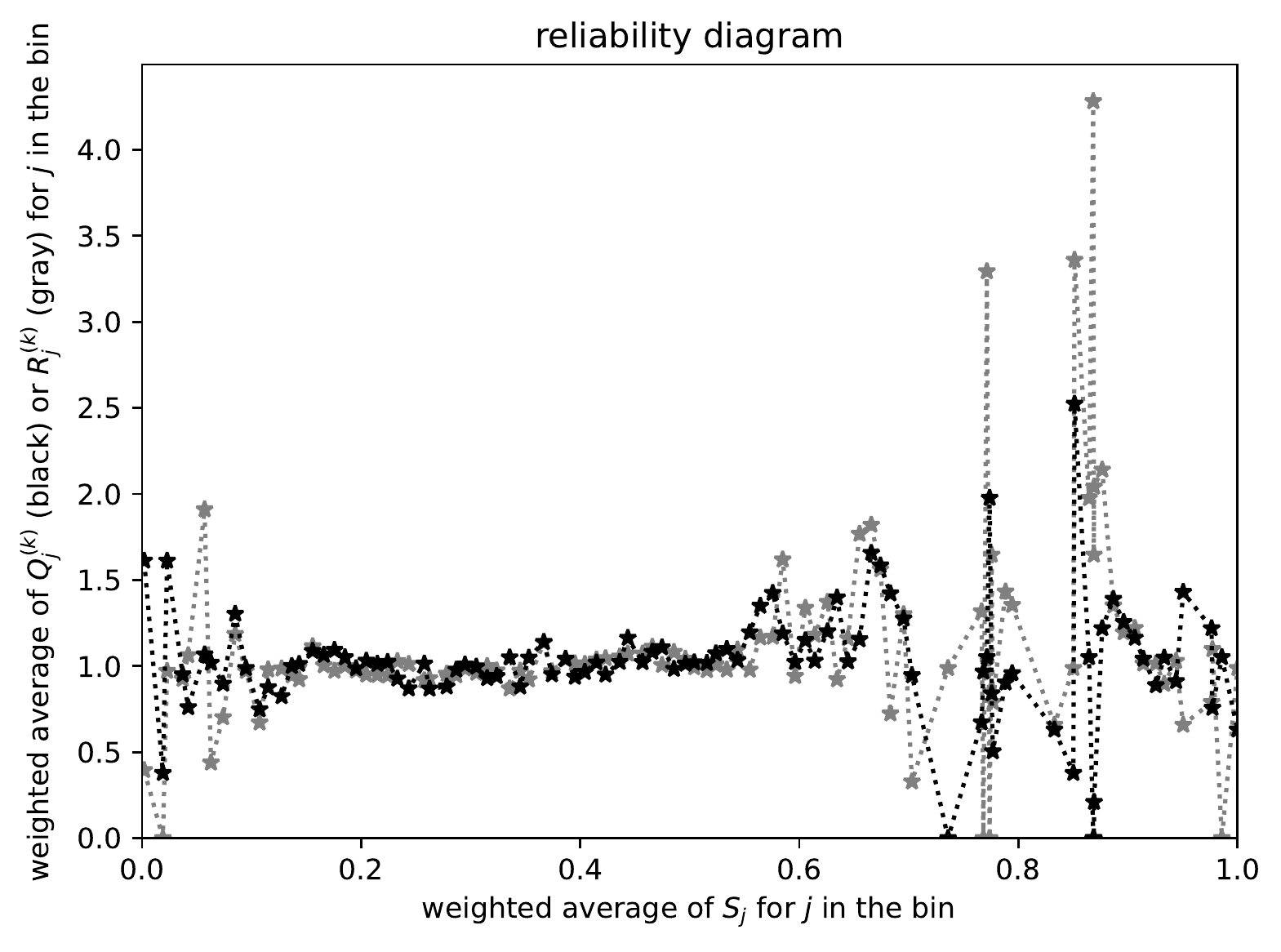}}
\quad\quad
\parbox{\imsize}{\includegraphics[width=\imsize]
                {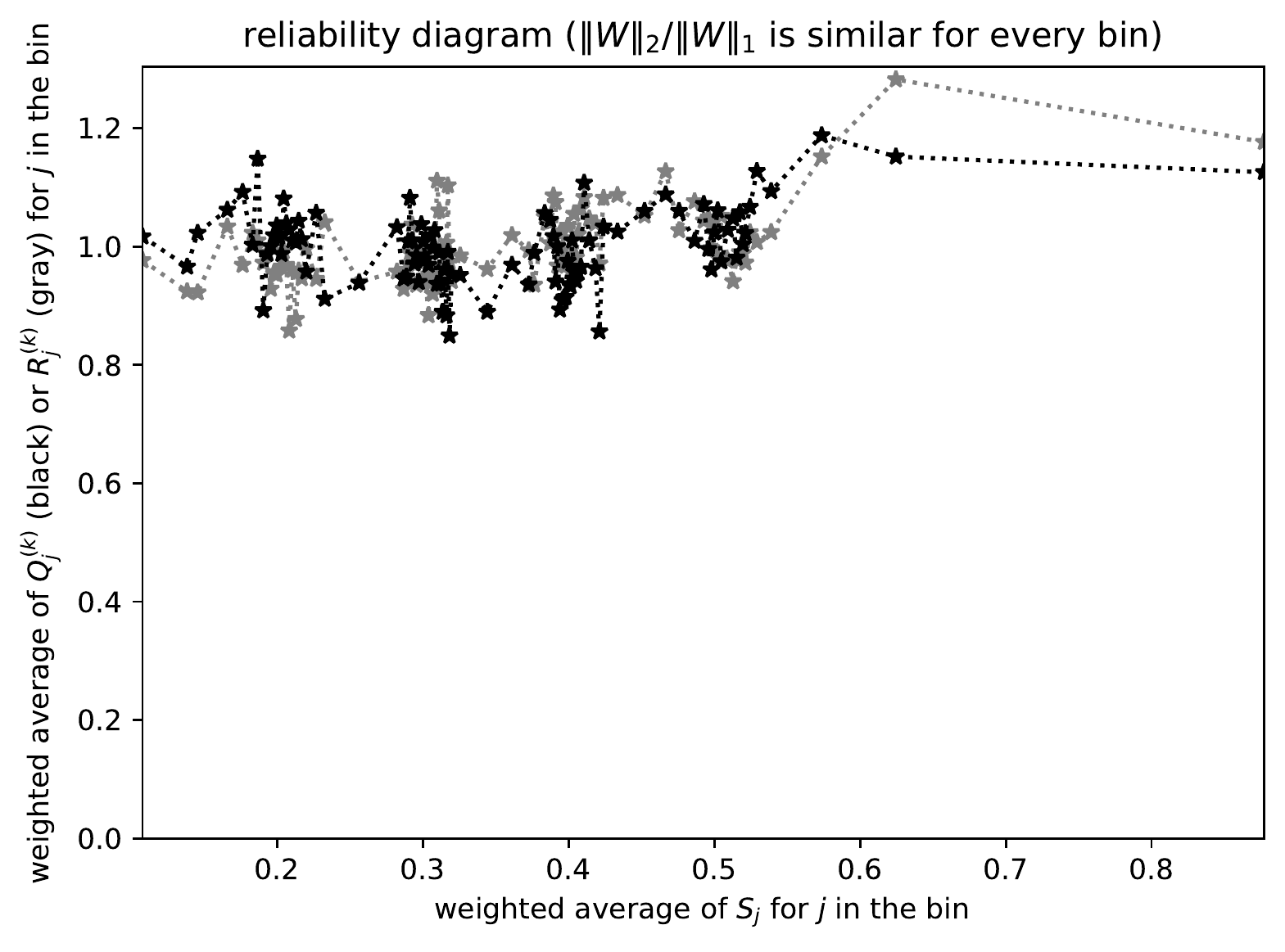}}

\end{centering}
\caption{$m = n =$ 63,826;
         Kuiper's statistic is $0.02246 / \sigma = 4.170$,
         Kolmogorov's and Smirnov's is $0.01813 / \sigma = 3.366$.
         This figure is the same as the previous figure, Figure~\ref{ex02},
         but with the order of the covariates reversed in the parameterization
         by the Hilbert curve. Both the present figure and the former figure
         reveal similarly steep slopes; the present figure reduces oscillations
         in the cumulative graph in comparison to the previous figure.
         The scalar statistics reflect the steeper slopes better
         in the present figure, due to the reduction in cancellation
         from oscillation.
}
\label{ex20}
\end{figure}

\begin{figure}
\begin{centering}

\parbox{\imsize}{\includegraphics[width=\imsize]
                {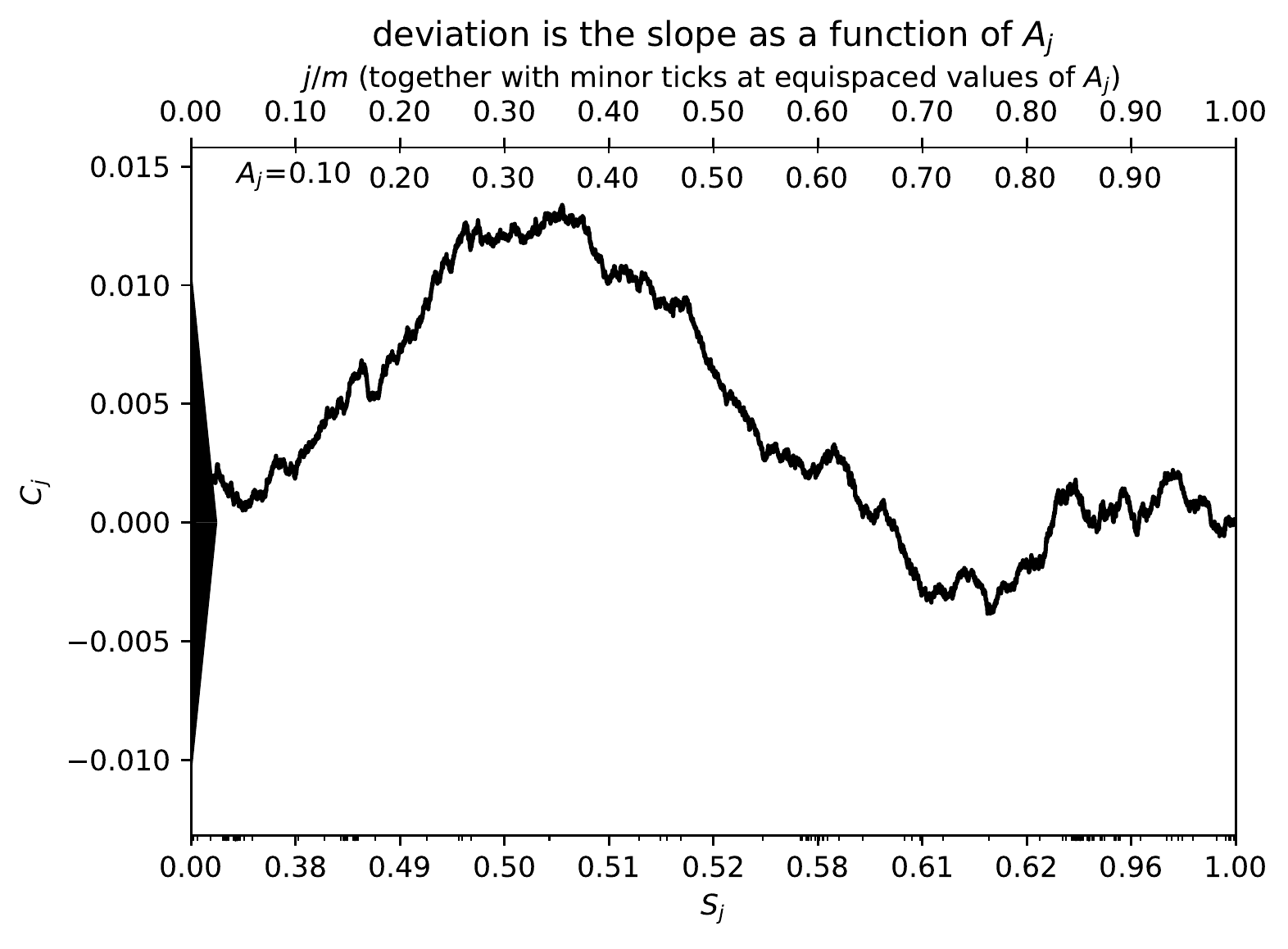}}

\vspace{\vertsep}

\parbox{\imsize}{\includegraphics[width=\imsize]
                {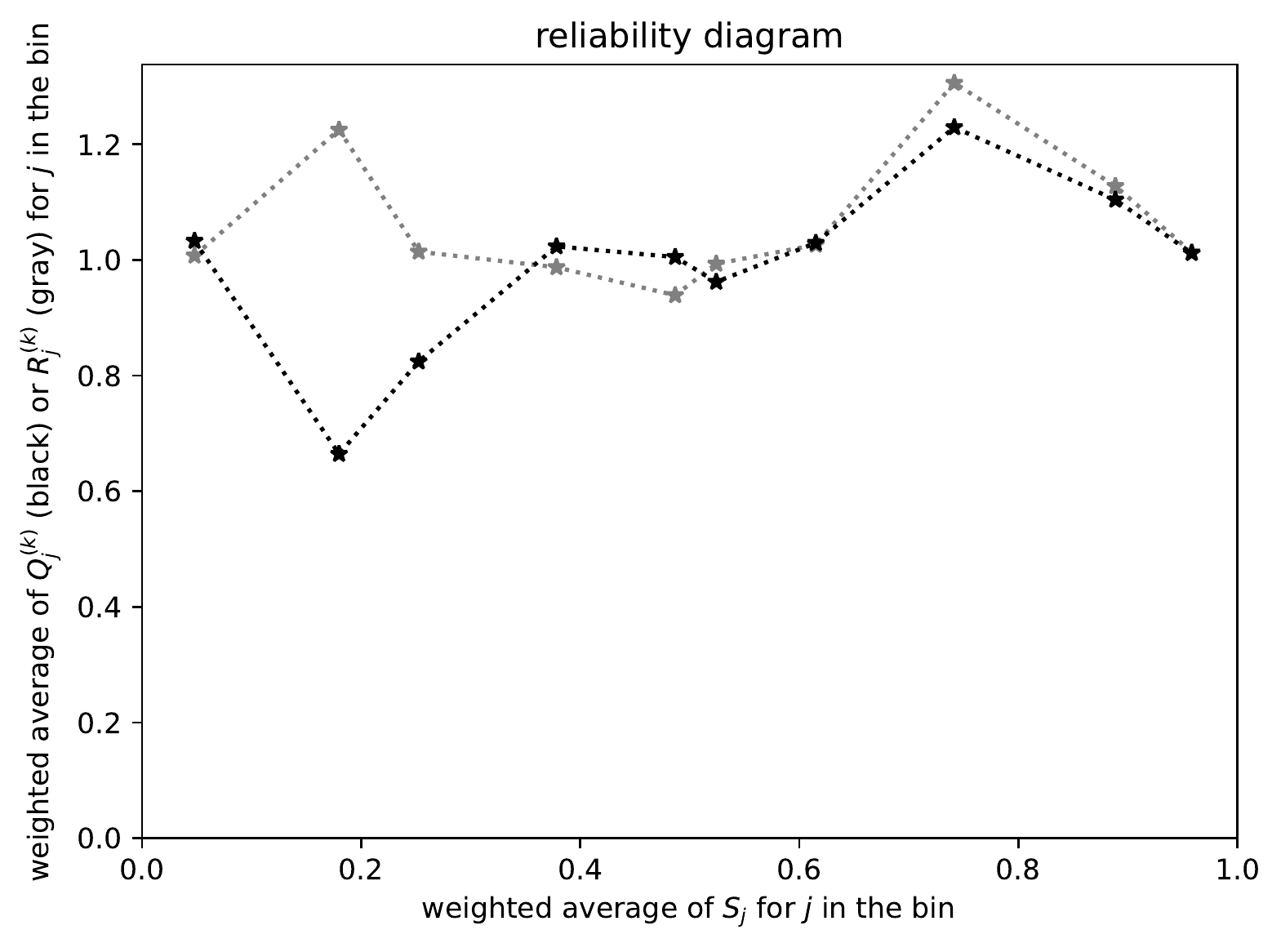}}
\quad\quad
\parbox{\imsize}{\includegraphics[width=\imsize]
                {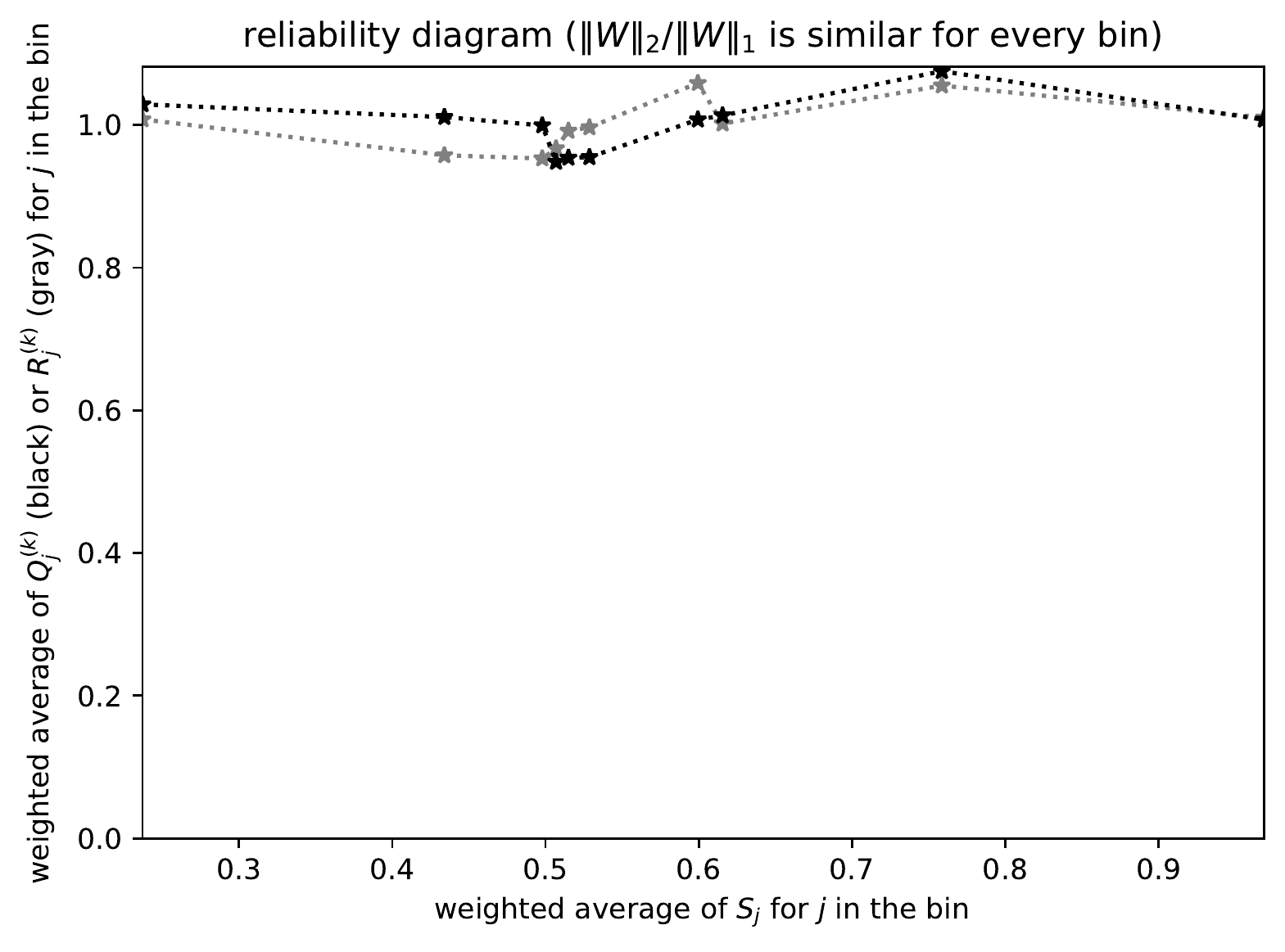}}

\vspace{\vertsep}

\parbox{\imsize}{\includegraphics[width=\imsize]
                {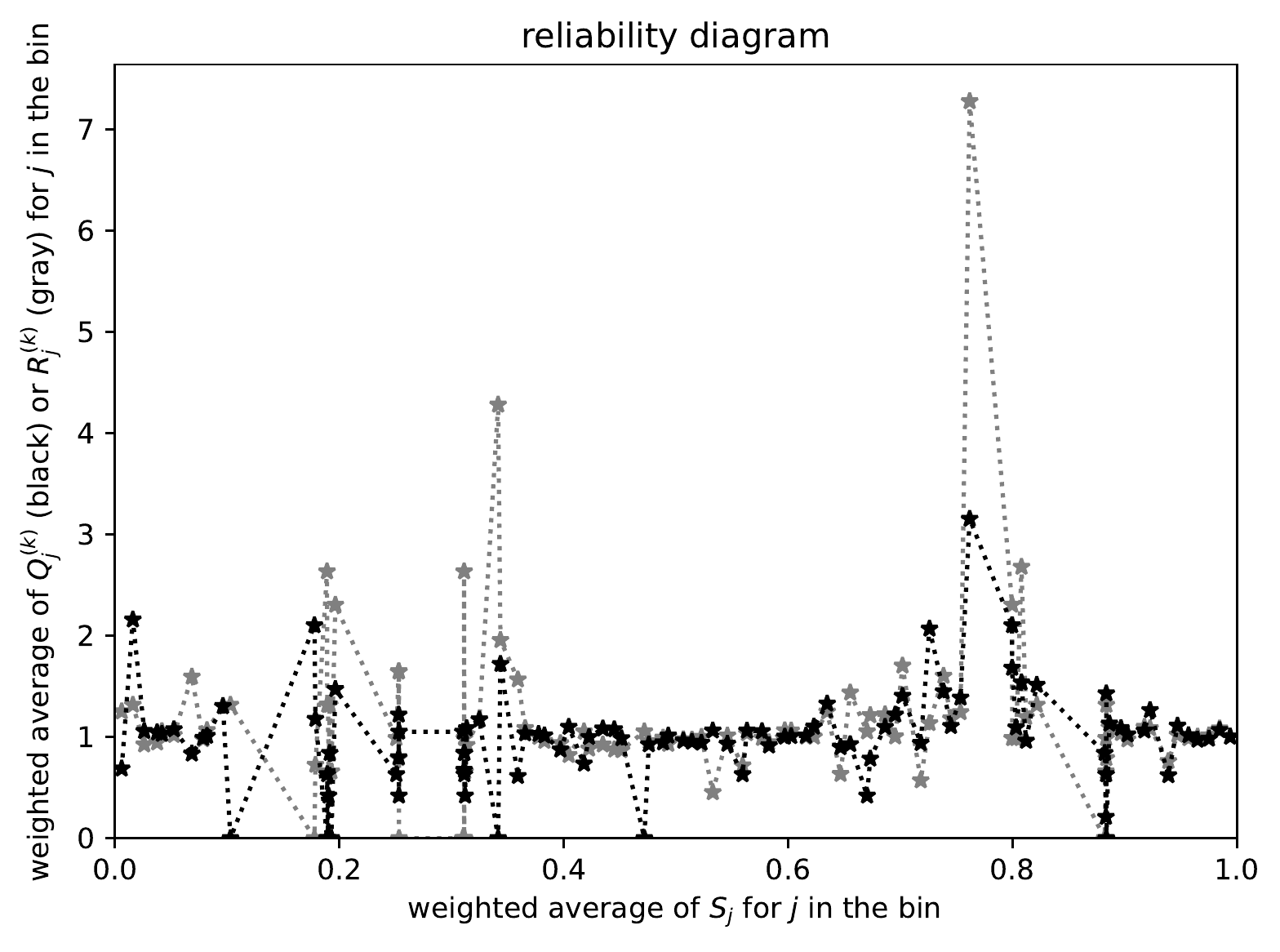}}
\quad\quad
\parbox{\imsize}{\includegraphics[width=\imsize]
                {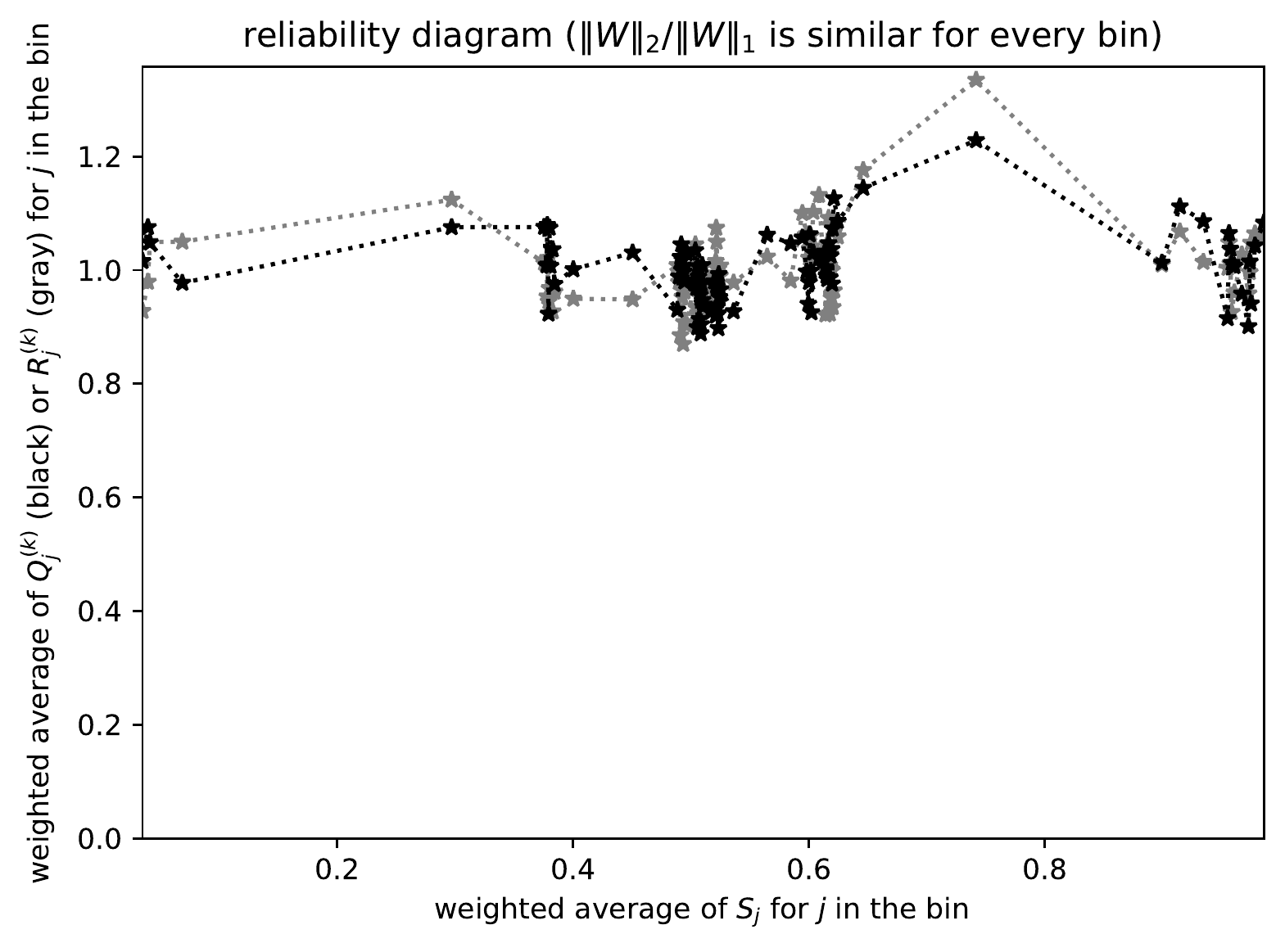}}

\end{centering}
\caption{$m = n =$ 63,826;
         Kuiper's statistic is $0.01723 / \sigma = 3.199$,
         Kolmogorov's and Smirnov's is $0.01338 / \sigma = 2.485$.
         This figure is the same as the previous two figures,
         Figures~\ref{ex02} and~\ref{ex20}, but now using all three covariates
         in the parameterization via the Hilbert curve.
         The covariates for the present figure are the normalized age
         of the donor, the normalized average fraction married
         in the Census block where the donor lives, and the normalized
         average household income in the Census block where the donor lives.
         The previous two figures omitted conditioning on the fraction married.
}
\label{ex012}
\end{figure}

\begin{figure}
\begin{centering}

\parbox{\imsize}{\includegraphics[width=\imsize]
                {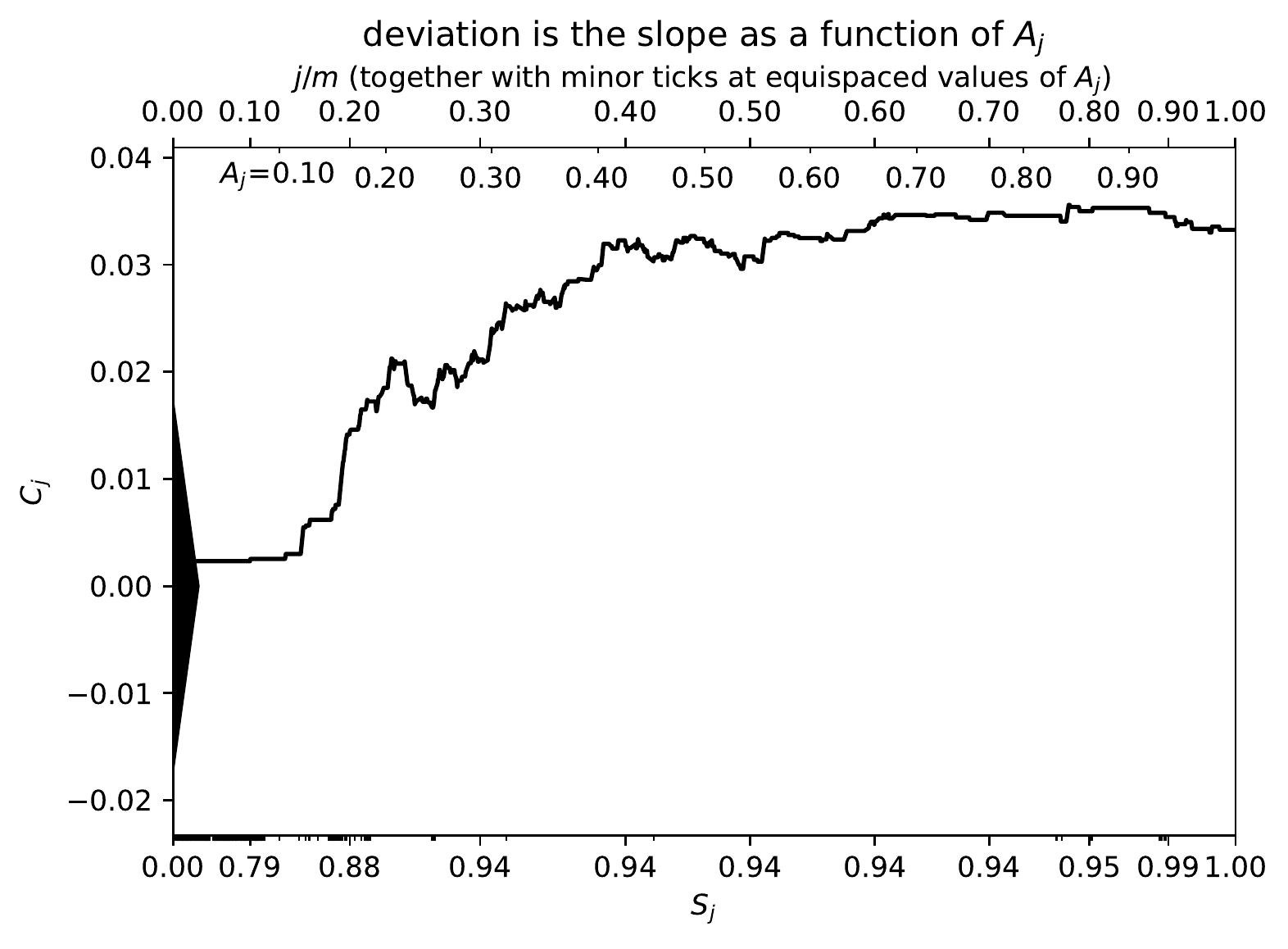}}
\quad\quad
\parbox{\imsize}{\includegraphics[width=\imsize]
                {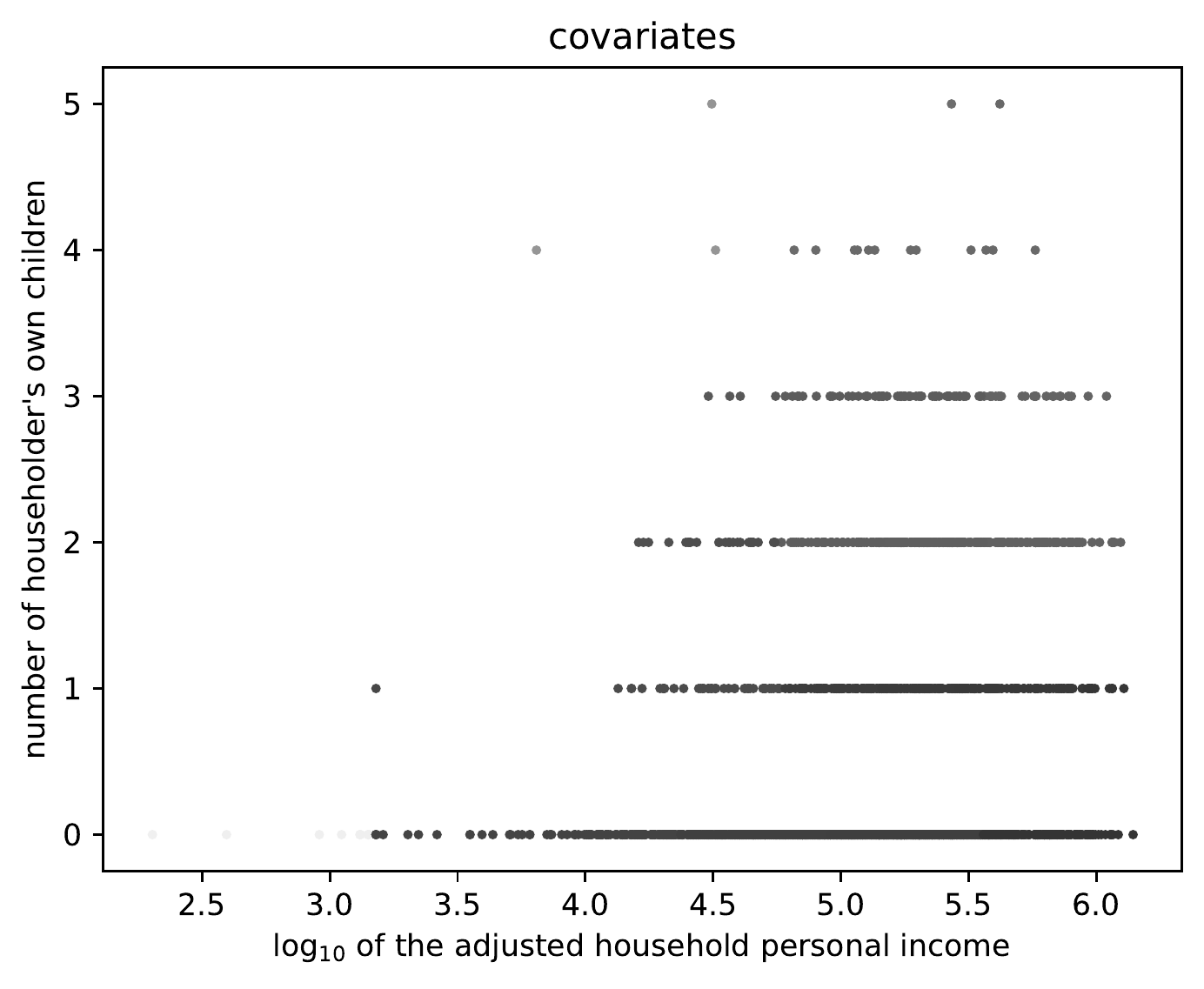}}

\vspace{\vertsep}

\parbox{\imsize}{\includegraphics[width=\imsize]
                {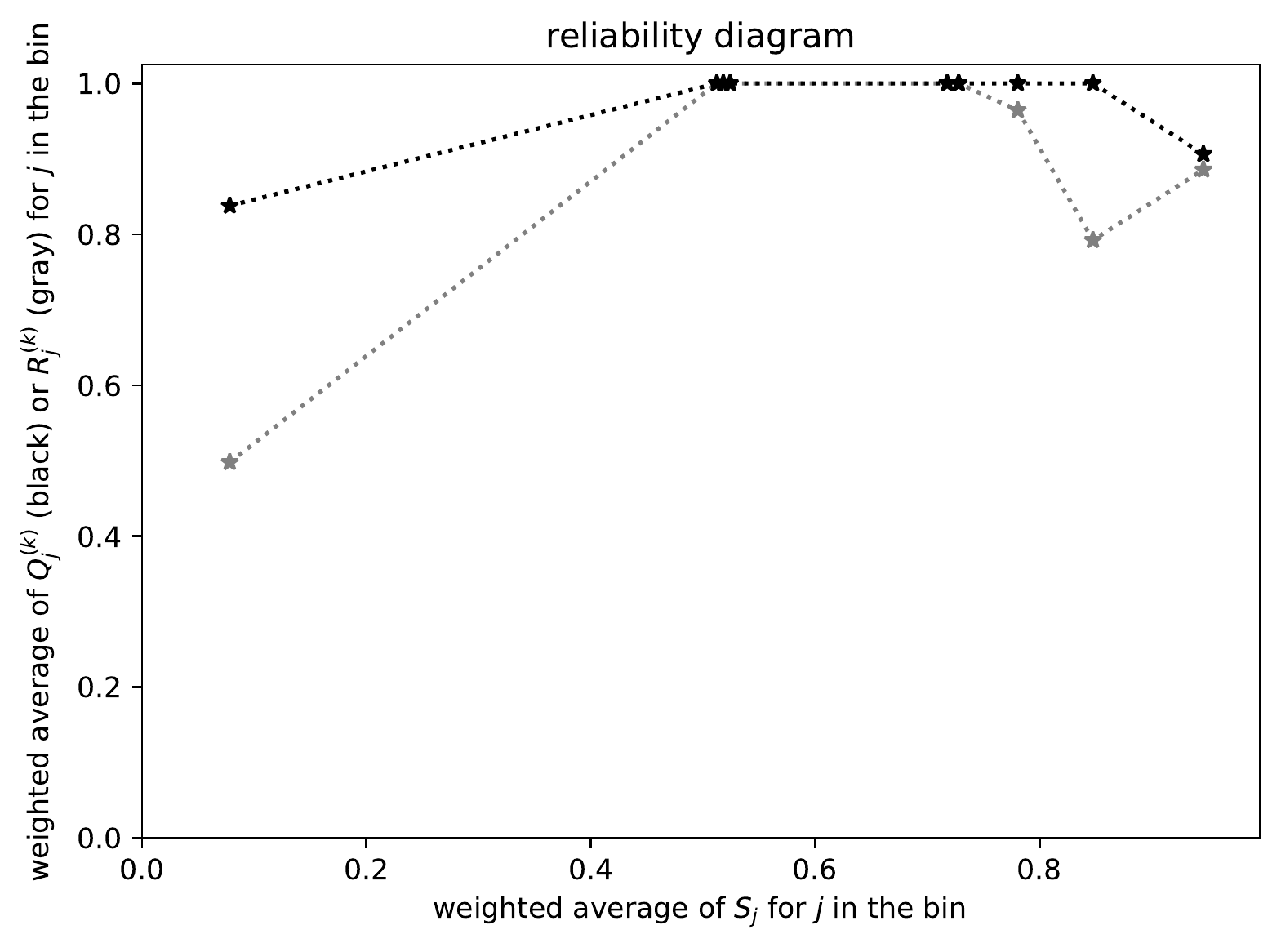}}
\quad\quad
\parbox{\imsize}{\includegraphics[width=\imsize]
                {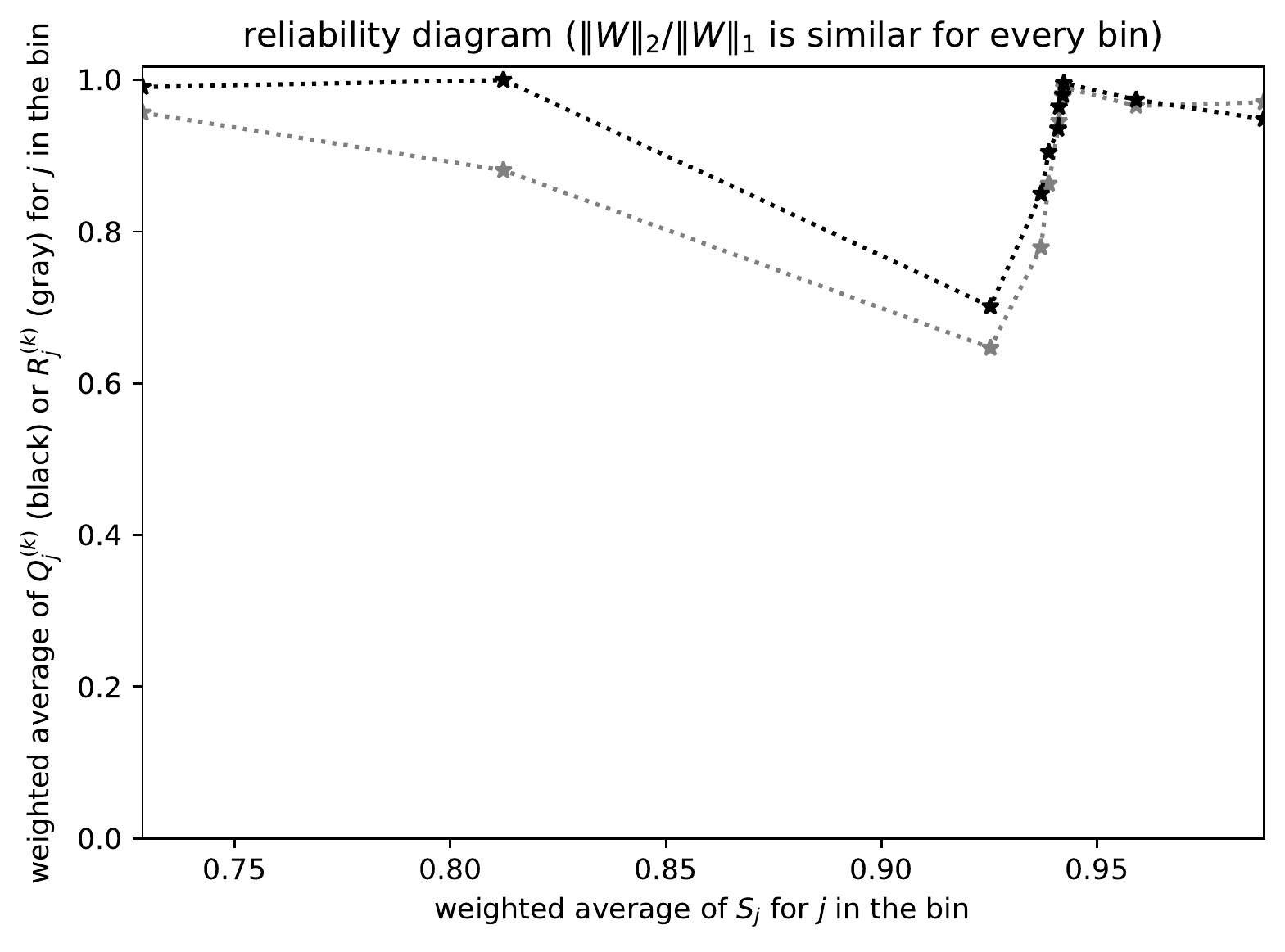}}

\vspace{\vertsep}

\parbox{\imsize}{\includegraphics[width=\imsize]
                {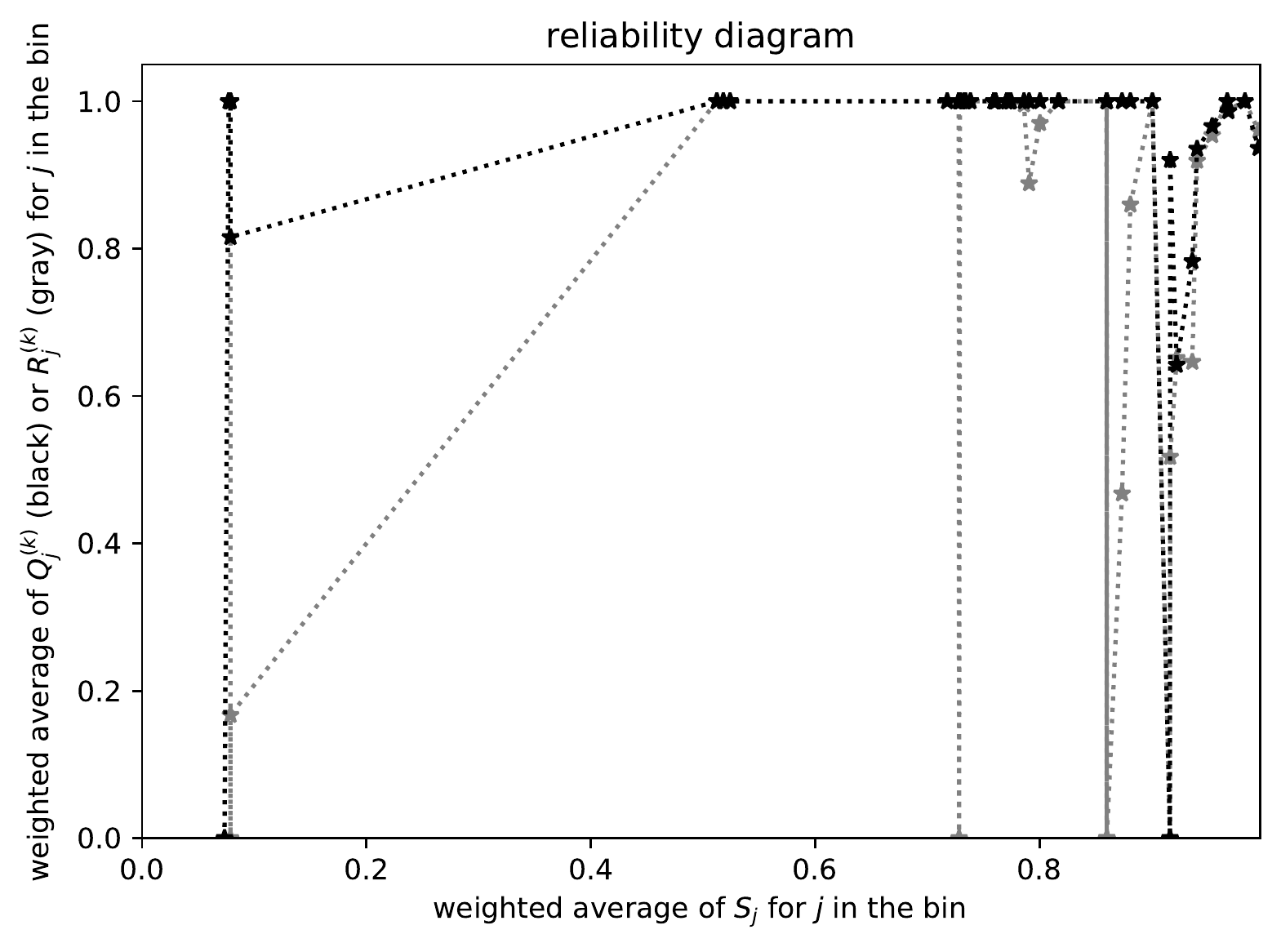}}
\quad\quad
\parbox{\imsize}{\includegraphics[width=\imsize]
                {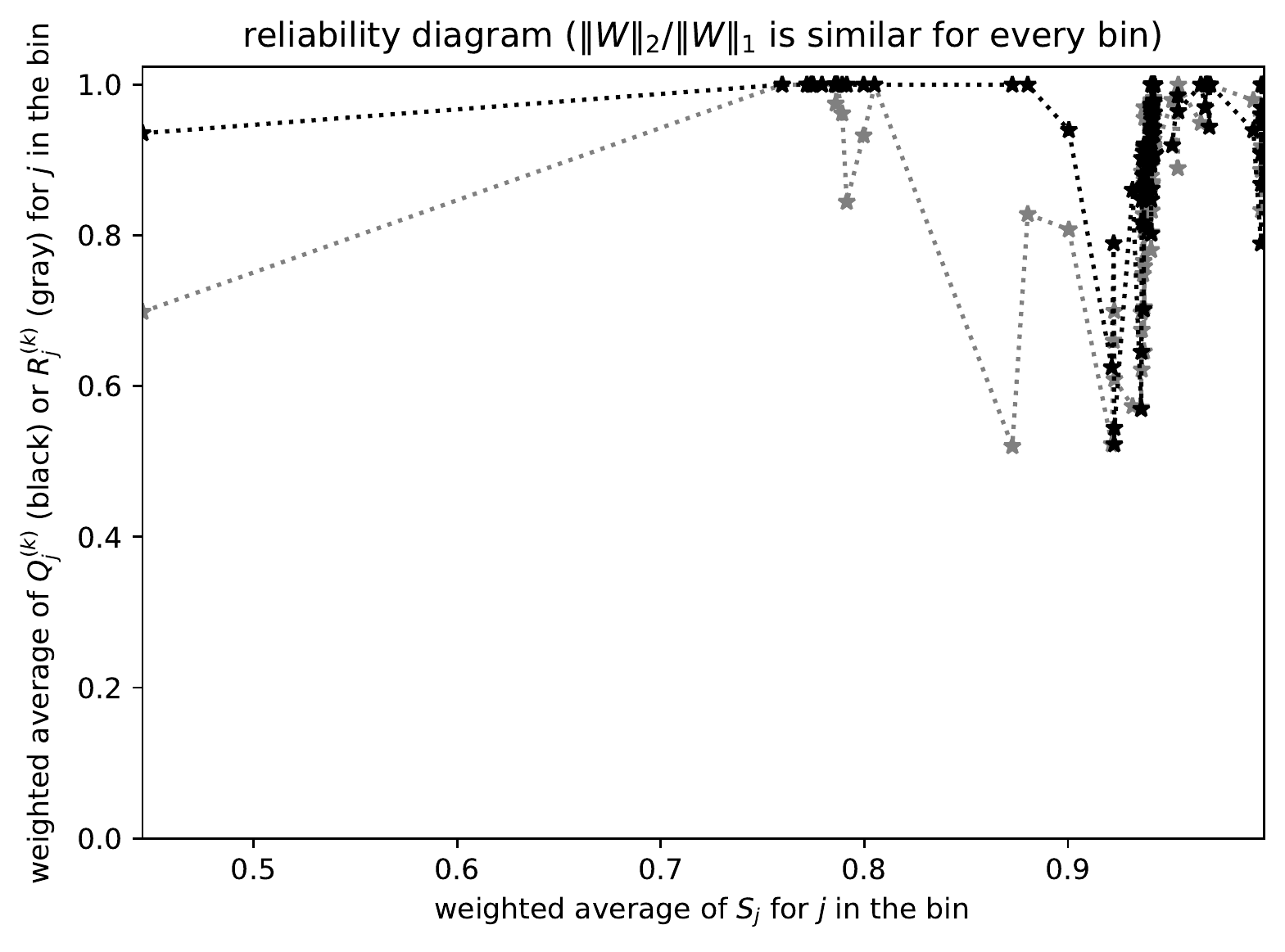}}

\end{centering}
\caption{County of San Mateo --- $m =$ 1,980, $n =$ 2,888;
         Kuiper's statistic is $0.03558 / \sigma = 3.97$,
         Kolmogorov's and Smirnov's is $0.03558 / \sigma = 3.97$.
         This figure uses two covariates in the parameterization
         via the Hilbert curve. The covariates are the number
         of the householder's own children and the logarithm
         of the adjusted household personal income, both normalized
         to range from 0 to 1 prior to parameterization via the Hilbert curve.
         The cumulative plot displays a stark jump in the range of scores
         from 0.8 to 0.92. The rightmost reliability diagrams
         display the same phenomenon, albeit less convincingly at first glance.
         This range of scores corresponds to lower-income households
         with multiple children (as easily discovered
         from the interactive versions of the plots).
}
\label{san_mateo}
\end{figure}

\begin{figure}
\captionsetup[subfigure]{justification=centering}
\begin{centering}

\subfloat[Napa][Napa\\ $m =$ 575, $n =$ 679\\
Kuiper = 0.05534 / $\sigma$ = 2.218\\
Kolmogorov-Smirnov = 0.05334 / $\sigma$ = 2.218]{
\includegraphics[width=\imsize]
{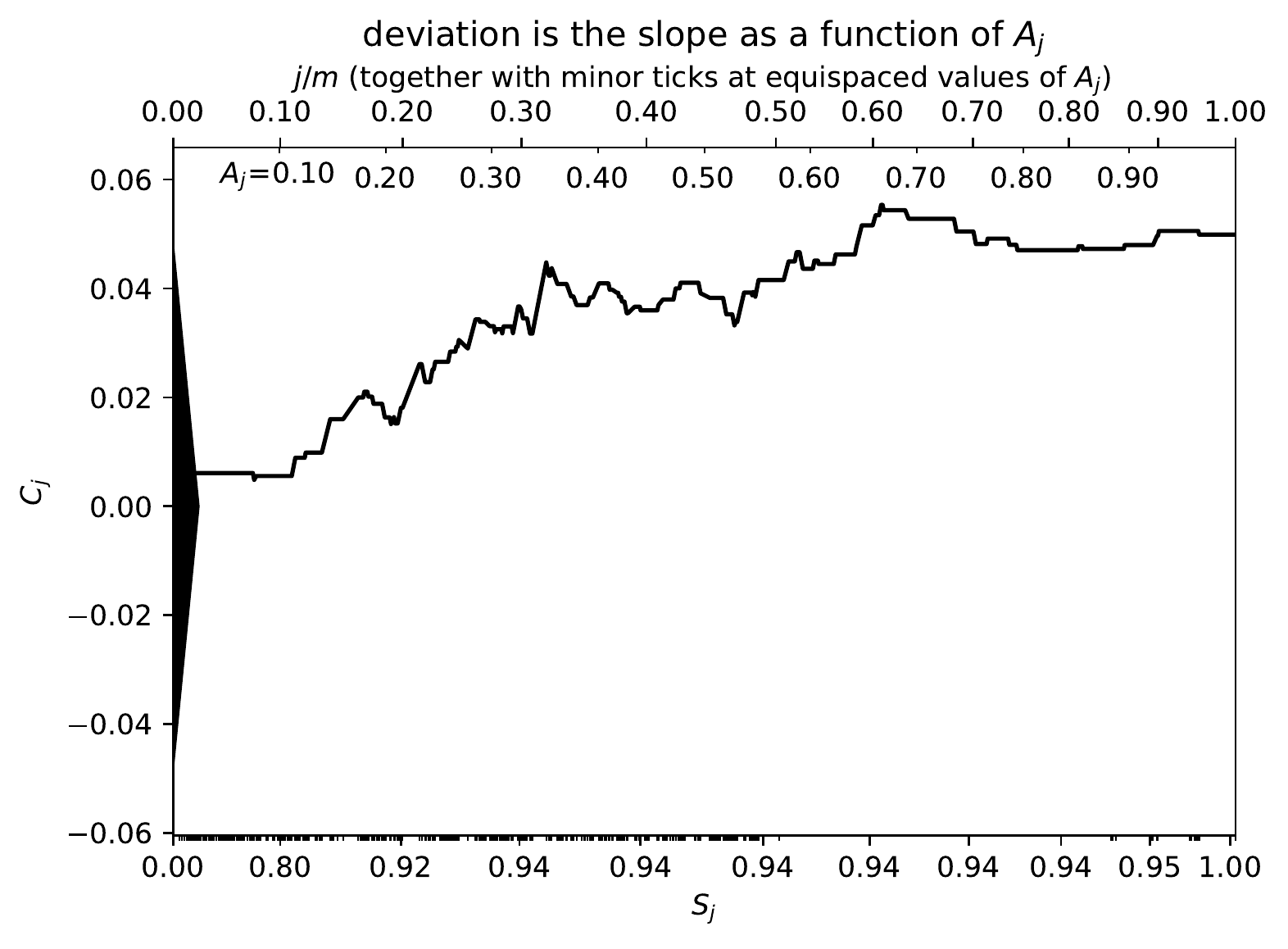}\quad}
\quad\quad
\subfloat[Santa Cruz][Santa Cruz\\ $m =$ 773, $n =$ 935\\
Kuiper = 0.05844 / $\sigma$ = 3.197\\
Kolmogorov-Smirnov = 0.05605 / $\sigma$ = 3.066]{
\includegraphics[width=\imsize]
{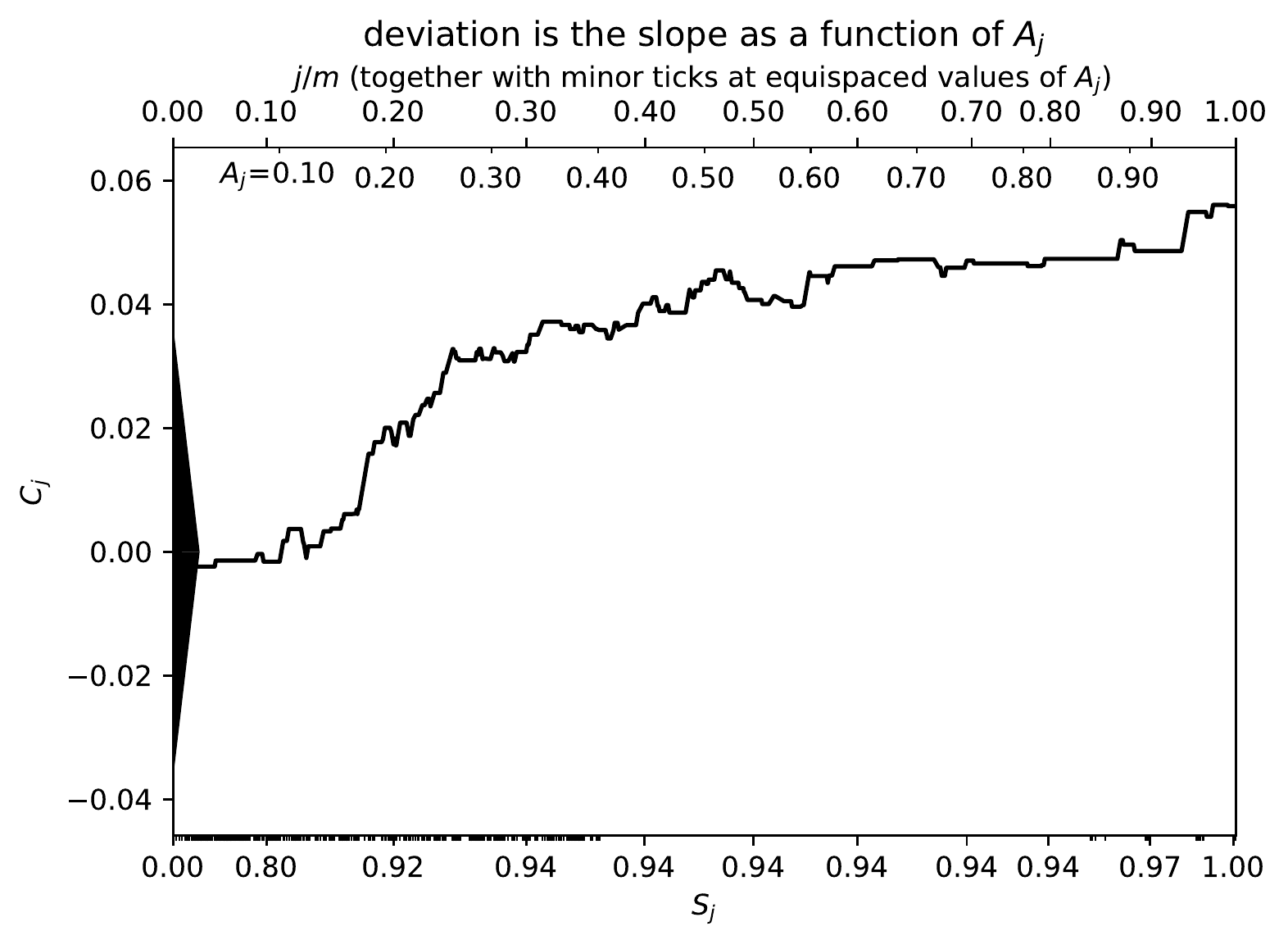}\quad}

\vspace{1.25\vertsep}

\subfloat[San Mateo][San Mateo\\ $m =$ 1,980, $n =$ 2,888\\
Kuiper = 0.03558 / $\sigma$ = 3.970\\
Kolmogorov-Smirnov = 0.03558 / $\sigma$ = 3.970]{
\includegraphics[width=\imsize]
{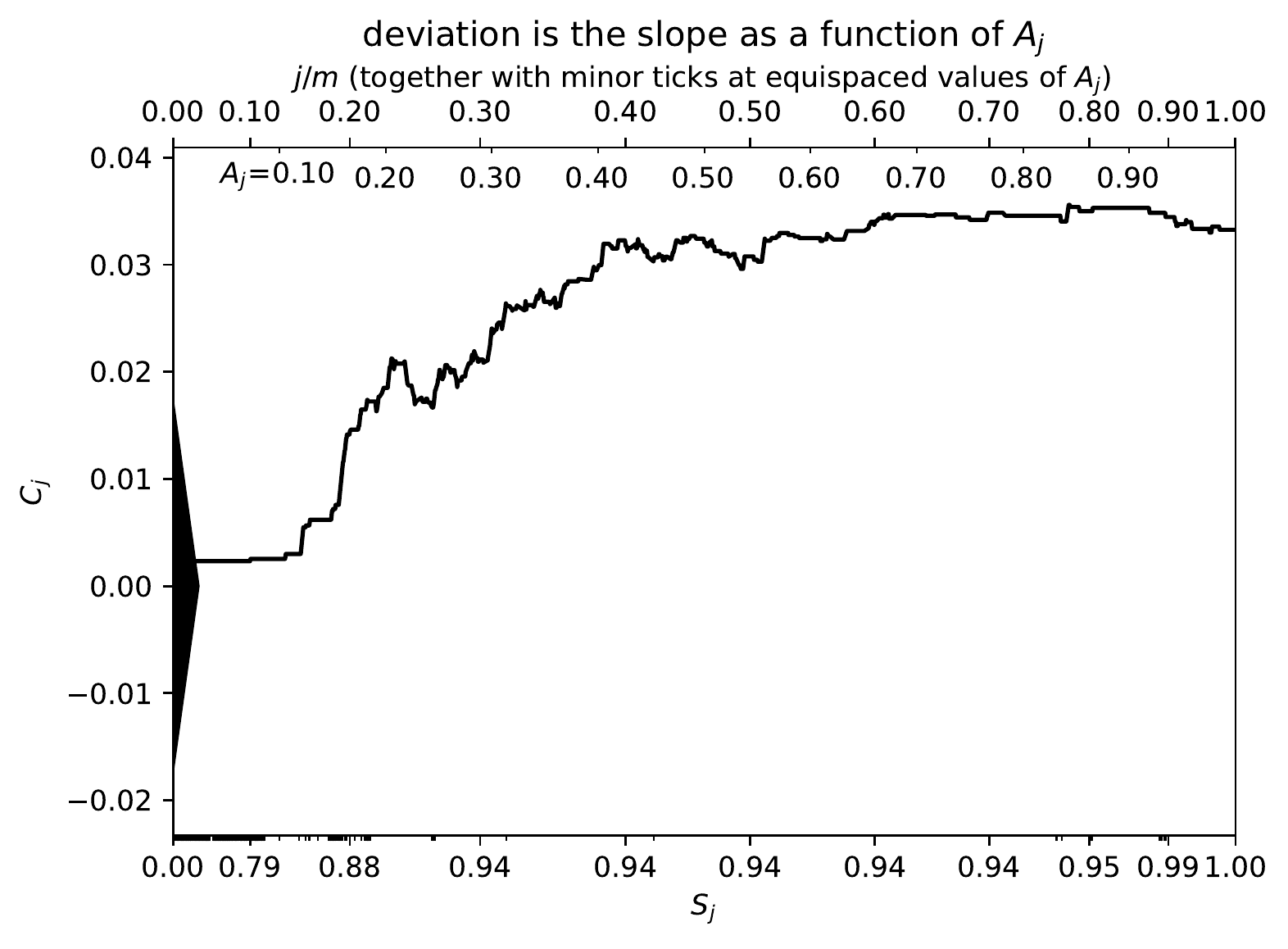}\quad}
\quad\quad
\subfloat[Santa Clara][Santa Clara\\ $m =$ 3,867, $n =$ 6,993\\
Kuiper = 0.03360 / $\sigma$ = 6.838\\
Kolmogorov-Smirnov = 0.03321 / $\sigma$ = 6.759]{
\includegraphics[width=\imsize]
{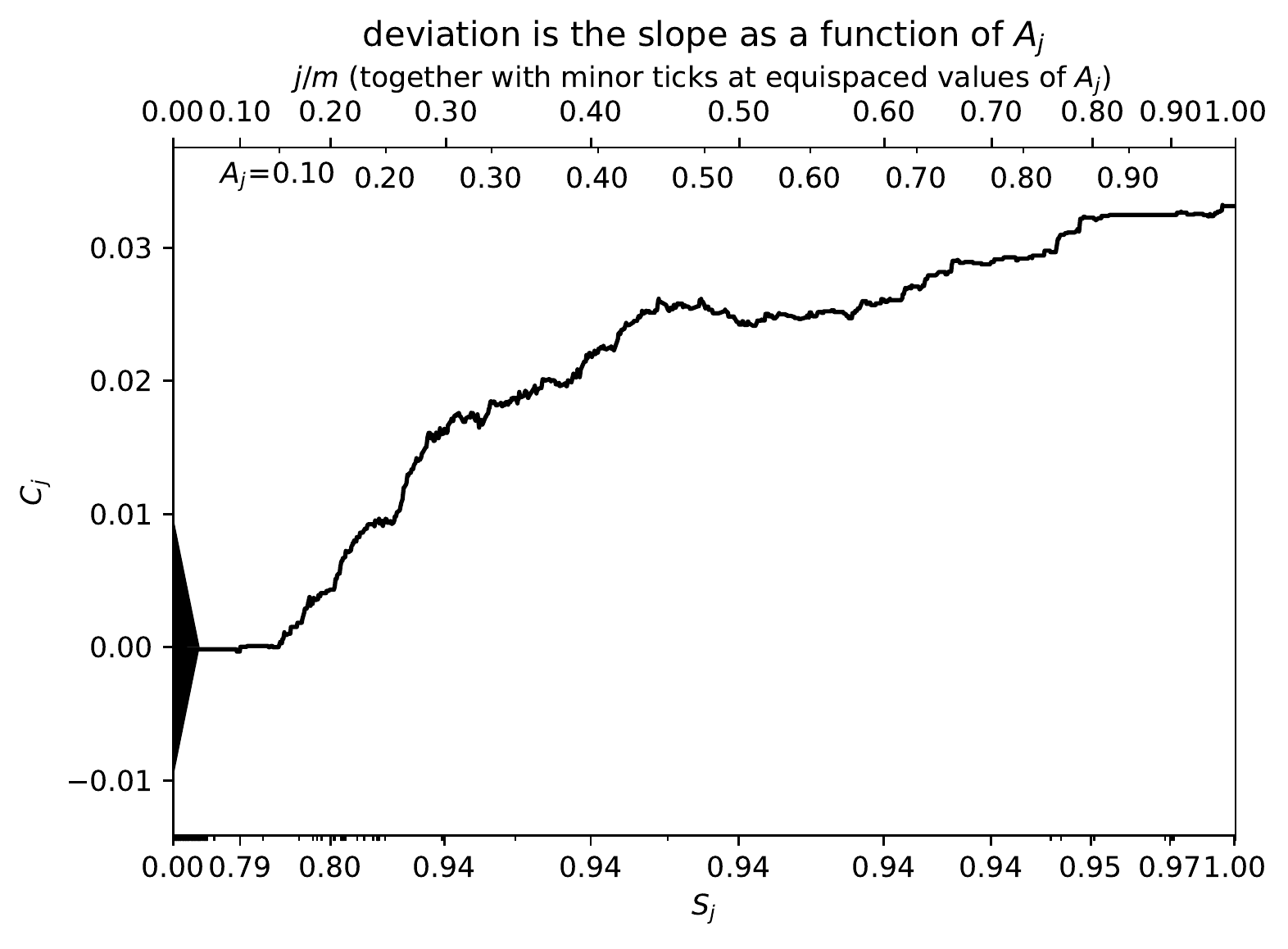}\quad}

\vspace{1.25\vertsep}

\subfloat[San Diego][San Diego\\ $m =$ 4,463, $n =$ 10,803\\
Kuiper = 0.05252 / $\sigma$ = 10.03\\
Kolmogorov-Smirnov = 0.05238 / $\sigma$ = 10.00]{
\includegraphics[width=\imsize]
{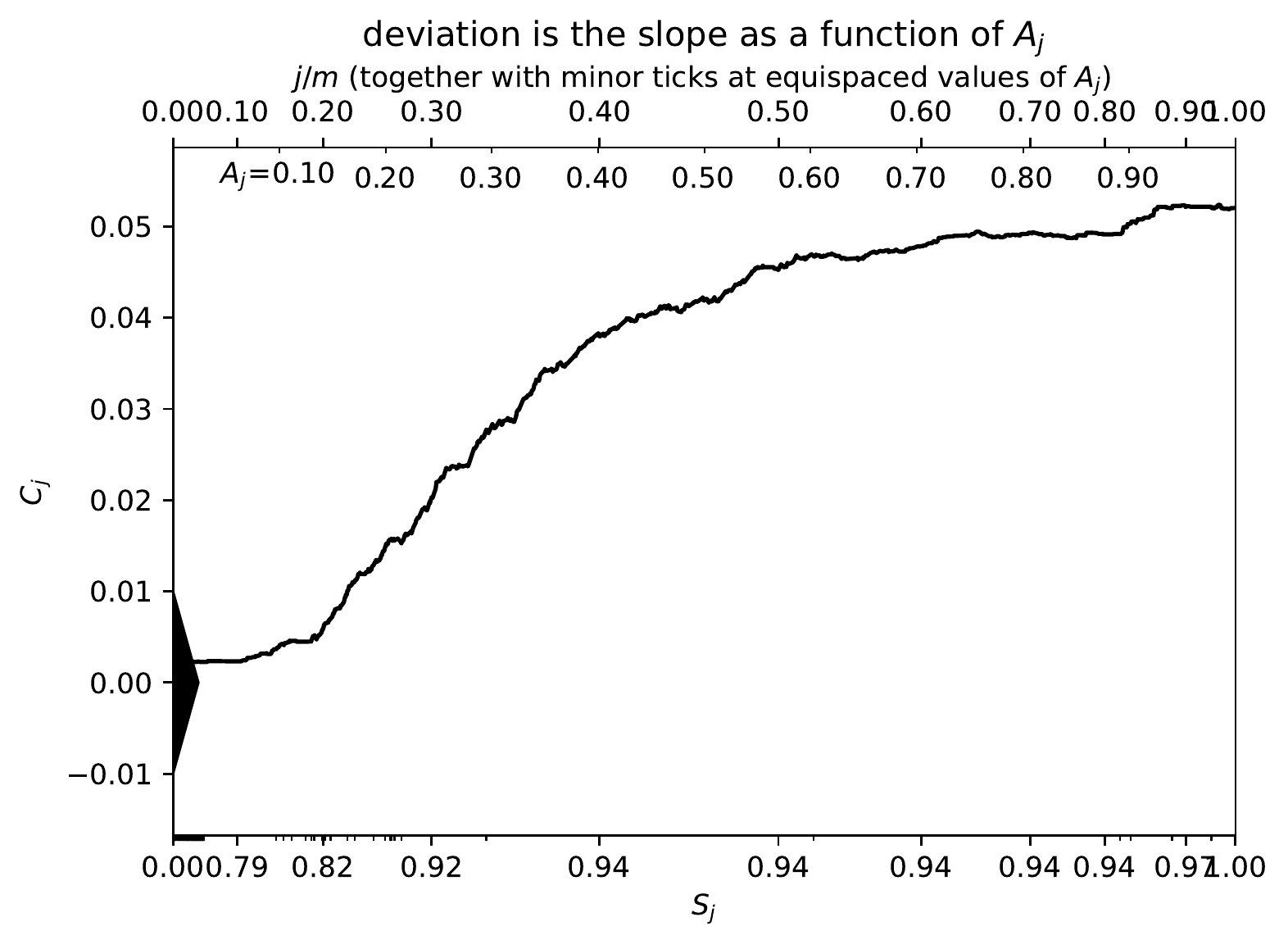}\quad}
\quad\quad
\subfloat[Los Angeles][Los Angeles\\ $m =$ 8,818, $n =$ 35,364\\
Kuiper = 0.1022 / $\sigma$ = 18.86\\
Kolmogorov-Smirnov = 0.1022 / $\sigma$ = 18.86]{
\includegraphics[width=\imsize]
{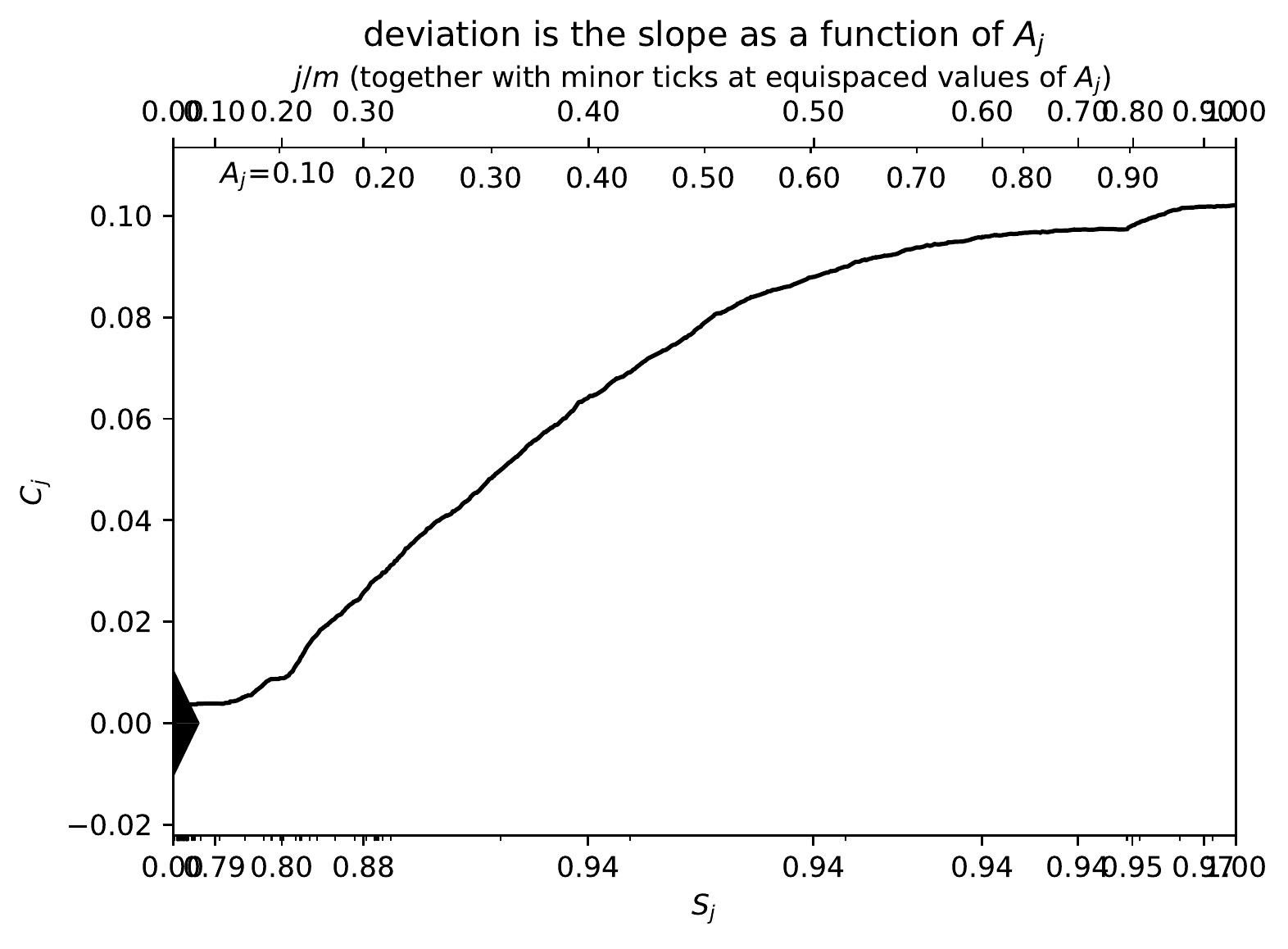}\quad}

\end{centering}

\vspace{.5\vertsep}

\caption{Counties of California.
         As in the previous figure, Figure~\ref{san_mateo},
         the present figure uses two covariates in the parameterization
         via the Hilbert curve. The covariates are the number
         of the householder's own children and the logarithm
         of the adjusted household personal income, both normalized
         to range from 0 to 1.
         The subfigures present the counties in the order of the sizes
         of their populations (which coincides with the order of their values
         for $m$ or $n$).
         The statistical significance of the prevalence of smartphones
         over laptop or desktop computers appears to be a monotonic function
         of the sample size; the Kuiper and Kolmogorov-Smirnov statistics,
         normalized by $\sigma$, increase monotonically from subfigures
         (a) to (b) to (c) to (d) to (e) to (f).
}
\label{counties}
\end{figure}

\clearpage

\appendix
\section{Review of space-filling curves}
\label{multidim}

This appendix reviews the properties and applications of Hilbert curves.
Hilbert curves reduce the problem of analyzing the differences in responses
while controlling for multiple covariates to the problem of analyzing
the differences while controlling for only a single scalar covariate.
The parts of the present paper prior to this appendix treat the latter problem
in detail.
This appendix reviews the reduction introduced by~\cite{tygert_multidim}.

Space-filling curves were introduced by~\cite{peano}.
A simplification by~\cite{hilbert} soon followed
and forms the basis for the method used in the present paper.
Related applications are detailed by~\cite{moon-jagadish-faloutsos-saltz}.

A space-filling curve in $p$ dimensions is a continuous function $h$
that maps the unit interval $(0, 1)$ onto the unit hypercube $(0, 1)^p$
(``onto'' means that the mapping is surjective, covering every point
in the hypercube).
Accompanying $h$ is a function $g$ mapping from the unit hypercube $(0, 1)^p$
to the unit interval $(0, 1)$ that is a right inverse for $h$, that is,
$h(g(x)) = x$ for any point $x$ in the unit hypercube $(0, 1)^p$.
As reviewed, for example, by~\cite{tygert_multidim},
$g$ cannot be continuous, unlike $h$.

The continuity of the function $h$ mapping from the unit interval $(0, 1)$
onto the unit hypercube $(0, 1)^p$ guarantees that,
for any points $t$ and $u$ from the unit interval $(0, 1)$,
if $t$ and $u$ are near to each other, then $h(t)$ and $h(u)$ are also close.
This ensures that, for any real-valued function $f$
on the unit hypercube $(0, 1)^p$,
local averages of $f \circ h$ will also be local averages of $f$;
here, ``local'' refers to proximity in the usual Euclidean metric
and $f \circ h$ denotes the composition of $f$ and $h$, that is,
$(f \circ h)(t) = f(h(t))$ for any point $t$ in the unit interval $(0, 1)$.
We can place a total order on the unit hypercube $(0, 1)^p$ by defining
that $x < y$ means $g(x) < g(y)$
for any two points $x$ and $y$ in the unit hypercube $(0, 1)^p$.
A local average of $f$ under this ordering
is also a local average of $f \circ h$ and thus is also a local average of $f$
under the usual Euclidean metric, as observed already
in the second sentence of this same paragraph.

\begin{figure}
\begin{centering}
\hfil\parbox{0.71\textwidth}
{\includegraphics[width=0.71\textwidth]{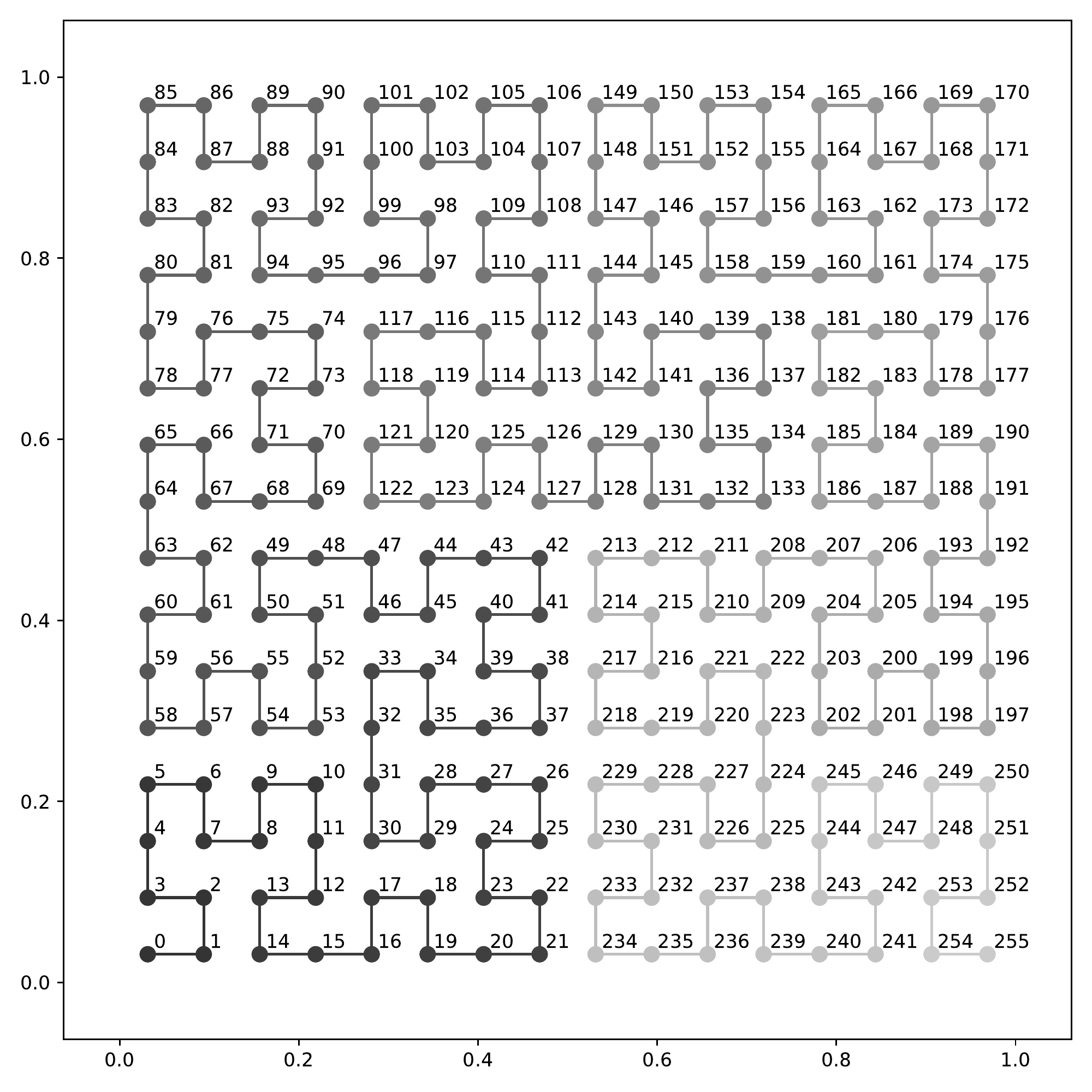}}
\end{centering}
\caption{An approximation with $255$ line segments to the Hilbert curve
in $p = 2$ dimensions; the numbers displayed near the points
specify their order in the Hilbert curve.}
\label{hilbert}
\end{figure}

A Hilbert curve $h$ is the particular type of space-filling curve
that orders points via a depth-first traversal
of the canonical $2^p$-ary (dyadic) tree on the unit hypercube $(0, 1)^p$.
Of course, the canonical dyadic tree in one dimension ($p = 1$)
is the binary tree, the canonical dyadic tree in two dimensions ($p = 2$)
is the quad-tree, and the canonical dyadic tree in three dimensions ($p = 3$)
is the oct-tree.
Figure~\ref{hilbert} approximates the image of a Hilbert curve
via $255$ line segments.
The full Hilbert curve is the limit of approximations
with increasingly many line segments, with every line segment
in an approximation being equally long as every other line segment
in the same approximation. Since the mapping from one dimension
to $p$ dimensions for each approximation is continuous
and the approximations converge uniformly, the limit --- the Hilbert curve ---
must exist and be continuous.
Our implementation uses an approximation with $2^{64} - 1$ line segments,
suitable for computation in 64-bit arithmetic with unsigned 8-byte integers.

To form the cumulative graphs and scalar summary statistics
of Section~\ref{methods} above, we need scalar scores.
To obtain the score associated with a given value of a vector of covariates,
we apply $g$ to the vector.
This reduces the problem of analyzing multiple covariates
to the problem of analyzing a single scalar covariate;
Section~\ref{methods} details a solution to the latter problem,
thus yielding a solution to the former problem (via the scalar scores
obtained from the Hilbert curve), too.
As discussed in the fourth paragraph of this appendix,
averaging responses locally at the scalar scores turns out
to average the same responses locally at the associated values
of the multiple covariates, where the ``associated values'' are those obtained
by applying $h$ to the scalar scores.
This ensures that the cumulative graphs and statistics
take meaningful averages.

(Only the ordering of the scores affects the shapes of the cumulative graphs
and the values of the scalar summary statistics.
For the numerical values of the scores used to label the lower horizontal axes,
we normalize the results to range from 0 to 1.
Such normalization has no effect on the ordering of the scores.
Specifically, we apply an affine transformation --- linear plus a constant ---
first subtracting the minimum of the results of $g$
applied to all observed vectors of covariates
and then dividing by the original maximum minus the original minimum.)

\clearpage

\bibliography{paired}

\begin{thebibliography}{}

\bibitem[\protect\citename{Austin, }2011]{austin}
Austin, Peter~C. 2011.
\newblock An introduction to propensity score methods for reducing the effects
  of confounding in observational studies.
\newblock {\em Multivar. Behav. Res.}, {\bf 43}(3), 399--424.

\bibitem[\protect\citename{Bai {\em et~al.}, }2022]{bai-romano-shaikh}
Bai, Yuehao, Romano, Joseph~P., \& Shaikh, Azeem~M. 2022.
\newblock Inference in experiments with matched pairs.
\newblock {\em J. Amer. Statist. Assoc.}, {\bf 117}(540), 1726--1737.

\bibitem[\protect\citename{Br\"ocker \& Smith, }2007]{brocker-smith}
Br\"ocker, Jochen, \& Smith, Leonard~A. 2007.
\newblock Increasing the reliability of reliability diagrams.
\newblock {\em Weather Forecast.}, {\bf 22}(3), 651--661.

\bibitem[\protect\citename{Brown \& Hollander, }2007]{brown-hollander}
Brown, Jr., Byron~William, \& Hollander, Myles. 2007.
\newblock {\em Statistics: A Biomedical Introduction}.
\newblock Wiley Classics Library.
\newblock Wiley.

\bibitem[\protect\citename{Caliendo \& Kopeinig, }2008]{caliendo-kopeinig}
Caliendo, Marco, \& Kopeinig, Sabine. 2008.
\newblock Some practical guidance for the implementation of propensity score
  matching.
\newblock {\em J. Econ. Surv.}, {\bf 22}(1), 32--72.

\bibitem[\protect\citename{Cox \& Reid, }2000]{cox-reid}
Cox, David~R., \& Reid, Nancy. 2000.
\newblock {\em The Theory of the Design of Experiments}.
\newblock Monographs on Statistics and Applied Probability, vol. 86.
\newblock Chapman \& Hall / CRC Press.

\bibitem[\protect\citename{Delgado, }1993]{delgado}
Delgado, Miguel~A. 1993.
\newblock Testing the equality of nonparametric regression curves.
\newblock {\em Statist. Probab. Lett.}, {\bf 17}(3), 199--204.

\bibitem[\protect\citename{Draper \& Smith, }1998]{draper-smith}
Draper, Norman~R., \& Smith, Harry. 1998.
\newblock {\em Applied Regression Analysis}. 3rd edn.
\newblock Wiley.

\bibitem[\protect\citename{Hilbert, }1891]{hilbert}
Hilbert, David. 1891.
\newblock Ueber die stetige abbildung einer line auf ein fl\"achenst\"uck.
\newblock {\em Math. Ann.}, {\bf 38}(3), 459--460.

\bibitem[\protect\citename{Hollander {\em et~al.},
  }2014]{hollander-wolfe-chicken}
Hollander, Myles, Wolfe, Douglas~A., \& Chicken, Eric. 2014.
\newblock {\em Nonparametric Statistical Methods}. 3rd edn.
\newblock Wiley.

\bibitem[\protect\citename{Hosmer {\em et~al.},
  }2013]{hosmer-lemeshow-sturdivant}
Hosmer, Jr., David~W., Lemeshow, Stanley, \& Sturdivant, Rodney~X. 2013.
\newblock {\em Applied Logistic Regression}. 3rd edn.
\newblock Wiley.

\bibitem[\protect\citename{Ibarra {\em et~al.},
  }2022]{arrieta-ibarra-gujral-tannen-tygert-xu}
Ibarra, Imanol~Arrieta, Gujral, Paman, Tannen, Jonathan, Tygert, Mark, \& Xu,
  Cherie. 2022.
\newblock Metrics of calibration for probabilistic predictions.
\newblock {\em J. Mach. Learn. Res.}, {\bf 23}, 1--54.

\bibitem[\protect\citename{Kuiper, }1962]{kuiper}
Kuiper, Nicolaas~H. 1962.
\newblock Tests concerning random points on a circle.
\newblock {\em Proc. Koninklijke Nederlandse Akademie van Wetenschappen Series
  A}, {\bf 63}, 38--47.

\bibitem[\protect\citename{Moon {\em et~al.},
  }2001]{moon-jagadish-faloutsos-saltz}
Moon, Bongki, Jagadish, H.~V., Faloutsos, Christos, \& Saltz, Joel~H. 2001.
\newblock Analysis of the clustering properties of the {H}ilbert space-filling
  curve.
\newblock {\em IEEE Trans. Knowl. Data Eng.}, {\bf 13}(1), 124--141.

\bibitem[\protect\citename{Peano, }1890]{peano}
Peano, Giuseppe. 1890.
\newblock Sur une courbe, qui remplit toute une aire plane.
\newblock {\em Math. Ann.}, {\bf 36}(1), 157--160.

\bibitem[\protect\citename{Rubin, }1973]{rubin}
Rubin, Donald~B. 1973.
\newblock Matching to remove bias in observational studies.
\newblock {\em Biometrics}, {\bf 29}, 159--183.

\bibitem[\protect\citename{Tygert, }2021a]{tygert_multidim}
Tygert, Mark. 2021a.
\newblock {\em Controlling for multiple covariates}.
\newblock Tech. rept. 2112.00672. arXiv.
\newblock Available at \url{https://arxiv.org/abs/2112.00672}.

\bibitem[\protect\citename{Tygert, }2021b]{tygert_full}
Tygert, Mark. 2021b.
\newblock Cumulative deviation of a subpopulation from the full population.
\newblock {\em J. Big Data}, {\bf 8}(117), 1--60.
\newblock Available at \url{https://arxiv.org/abs/2008.01779}.

\bibitem[\protect\citename{Tygert, }2021c]{tygert_two}
Tygert, Mark. 2021c.
\newblock A graphical method of cumulative differences between two
  subpopulations.
\newblock {\em J. Big Data}, {\bf 8}(158), 1--29.
\newblock Available at \url{https://arxiv.org/abs/2108.02666}.

\bibitem[\protect\citename{Tygert, }2023]{tygert_pvals}
Tygert, Mark. 2023.
\newblock Calibration of {P}-values for calibration and for deviation of a
  subpopulation from the full population.
\newblock {\em Adv. Comput. Math.}, {\bf 49}(70), 1--22.
\newblock Available at \url{https://arxiv.org/abs/2202.00100}.

\bibitem[\protect\citename{Wilcoxon, }1945]{wilcoxon}
Wilcoxon, Frank. 1945.
\newblock Individual comparisons by ranking methods.
\newblock {\em Biometrics Bulletin}, {\bf 1}(6), 80--83.

\end{thebibliography}
\bibliographystyle{authordate1}

\end{document}